\documentclass[aps,prl,reprint,amsmath,amssymb]{revtex4-2}
\usepackage{bm}
\usepackage[dvipsnames]{xcolor}
\usepackage{graphicx}
\usepackage{hyperref} 
\usepackage{cleveref} 
\newcommand{\sectionprl}[1]{{\par\it #1.---}}

\begin{document}

\title{Inverse thermodynamic uncertainty relation and entropy production}
\author{Van Tuan Vo}
\email{vantuanvo.vtv@gmail.com}
\affiliation{Department of Physics, Kyoto University, Kyoto 606-8502, Japan}
\author{Andreas Dechant}
\email{andreas.dechant@outlook.com}
\affiliation{Department of Physics, Kyoto University, Kyoto 606-8502, Japan}
\author{Keiji Saito}
\email{keiji.saitoh@scphys.kyoto-u.ac.jp}
\affiliation{Department of Physics, Kyoto University, Kyoto 606-8502, Japan}

\begin{abstract}
Nonequilibrium current fluctuations represent one of the central topics in nonequilibrium physics. The thermodynamic uncertainty relation (TUR) is widely acclaimed for rigorously establishing a lower bound on current fluctuations, expressed in terms of the entropy production rate and the average current. In this study, we focus on an upper bound for the fluctuations, referred to as the inverse thermodynamic uncertainty relation (iTUR). We derive a universal iTUR expression in terms of the entropy production rate for continuous-variable systems governed by overdamped Langevin equations, as well as for discrete-variable systems described by Markov jump processes. The iTUR establishes a no-go theorem prohibiting perpetual superdiffusion in systems with a finite entropy production rate and a finite spectral gap. The divergence of the variance of any current becomes possible only when the spectral gap of the symmetrized time-evolution operator closes or the entropy production rate diverges. As a relevant experimental scenario, we apply the iTUR to the phenomenon of giant diffusion, emphasizing the pivotal roles of the spectral gap and entropy production.
\end{abstract}

\maketitle

\sectionprl{Introduction}
Understanding nonequilibrium fluctuations is a cornerstone of nonequilibrium statistical physics. Among the numerous breakthroughs in the field of stochastic thermodynamics \cite{Sekimoto.2010,Seifert.2008.EPJB}, the thermodynamic uncertainty relation (TUR) stands out as a landmark result, establishing a fundamental constraint on nonequilibrium current fluctuations and entropy production \cite{Barato_PRL_2015, Gingrich_PRL_2016, Pietzonka_PRE_2016, Koyuk_PRL_2020, Dechant_PhysRevX_2021, Liu_PRL_2020, Pietzonka_PRL_2022, Hasegawa_PhysRevLett_2020, Miller_PhysRevLett_2021, Tan_PhysRevLett_2022, Barato.2019.JSM, Chun.2019.PRE, Dechant.2018.JSM, Dechant.2019.JPA, Dechant.2020.PNAS, Dechant.2021.PRR,Gupta.2020.EPJB, Hiura.2021.PRE, Koyuk.2019.JPA, Manikandan.2018.JPA, Park.2021.PRR, Pietzonka.2017.PRE, Proesmans.2017.EPL, Wolpert.2020.PRL, Kalaee.2021.PRE, brandner2018thermodynamic, Macieszczak.2018.PRL, Agarwalla.2018.PRB, Menczel.2021.JPA, Potts.2019.PRE, Saryal.2019.PRE}. The TUR is valid for overdamped dynamics, and provides critical insights and applications in nonequilibrium thermodynamics \cite{Polettini2016PRE, Gingrich_2017, Pietzonka.2018.PRL,Pietzonka.2016.JSM}, such as a trade-off between precision and thermodynamic cost, and a measurable lower bound for entropy production, inspiring studies aimed at inferring entropy production from observable quantities \cite{Seifert.2019.ARCMP,Li.2019.NC,Manikandan.2020.PRL,Martinez.2019.NC,Otsubo.2020.PRE,van2022thermodynamic}. In particular, when interpreted as a lower bound on nonequilibrium fluctuations, the TUR takes the following form \cite{Barato_PRL_2015, Gingrich_PRL_2016, Pietzonka_PRE_2016}:
\begin{align}
D_{\infty} \geq \frac{j_{\rm st}^2}{\sigma_{\text{st}}}\, , \label{tur-first}
\end{align}
where $D_{\infty}$ is the long-time fluctuations of the current defined as in the same way as the diffusion coefficient (detailed below in Eq.~\eqref{eq:diffusion_coefficient}), $j_{\rm st}$ is its steady state average, and $\sigma_{\text{st}}$ is the entropy production rate. 

Beyond the TUR's lower bound on nonequilibrium fluctuations, a significant advancement has recently been achieved in identifying the upper bound on the fluctuations. Ref.~\cite{Garrahan_PRL_2023} introduced such a constraint---dubbed the inverse Thermodynamic Uncertainty Relation (iTUR)---which depends solely on the instantaneous fluctuations and the spectral gap of the symmetrized rate matrix, without involving entropy production. This complementary result has been extended to the quantum regime for counting observables and their first passage times \cite{vanvu2024, bakewellsmith2024}.
 However, existing upper bounds, which are limited to discrete systems, do not explicitly account for the role of thermodynamic cost. Nonequilibrium driving can actively enhance fluctuations beyond their equilibrium limits, such as in a phenomenon exemplified by giant diffusion in tilted periodic potentials \cite{Reimann_PRL_2001, Reimann_PRE_2002,Marchenko_PRE_2023}. Such enhancement necessarily entails dissipation of energy and is inextricably linked to the entropy production required to maintain the system away from equilibrium. These observations give rise to three pressing questions that motivate our work: Can we upper-bound the fluctuations in terms of entropy production to elucidate the role of thermodynamic costs? Is such an iTUR applicable to continuous-variable systems governed by overdamped Langevin dynamics? Can we use iTUR to obtain new insights into the mechanisms of physical phenomena?

In this study, we address these questions. Employing the variational principle for the cumulant generating function of the current, we derive a concise expression of the iTUR that incorporates the entropy production rate for continuous variables (Eqs. \eqref{eq:giant_diffusion_bound} and \eqref{eq:long_time_bound} below), and for discrete variables (Eq. \eqref{eq:current_iTUR}). These iTURs elucidate the interplay between the entropy production rate and the spectral gap in governing current fluctuations. Large current fluctuations necessitate either a high entropy production rate or a small spectral gap. In particular, the iTUR establishes a no-go theorem on perpetual superdiffusion in systems with a finite entropy production rate and a non-zero spectral gap. Furthermore, the relations show that in equilibrium states, instantaneous fluctuations ($D_0$ in Eq.~\eqref{eq:diffusion_coefficient}) always exceed the long-time fluctuations ($D_{\infty}$). 

We apply the iTUR to the phenomenon of giant diffusion observed in particles subjected to nonconservative forces within a periodic potential \cite{Reimann_PRL_2001, Reimann_PRE_2002,Marchenko_PRE_2023}. This phenomenon arises from the competition between local diffusion near potential minima and transitions across potential barriers driven by nonconservative forces, leading to a significant enhancement of the long-time diffusion constant at a critical nonconservative force. Experimentally, giant diffusion has been observed in systems such as biological rotary motors, such as F1-ATPase \cite{Hayashi_PRL_2015, Shinagawa2016}, colloidal particles \cite{Lee_PRL_2006} and DNA molecules \cite{Kim_PRL_2017, Suma_PRL_2023}. The iTUR provides a perspective on this phenomenon through the interplay of entropy production rate and spectral gap. 
Driving the system further out of equilibrium through an increased entropy production rate consistently enhances diffusion.
This enhancement becomes more pronounced when the spectral gap is small, allowing fluctuations to persist longer. 
Our results illuminate how nonequilibrium driving enhances diffusion within the constraints of entropy production and relaxation properties.

The iTURs presented in this paper offer profound insights into the physical principles underlying nonequilibrium current fluctuations, framed through the lens of thermodynamic costs, particularly entropy production. Furthermore, it holds significant potential for applications that span a broad spectrum of nonequilibrium physical phenomena.

\sectionprl{Setup for continuous variables}
We consider particles undergoing overdamped Langevin dynamics in $ \mathbb{R}^n $ with a uniform temperature, described by the following stochastic differential equation
\begin{equation}
\dot{\bm{x}}_t = \bm{A}(\bm{x}_t) + \sqrt{2\bm{B}}\, \bm{\xi}(t) \,,
\label{eq:langevin}
\end{equation}
where $\dot{\bm{x}}_t $ denotes the velocity vector of the particle, $ \bm{A}(\bm{x}) $ is the drift vector field, $ \bm{B} $ is a positive-definite diffusion matrix, and $ \boldsymbol{\xi}(t) $ represents Gaussian white noise with zero mean and correlations $ \langle \xi_i(t) \xi_j(t') \rangle = \delta_{ij} \delta(t - t') $. The corresponding probability density function $ p(\bm{x}, t) $ evolves in time according to the Fokker--Planck equation \cite{risken1996fokker, gardiner2009stochastic}:
\begin{align}
\begin{split}
\partial_t p(\bm{x}, t) &= \mathbb{L} p(\bm{x},t) \, , \\
\mathbb{L}p(\bm{x},t)&=-\nabla \cdot \left[ \bm{A}(\bm{x}) p(\bm{x}, t) - \bm{B} \nabla p(\bm{x}, t) \right].
\end{split}
\label{overdamp}
\end{align}
In steady state, $ \partial_t p_{\text{st}}(\bm{x}) = 0 $, and the system may sustain persistent currents due to nonequilibrium conditions. 

We consider the accumulation of the general current $\hat{\cal J}$ linearly dependent on the velocity with the weight function $\boldsymbol{\theta}(\bm{x}) $ as
\begin{equation}
\hat{\cal J} = \int_0^\tau \bm{\theta}(\bm{x}_t) \circ \dot{\bm{x}}_t \, dt \,,
\label{eq:stochastic_current}
\end{equation}
where $ \circ $ denotes the Stratonovich product. The steady-state average current $j_{\rm st}$ is computed as $j_{\rm st}= \langle \hat{\cal J}\rangle / \tau$ within the framework of the Langevin dynamics. In the Fokker-Planck picture, an equivalent expression is
\begin{align}
j_{\text{st}} &= \int d\bm{x} \, \bm{\theta}(\bm{x})\cdot \bm{\nu}_{\text{st}}(\bm{x}) p_{\text{st}} ({\bm x}) \, ,
\end{align}
where $ \bm{\nu}_{\text{st}}({\bm x}) := \bm{A}({\bm x}) - \bm{B} \nabla \ln p_{\text{st}}({\bm x}) $ is the local mean velocity. Note that the product ${\bm \nu}_{\text{st}} ({\bm x})p_{\text{st}}({\bm x})$ is the probability current at the steady state.
We also introduce the fluctuation of the current over a time interval $\tau$, defined as
\begin{equation}
D_\tau = \frac{\text{Var}(\hat{\cal J})}{2\tau} \,.
\label{eq:diffusion_coefficient}
\end{equation}
Here, ${\rm Var}(\hat{\cal J})$ represents the variance of the accumulated stochastic current given by Eq.~\eqref{eq:stochastic_current}. Inclusion of the factor $2$ in the denominator ensures consistency with the standard definition of the diffusion coefficient, particularly when the weight function ${\bm \theta}$ is set to the all-ones vector.
 When $\tau$ approaches zero, $D_0$ denotes the instantaneous fluctuation of the current, given by the steady-state average $D_0 = \int d\bm{x} \, \bm{\theta}(\bm{x}) \cdot \mathbf{B}\bm{\theta}(\bm{x}) \, p_{\text{st}}(\bm{x})$.

As a fundamental measure of thermodynamic cost, we consider the entropy production rate $\sigma_{\text{st}}$ at the steady state, expressed as \cite{Seifert_2012}:
\begin{equation}
\sigma_{\text{st}} = \int d\bm{x} \, \bm{\nu}_{\text{st}}({\bm x})\cdot \bm{B}^{-1} \bm{\nu}_{\text{st}} ({\bm x}) \, p_{\text{st}}({\bm x}) \, .
\label{eq:entropy_production}
\end{equation}
This expression explicitly demonstrates the nonnegativity of the entropy production rate. For steady-state thermal systems, the entropy production rate expresses the rate of heat dissipation. Throughout this paper, we set the Boltzmann constant to unity and assume local detailed balance (LDB). Our results remain mathematically valid without this assumption; however, interpreting \(\sigma_{\text{st}}\) as the thermodynamic entropy production rate requires LDB.

The TUR establishes a lower bound on fluctuations $D_{\tau}$ in terms of the entropy production rate and the steady-state current. A more general result than Eq.~\eqref{tur-first} is the finite-time TUR, which provides a bound on the fluctuations in the steady state at arbitrary time interval \cite{horowitz2017proof,Dechant.2018.JSM}:
\begin{equation}
D_\tau \geq \frac{j_{\text{st}}^2}{ \sigma_{\text{st}}} =:D_{\tau,{\bm \theta}}^{\textrm{TUR}} \, . 
\label{eq:TUR_diffusion}
\end{equation}

To establish an upper bound on fluctuations, the previous iTUR was formulated using the spectral gap of a symmetrized generator of the original Markov jump process \cite{Garrahan_PRL_2023}.
In this work, we extend this concept to the continuous-variable setting by considering the spectral gap $\lambda$ of the symmetrized Fokker--Planck operator, which is given by replacing the drift term ${\bm A} ({\bm x})$ by $\tilde{\bm{A}}(\bm{x}) = \bm{B} \nabla \ln p_{\text{st}}(\bm{x})$ in the operator $\mathbb{L}$. Importantly, the dynamics governed by this operator preserves the same steady state $p_{\text{st}}(\bm{x})$. 
The spectral gap is related to the slowest timescale of the equilibrium relaxation with the same diffusion matrix $\bm{B}$ and the steady state $p_{\text{st}}(\bm{x})$.
While symmetrization alters the dynamics, the spectral gap remains closely linked to the relaxation properties of the original dynamics (see, for example, \cite{mori2023symmetrized}).

\sectionprl{The iTUR for the continuous variable case}
We present here the main results for continuous variables governed by dynamics (\ref{overdamp}). An outline of the derivation is provided in Appendix A of the end matter, with details presented in the supplementary material (SM)\cite{supple}.
First, we present the result for the most typical and experimentally relevant scenario, namely, the standard diffusion case with ${\bm \theta}$ set to the all-ones vector. We derive the bound on the long-time diffusion coefficient as
\begin{equation} 
D_\infty \leq D_0 + \frac{D_0 \sigma_{\text{st}} - j_{\text{st}}^2}{\lambda}=: D_{\infty}^{\rm iTUR}\, , ~~(\bm{\theta} ({\bm x})=1 \, , \forall {\bm x}) \,, \label{eq:giant_diffusion_bound} 
\end{equation}
where the non-negativity of $D_0 \sigma_{\text{st}} - j_{\text{st}}^2$ follows directly from the short-time limit of the TUR (\ref{eq:TUR_diffusion}).
The iTUR yields $D_\infty / D_0 \leq 1 + \sigma_{\text{st}} / \lambda$, elucidating how the thermodynamic cost $\sigma_{\rm st}$ and the relaxation rate through the spectral gap $\lambda$ jointly constrain the long-time diffusion coefficient. A large long-time diffusion constant can only arise either through a high entropy production rate or a small spectral gap. A smaller $\lambda$ physically allows correlations to persist for longer durations, thereby increasing the upper bound on the long-time diffusion constant. In the limit of very large $\lambda$, the bound approaches $D_0$, demonstrating that rapid relaxation negates enhanced diffusion.

Second, we derive the result for finite measurement times $\tau$ at the steady state for a general current, i.e., a current with a general weight function:
\begin{equation} D_\tau \leq D_0 + g(\bar{\lambda} \tau)\frac{D_{\text{max}} \sigma_{\text{st}} - j_{\text{st}}^2 }{\lambda} =: D_{\tau,{\bm\theta}}^{\rm iTUR}, ~~~~~({\bm \theta}: {\rm general}). \label{eq:long_time_bound} 
\end{equation}
The detailed proof is given in the SM \cite{supple}. 
Here, $D_{\text{max}}$ and $\sigma_{\text{max}}$ represent the maximum values of the local current fluctuation and entropy production, satisfying $D_{\text{max}} := \sup_{\bm{x}} [\bm{\theta}(\bm{x})\cdot \mathbf{B} \bm{\theta}(\bm{x})]$ and $\sigma_{\text{max}} := \sup_{\bm{x}} [\bm{\nu}_{\mathrm{st}}(\bm{x}) \cdot \mathbf{B}^{-1} \bm{\nu}_{\mathrm{st}}(\bm{x})]$, respectively. 
The characteristic rate $\bar{\lambda} := \sqrt{\lambda(\lambda + \sigma_{\text{max}})}$ sets a timescale.
The function $g(x)$ is defined as
\begin{equation} g(x) = \frac{2}{\pi x} \left(\int_0^1 dz \frac{1-\cos(zx)}{z^2(1+z^2)} + \int_1^\infty dz \frac{1-\cos(zx)}{2z^3}\right) \, , \nonumber 
\end{equation}
which is nonnegative, monotonically increasing, and smoothly interpolates between $0$ and $1$ as $x$ changes from $0$ to $+\infty$. 
In the limit of the vanishing measurement time $\tau \to 0$ with $\sigma_{\text{max}}$ finite, the bound simplifies to $D_\tau \leq D_0$, which is achieved with equality. In contrast, as $\tau \to \infty$, the function $g(\bar{\lambda} \tau)$ approaches $1$, and hence the inequality (\ref{eq:long_time_bound}) covers the specific case (\ref{eq:giant_diffusion_bound}) 
by setting the weight function to the all-ones vector.

The iTUR for the general case (\ref{eq:long_time_bound}) elucidates similar physics to the specific case (\ref{eq:giant_diffusion_bound}) regarding the roles of the entropy production rate and spectral gap for large current fluctuations. Moreover, the rigorous relation $D_\tau \leq D_0$ is established in the equilibrium states for any measurement time. The TUR and iTUR constrain the range of current fluctuations for arbitrary measurement time as 
\begin{align}
D_{\tau, {\bm \theta}}^{\rm TUR} &\leq D_\tau \leq D_{\tau, {\bm \theta}}^{\rm iTUR}\, .
\end{align} 
When considering particle diffusion setting $\bm{\theta} ({\bm x})=1$, $D_{\tau}$ behaves as $D_{\tau}\sim \bar{D} \tau^{\nu}$ at large scale $\tau$, where $\nu <0\, (>0)$ implies sub(super) diffusion. The TUR imposes a constraint on the cutoff time marking the end of subdiffusion~\cite{Hartich_PRL_2021}, while the iTUR similarly limits the duration of superdiffusive behavior. Notably, the iTUR establishes a no-go theorem, ruling out the possibility of perpetual superdiffusion in systems with a finite entropy production rate and a finite spectral gap. 

\begin{figure}[tbp]
 \centering
 \includegraphics[width=\linewidth]{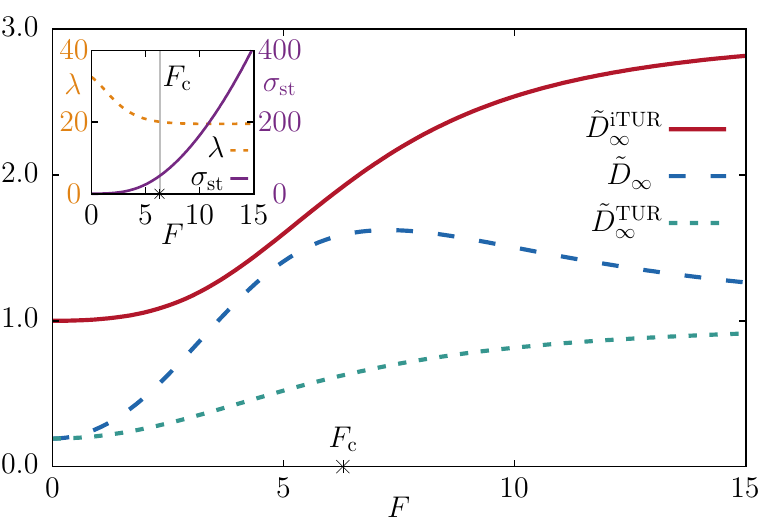}
 \caption{
  Diffusion coefficient, iTUR, and TUR as functions of the external force $ F $ in a tilted periodic potential. Parameters are $ V_0 = 1 $, $ L = 1 $, and $ D_0 = 0.5$ $(T= 0.5) $. The dashed line represents $ \tilde{D}_\infty $, showing a significant increase near the critical force $ F_{\mathrm{c}} = 2\pi $, indicative of giant diffusion. The solid line denotes the iTUR upper bound, closely aligning with $ \tilde{D}_\infty $, while the dotted line represents the TUR lower bound. The inset shows the behavior of the entropy production rate $ \sigma_{\text{st}} $ and spectral gap $ \lambda $ with $ F $. The left y-axis corresponds to the spectral gap, while the right y-axis represents the entropy production rate.}
 \label{fig:giant_diffusion}
\end{figure}

\sectionprl{Application to the giant diffusion}
We apply the iTUR for the diffusion coefficient (\ref{eq:giant_diffusion_bound}) to the giant diffusion phenomenon, which is characterized by the emergence of anomalously large diffusion coefficients exceeding the instantaneous diffusion constant in systems governed by periodic potentials under the influence of nonconservative forces \cite{Reimann_PRL_2001, Reimann_PRE_2002}. Giant diffusion has been experimentally observed in various biological and soft matter systems \cite{Hayashi_PRL_2015, Shinagawa2016,Lee_PRL_2006,Kim_PRL_2017, Suma_PRL_2023}. The iTUR (\ref{eq:giant_diffusion_bound}) establishes that such enhanced diffusivity is achievable only under conditions of either a substantial entropy production rate or a small spectral gap. Within this theoretical framework, we consider this phenomenon. The dynamics of a particle in this system is governed by the stochastic differential equation
\begin{equation}
\dot{x}_t = -V'(x_t) + F + \sqrt{2D_0} \, \xi(t) \,,
\label{eq:tilted_langevin}
\end{equation}
where $ V(x) = V_0 \sin\bigl[ (2\pi x)/L \bigr]$ is a periodic potential with amplitude $ V_0 $ and period $ L $, $ F $ is a constant nonconservative force and $ D_0 = T/\gamma $ is the instantaneous diffusion coefficient determined by the temperature $ T $ and friction coefficient $ \gamma $. In our analysis, we set the friction coefficient $ \gamma $ to unity. To discuss the long-time diffusion constant, we define the normalized diffusion constants:
\begin{align}
\begin{split}
\tilde{D}_{\infty} &:= D_{\infty} D_0^{-1} \, , \\
\tilde{D}_{\infty}^{{\rm iTUR}/{\rm TUR}} &:= D_{\infty}^{{\rm iTUR}/{\rm TUR}} D_0^{-1} \, ,
\end{split}
\nonumber
\end{align}
where $D_{\infty}^{\rm TUR} = D_{\infty, { \theta}}^{\rm TUR} $ with $\theta ({ x})=1, \forall { x}$.

Fig.~\ref{fig:giant_diffusion} shows the dependence of the normalized diffusion coefficients on the external force $F$. As $F$ increases, $\tilde{D}_\infty$ exhibits a pronounced enhancement near the critical force $F_{\mathrm{c}} = (2\pi V_0)/L$, corresponding to the point where the potential barrier vanishes in the infinite-line picture. The iTUR provides an upper bound for $\tilde{D}_\infty$, as evidenced by the close agreement between the theoretical prediction of the iTUR (solid line) and the observed diffusion coefficients (dashed line). In contrast, the TUR establishes a complementary lower bound based on the entropy production rate.
The inset illustrates the $F$-dependence of the spectral gap $\lambda$ and the entropy production rate $\sigma_{\text{st}}$. As $F$ approaches $F_{\mathrm{c}}$, the spectral gap decreases, indicating slower relaxation dynamics, while the entropy production rate begins to increase near this point. Slower relaxation (smaller $\lambda$) allows fluctuations to persist for longer durations, whereas a higher entropy production rate drives enhanced transport. This interplay aligns with the behavior predicted by the upper bound of the iTUR. 
In the high-force limit ($F \to \infty$), both $D_\infty$ and the TUR lower bound $D_\infty^{\text{TUR}}$ converge to $D_0$, while the iTUR upper bound $D_\infty^{\text{iTUR}}$ approaches a finite value strictly greater than $D_0$ (see SM).
For this specific potential, the iTUR accurately captures diffusion enhancement up to the critical force $F_{\mathrm{c}}$ but does not explain the subsequent decrease beyond it, whereas in a generic tilted potential, the iTUR might predict a peak at $F_{\mathrm{c}}$.

\begin{figure}[tbp]
 \centering
 \includegraphics[width=\linewidth]{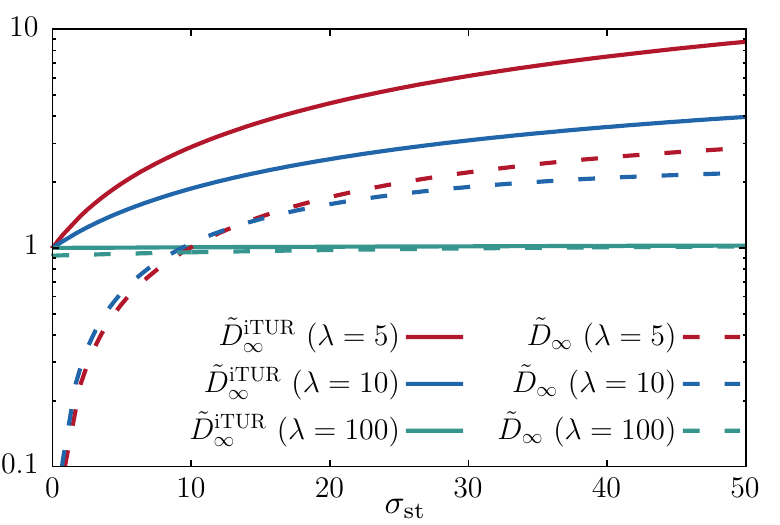}
 \caption{
Dependence of the diffusion coefficient $ D_\infty $ and the iTUR upper bound on the entropy production rate $ \sigma_{\text{st}} $ for fixed spectral gaps $( \lambda = 5, 10, 100 )$. Dashed lines represent $ D_\infty/D_0 $, and solid lines denote the iTUR upper bound.
 }
 \label{fig:spectral_gap_entropy}
\end{figure}

Fig.~\ref{fig:spectral_gap_entropy} shows the normalized diffusion constant as a function of entropy production rate $\sigma_{\text{st}}$ for a fixed spectral gap $\lambda$. We observe that $\tilde{D}_\infty$ increases monotonically with an increasing entropy production rate. 
The figure demonstrates that pushing the system further out of equilibrium via an increased entropy production rate consistently amplifies diffusion, and this effect is more pronounced at smaller spectral gaps.

\sectionprl{The iTUR for the discrete variable case}
We consider a continuous-time Markov jump process that describes transitions between discrete states. The transition rates $W_{km}$ from state $m$ to state $k$ define these transitions. 
To ensure thermodynamic consistency, we assume that the transition rates satisfy the local detailed balance condition. In the steady state, the state probabilities $\pi_k$ remain constant over time.
Let $ \Gamma_\tau = \{n_0, (n_1, t_1), \ldots, (n_N, t_N)\} $ be a stochastic trajectory of the system during the time interval $[0, \tau]$, where the system is initially in state $ n_0 $ and a transition from state $ n_{i-1} $ to state $ n_i $ occurs at time $ t_i $ for each $ 1 \leq i \leq N $. For each trajectory, we consider an arbitrary current observable $ \hat{\cal J}(\Gamma_\tau) = \sum_{i=1}^N d_{n_i n_{i-1}}$, where the increment $d_{mn}$ associated with the transition $ n \to m $, satisfying $d_{mn}=-d_{nm}$. The steady-state time average of this observable is then given by $ j_{\text{st}} = \sum_{k \neq m}d_{km} W_{km} \pi_m $.
The entropy production rate is expressed as: $\sigma_{\text{st}} = \sum_{k > m} J_{km} \ln \bigl[(W_{km} \pi_m)/(W_{mk} \pi_k) \bigr]$, where $J_{km} = W_{km}\pi_m - W_{mk}\pi_k$ are the steady-state fluxes. 
Let $\kappa = \max_k \sum_{m \neq k} W_{km} $ be the maximum escape rate. Let $\lambda$ be the spectral gap of the symmetrized matrix $(W^\top + W^\dagger)/2$ where $W^\dagger$ is the adjoint of $W^\top$ taken with respect to the inner product induced by the stationary state $\pi$, i.e., \((W^\dagger)_{i,j} = \pi_j^{-1} W_{j,i} \pi_i\). For a generalized current with $d_{km} = -d_{mk} \leq 1$, we derive the iTUR for the Markov jump process:
\begin{equation}
\label{eq:current_iTUR}
D_\tau\leq D_0 + g(\tilde{\lambda} \tau)\frac{ \kappa \sigma_{\text{st}} - 2j_{\text{st}} ^2}{2\lambda} \, ,
\end{equation}
where $D_0 := \sum_{k,m} \pi_m W_{mk} d_{mk}^2/2$ and $\tilde{\lambda} := \sqrt{\lambda(\lambda + 2\kappa)}$.
Note that the iTURs for the discrete variable case cannot reproduce the iTURs for the continuous variable case since the escape rate $\kappa$ diverges in the continuous variable limit. In Appendix B of the end matter and SM \cite{supple}, we provide a refined iTUR and the proof.

\sectionprl{Summary and Discussion}
We have derived upper bounds on current fluctuations at the steady state (iTUR) by utilizing both the entropy production rate and the spectral gap. By integrating the thermodynamic uncertainty relation (TUR), our results constrain the range of current fluctuations.
While the TUR suggests that current fluctuations relative to the average current can be reduced by increasing the entropy production rate, our iTUR involving entropy production implies that the overall extent of fluctuations tends to increase. Crucially, this effect is not apparent in the earlier version of iTUR derived in Ref.~\cite{Garrahan_PRL_2023}.
 We apply the derived iTUR to the phenomenon of giant diffusion, demonstrating the critical roles of both the spectral gap and the entropy production rate.

Although the iTUR provides a concise upper bound on fluctuations, direct experimental validation is challenging because it depends on the global suprema of local current fluctuations and on the symmetrized spectral gap—quantities that typically require detailed knowledge of the system. These challenges can be addressed by practical estimation strategies using trajectory data: supremum terms can be replaced with moment-based or probabilistic bounds, and the symmetrized spectral gap can be estimated variationally via a generalized eigenvalue procedure (see SM~\cite{supple}). These strategies enable targeted tests in controlled experimental platforms such as colloidal particles in optical potentials, microfluidic devices, and single-molecule motors.

The iTURs introduced in this paper have significant potential for broad applications across relevant systems. For instance, the iTUR provides thermodynamic concentration inequalities, imposing upper limits on the probability of observing current deviations significantly far from their mean. 
Furthermore, the iTUR can be inverted to establish a new lower bound on the entropy production rate. This bound is complementary to the conventional TUR, proving particularly powerful for thermodynamic inference in the regime of enhanced diffusion, where the standard TUR is often less informative.
The iTUR may also be useful to discuss phase transitions where current fluctuations diverge. It highlights the interplay between entropy production and the spectral gap near the transition point. Applications extend beyond theoretical insights, enabling the estimation of maximum fluctuations and providing quantitative design limits for fluctuation enhancement versus thermodynamic cost in nanotechnology and synthetic biology. Future work includes extending the iTUR beyond steady states by defining a suitable analogue of the spectral gap for time-dependent dynamics (e.g., instantaneous spectra or Floquet exponents) and formulating a quantum iTUR for open quantum systems, incorporating coherence and entropy production.

\section*{Acknowledgments}
The authors thank T. Van Vu and J. P. Garrahan for
the insightful discussion. This work is supported by JSPS KAKENHI Grant No. JP23K25796. 
AD is supported by JSPS KAKENHI (Grant No. 22K13974 and 24H00833).

\noindent 
\\

\section*{End Matter}
\sectionprl{Appendix A: Outline of the proof}
We here outline the proof, while the details are presented in the SM \cite{supple}.
Utilizing a variational principle for diffusion coefficients, we can express $ D_\infty $ as
\begin{equation}
\begin{aligned}
	D_\infty &= \sup_{\chi} \inf_{\eta} \bigg[ 2\, \text{Cov}_{\text{st}} \left( (\bm{\theta} + \nabla \eta) \cdot \bm{\nu}_{\text{st}}, \chi \right) \\
	&\quad + \langle (\bm{\theta} + \nabla \eta)\cdot \bm{B} (\bm{\theta} + \nabla \eta) \rangle_{\text{st}} - \langle \nabla \chi \cdot \bm{B} \nabla \chi \rangle_{\text{st}} \bigg],
\end{aligned}
\label{eq:variational_expression}
\end{equation}
where $\langle \diamond \rangle_{\text{st}}$ and $\text{Cov}_{\text{st}}( \diamond , \diamond)$ denote the steady-state average and covariance, respectively.
Setting $ \eta(\bm{x}) = \bm{0} $ simplifies Eq.~\eqref{eq:variational_expression} to
\begin{equation}
\begin{aligned}
	D_\infty &\leq D_0 + \sup_{\chi} \left[ 2\, \text{Cov}_{\text{st}}(\bm{\theta} \cdot \bm{\nu}_{\text{st}}, \chi) - \langle \nabla \chi \cdot \bm{B} \nabla \chi \rangle_{\text{st}} \right],
\end{aligned}
\label{eq:upper_bound_step1}
\end{equation}
where $ D_0 = \langle \bm{\theta}\cdot \bm{B} \bm{\theta} \rangle_{\text{st}} $.
Using the Cauchy--Schwarz inequality, we obtain
\begin{equation}
D_\infty \leq D_0 + \sup_{\chi} \left[ \frac{ \left[ \text{Cov}_{\text{st}}(\bm{\theta} \cdot \bm{\nu}_{\text{st}}, \chi) \right]^2 }{ \langle \nabla \chi \cdot \bm{B} \nabla \chi \rangle_{\text{st}} } \right].
\label{eq:upper_bound_step2}
\end{equation}
We define the symmetrized Fokker--Planck operator, \(\mathbb{L}_{\rm sym}\), by substituting the original drift in \(\mathbb{L}\) with the modified field \(\tilde{\bm{A}}(\bm{x}) = \bm{B}\nabla\ln p_{\text{st}}(\bm{x})\). This substitution renders \(\mathbb{L}_{\rm sym}\) symmetric with respect to a particular inner product weighted by the steady-state distribution. Its spectral gap, \(\lambda\), can then be determined via the variational formula:
\begin{equation}
\lambda = \inf_{\chi} \left[ \frac{\langle \nabla \chi \cdot \bm{B} \nabla \chi \rangle_{\text{\rm st}}}{\text{Var}_{\text{\rm st}}(\chi)} \right].
\end{equation}
We can deduce that $\lambda$ is upper bounded by the ratio $\langle \nabla \chi \cdot \bm{B} \nabla \chi \rangle_{\text{st}} / \text{Var}_{\text{st}}(\chi)$.
This leads to the bound:
\begin{equation}
D_\infty \leq D_0 + \frac{1}{\lambda} \text{Var}_{\text{st}}(\bm{\theta} \cdot \bm{\nu}_{\text{st}} )\,.
\label{eq:upper_bound_final}
\end{equation}
We now focus on bounding $\text{Var}_{\text{st}}(\bm{\theta} \cdot \bm{\nu}_{\text{st}})$. By definition, we have
\begin{equation}
\text{Var}_{\text{\rm st}}(\bm{\theta} \cdot \bm{\nu}_{\text{\rm st}} ) = \langle (\bm{\theta} \cdot \bm{\nu}_{\text{\rm st}} )^2 \rangle_{\text{\rm st}} - j^2_{\text{\rm st}} \,.
\end{equation}
The first term can be bounded as follows:
\begin{equation}
	\langle (\bm{\theta} \cdot \bm{\nu}_{\text{\rm st}} )^2 \rangle_{\text{\rm st}} \leq \langle (\bm{\theta}\cdot \bm{B} \bm{\theta}) (\bm{\nu}_{\text{\rm st}}\cdot \bm{B}^{-1} \bm{\nu}_{\text{\rm st}} ) \rangle_{\text{\rm st}} \,.
\end{equation}
Introducing $ D_{\text{max}} $, satisfying $\bm{\theta}\cdot \bm{B} \bm{\theta} \leq D_{\text{max}}$, and noting $\sigma_{\text{st}} =\langle \bm{\nu}_{\text{st}}\cdot \bm{B}^{-1} \bm{\nu}_{\text{st}} \rangle_{\text{st}}$, we obtain
\begin{equation}
\text{Var}_{\text{st}}(\bm{\theta} \cdot \bm{\nu}_{\text{st}}) \leq D_{\text{max}} \sigma_{\text{st}} - j_{\text{st}}^2 \,.
\label{eq:variance_upper_bound}
\end{equation}
Substituting this bound into Eq.~\eqref{eq:upper_bound_final}, we arrive at the iTUR bound (\ref{eq:long_time_bound}) with $\tau\to \infty$.

\sectionprl{Appendix B: Comparison with previous iTUR}
Here, we compare our iTUR with the formulation previously established in Ref.~\cite{Garrahan_PRL_2023}. Our iTUR offers several advantages: it provides a tighter bound, explicitly incorporates the entropy production rate, and correctly captures both the short-time and near-equilibrium limits.
The additional auxiliary fields in the variation expression of the variance offer the flexibility to incorporate antisymmetric contributions and thus the upper bound explicitly includes the entropy production.
Concretely, the iTUR in Ref.~\cite{Garrahan_PRL_2023} for a continuous-time Markov jump process states: 
\begin{equation}
 D_\tau \le D_0\left(1 + \frac{2\kappa}{\lambda}\right) \, .
\end{equation}
Our work establishes the following refined bound (see SM \cite{supple} for proof):
\begin{equation}
D_\tau \le D_0 + g(\tilde{\lambda}\tau) \frac{\kappa\, f(d_{\text{max}}^2\sigma_{\text{st}},\, 2D_0) - j_{\text{st}}^2}{\lambda} \, ,
\end{equation}
where $g(\tilde{\lambda}\tau) \leq 1$ and $f(x,y) = \frac{x^2}{4y} [\mathfrak{h}(x/2y)]^{-2}$. Here, $\mathfrak{h}$ is the inverse of $z \tanh(z)$.
Using the inequality $f(x,y) \le y$, our bound is at most $D_0 + (2D_0\kappa - j_{\text{st}}^2)/\lambda$. This is tighter than the previous result, $D_\tau \leq D_0(1 + 2\kappa/\lambda)$, by the term $j_{\text{st}}^2/\lambda$.
Furthermore, the modulator $f(x,y)$ effectively interpolates between physical regimes. Near equilibrium ($\sigma_{\text{st}} \to 0$), $f(x,y) \to x/2$, and the iTUR reveals a physical limit $D_\tau \leq D_0$. Moreover, when $\tau \rightarrow 0$, $g(\tilde{\lambda}\tau) \rightarrow 0$, the upper bound iTUR becomes an equality: $D_0 = D^{\text{iTUR}}_0$. In stark contrast, the previous version of iTUR predicts a finite enhancement, an unphysical artifact.


\begin{thebibliography}{61}%
\makeatletter
\providecommand \@ifxundefined [1]{%
 \@ifx{#1\undefined}
}%
\providecommand \@ifnum [1]{%
 \ifnum #1\expandafter \@firstoftwo
 \else \expandafter \@secondoftwo
 \fi
}%
\providecommand \@ifx [1]{%
 \ifx #1\expandafter \@firstoftwo
 \else \expandafter \@secondoftwo
 \fi
}%
\providecommand \natexlab [1]{#1}%
\providecommand \enquote [1]{``#1''}%
\providecommand \bibnamefont [1]{#1}%
\providecommand \bibfnamefont [1]{#1}%
\providecommand \citenamefont [1]{#1}%
\providecommand \href@noop [0]{\@secondoftwo}%
\providecommand \href [0]{\begingroup \@sanitize@url \@href}%
\providecommand \@href[1]{\@@startlink{#1}\@@href}%
\providecommand \@@href[1]{\endgroup#1\@@endlink}%
\providecommand \@sanitize@url [0]{\catcode `\\12\catcode `\$12\catcode `\&12\catcode `\#12\catcode `\^12\catcode `\_12\catcode `\%12\relax}%
\providecommand \@@startlink[1]{}%
\providecommand \@@endlink[0]{}%
\providecommand \url [0]{\begingroup\@sanitize@url \@url }%
\providecommand \@url [1]{\endgroup\@href {#1}{\urlprefix }}%
\providecommand \urlprefix [0]{URL }%
\providecommand \Eprint [0]{\href }%
\providecommand \doibase [0]{https://doi.org/}%
\providecommand \selectlanguage [0]{\@gobble}%
\providecommand \bibinfo [0]{\@secondoftwo}%
\providecommand \bibfield [0]{\@secondoftwo}%
\providecommand \translation [1]{[#1]}%
\providecommand \BibitemOpen [0]{}%
\providecommand \bibitemStop [0]{}%
\providecommand \bibitemNoStop [0]{.\EOS\space}%
\providecommand \EOS [0]{\spacefactor3000\relax}%
\providecommand \BibitemShut [1]{\csname bibitem#1\endcsname}%
\let\auto@bib@innerbib\@empty
\bibitem [{\citenamefont {Sekimoto}(2010)}]{Sekimoto.2010}%
 \BibitemOpen
 \bibfield {author} {\bibinfo {author} {\bibfnamefont {K.}~\bibnamefont {Sekimoto}},\ }\href@noop {} {\emph {\bibinfo {title} {{Stochastic Energetics}}}},\ Vol.\ \bibinfo {volume} {799}\ (\bibinfo {publisher} {Springer},\ \bibinfo {address} {Berlin},\ \bibinfo {year} {2010})\BibitemShut {NoStop}%
\bibitem [{\citenamefont {Seifert}(2008)}]{Seifert.2008.EPJB}%
 \BibitemOpen
 \bibfield {author} {\bibinfo {author} {\bibfnamefont {U.}~\bibnamefont {Seifert}},\ }\bibfield {title} {\bibinfo {title} {{Stochastic thermodynamics: principles and perspectives}},\ }\href {https://doi.org/10.1140/epjb/e2008-00001-9} {\bibfield {journal} {\bibinfo {journal} {Eur. Phys. J. E}\ }\textbf {\bibinfo {volume} {64}},\ \bibinfo {pages} {423} (\bibinfo {year} {2008})}\BibitemShut {NoStop}%
\bibitem [{\citenamefont {Barato}\ and\ \citenamefont {Seifert}(2015)}]{Barato_PRL_2015}%
 \BibitemOpen
 \bibfield {author} {\bibinfo {author} {\bibfnamefont {A.~C.}\ \bibnamefont {Barato}}\ and\ \bibinfo {author} {\bibfnamefont {U.}~\bibnamefont {Seifert}},\ }\bibfield {title} {\bibinfo {title} {Thermodynamic uncertainty relation for biomolecular processes},\ }\href {https://doi.org/10.1103/PhysRevLett.114.158101} {\bibfield {journal} {\bibinfo {journal} {Phys. Rev. Lett.}\ }\textbf {\bibinfo {volume} {114}},\ \bibinfo {pages} {158101} (\bibinfo {year} {2015})}\BibitemShut {NoStop}%
\bibitem [{\citenamefont {Gingrich}\ \emph {et~al.}(2016)\citenamefont {Gingrich}, \citenamefont {Horowitz}, \citenamefont {Perunov},\ and\ \citenamefont {England}}]{Gingrich_PRL_2016}%
 \BibitemOpen
 \bibfield {author} {\bibinfo {author} {\bibfnamefont {T.~R.}\ \bibnamefont {Gingrich}}, \bibinfo {author} {\bibfnamefont {J.~M.}\ \bibnamefont {Horowitz}}, \bibinfo {author} {\bibfnamefont {N.}~\bibnamefont {Perunov}},\ and\ \bibinfo {author} {\bibfnamefont {J.~L.}\ \bibnamefont {England}},\ }\bibfield {title} {\bibinfo {title} {Dissipation bounds all steady-state current fluctuations},\ }\href {https://doi.org/10.1103/PhysRevLett.116.120601} {\bibfield {journal} {\bibinfo {journal} {Phys. Rev. Lett.}\ }\textbf {\bibinfo {volume} {116}},\ \bibinfo {pages} {120601} (\bibinfo {year} {2016})}\BibitemShut {NoStop}%
\bibitem [{\citenamefont {Pietzonka}\ \emph {et~al.}(2016{\natexlab{a}})\citenamefont {Pietzonka}, \citenamefont {Barato},\ and\ \citenamefont {Seifert}}]{Pietzonka_PRE_2016}%
 \BibitemOpen
 \bibfield {author} {\bibinfo {author} {\bibfnamefont {P.}~\bibnamefont {Pietzonka}}, \bibinfo {author} {\bibfnamefont {A.~C.}\ \bibnamefont {Barato}},\ and\ \bibinfo {author} {\bibfnamefont {U.}~\bibnamefont {Seifert}},\ }\bibfield {title} {\bibinfo {title} {Universal bounds on current fluctuations},\ }\href {https://doi.org/10.1103/PhysRevE.93.052145} {\bibfield {journal} {\bibinfo {journal} {Phys. Rev. E}\ }\textbf {\bibinfo {volume} {93}},\ \bibinfo {pages} {052145} (\bibinfo {year} {2016}{\natexlab{a}})}\BibitemShut {NoStop}%
\bibitem [{\citenamefont {Koyuk}\ and\ \citenamefont {Seifert}(2020)}]{Koyuk_PRL_2020}%
 \BibitemOpen
 \bibfield {author} {\bibinfo {author} {\bibfnamefont {T.}~\bibnamefont {Koyuk}}\ and\ \bibinfo {author} {\bibfnamefont {U.}~\bibnamefont {Seifert}},\ }\bibfield {title} {\bibinfo {title} {Thermodynamic uncertainty relation for time-dependent driving},\ }\href {https://doi.org/10.1103/PhysRevLett.125.260604} {\bibfield {journal} {\bibinfo {journal} {Phys. Rev. Lett.}\ }\textbf {\bibinfo {volume} {125}},\ \bibinfo {pages} {260604} (\bibinfo {year} {2020})}\BibitemShut {NoStop}%
\bibitem [{\citenamefont {Dechant}\ and\ \citenamefont {Sasa}(2021{\natexlab{a}})}]{Dechant_PhysRevX_2021}%
 \BibitemOpen
 \bibfield {author} {\bibinfo {author} {\bibfnamefont {A.}~\bibnamefont {Dechant}}\ and\ \bibinfo {author} {\bibfnamefont {S.-i.}\ \bibnamefont {Sasa}},\ }\bibfield {title} {\bibinfo {title} {Improving thermodynamic bounds using correlations},\ }\href {https://doi.org/10.1103/PhysRevX.11.041061} {\bibfield {journal} {\bibinfo {journal} {Phys. Rev. X}\ }\textbf {\bibinfo {volume} {11}},\ \bibinfo {pages} {041061} (\bibinfo {year} {2021}{\natexlab{a}})}\BibitemShut {NoStop}%
\bibitem [{\citenamefont {Liu}\ \emph {et~al.}(2020)\citenamefont {Liu}, \citenamefont {Gong},\ and\ \citenamefont {Ueda}}]{Liu_PRL_2020}%
 \BibitemOpen
 \bibfield {author} {\bibinfo {author} {\bibfnamefont {K.}~\bibnamefont {Liu}}, \bibinfo {author} {\bibfnamefont {Z.}~\bibnamefont {Gong}},\ and\ \bibinfo {author} {\bibfnamefont {M.}~\bibnamefont {Ueda}},\ }\bibfield {title} {\bibinfo {title} {Thermodynamic uncertainty relation for arbitrary initial states},\ }\href {https://doi.org/10.1103/PhysRevLett.125.140602} {\bibfield {journal} {\bibinfo {journal} {Phys. Rev. Lett.}\ }\textbf {\bibinfo {volume} {125}},\ \bibinfo {pages} {140602} (\bibinfo {year} {2020})}\BibitemShut {NoStop}%
\bibitem [{\citenamefont {Pietzonka}(2022)}]{Pietzonka_PRL_2022}%
 \BibitemOpen
 \bibfield {author} {\bibinfo {author} {\bibfnamefont {P.}~\bibnamefont {Pietzonka}},\ }\bibfield {title} {\bibinfo {title} {Classical pendulum clocks break the thermodynamic uncertainty relation},\ }\href {https://doi.org/10.1103/PhysRevLett.128.130606} {\bibfield {journal} {\bibinfo {journal} {Phys. Rev. Lett.}\ }\textbf {\bibinfo {volume} {128}},\ \bibinfo {pages} {130606} (\bibinfo {year} {2022})}\BibitemShut {NoStop}%
\bibitem [{\citenamefont {Hasegawa}(2020)}]{Hasegawa_PhysRevLett_2020}%
 \BibitemOpen
 \bibfield {author} {\bibinfo {author} {\bibfnamefont {Y.}~\bibnamefont {Hasegawa}},\ }\bibfield {title} {\bibinfo {title} {Quantum thermodynamic uncertainty relation for continuous measurement},\ }\href {https://doi.org/10.1103/PhysRevLett.125.050601} {\bibfield {journal} {\bibinfo {journal} {Phys. Rev. Lett.}\ }\textbf {\bibinfo {volume} {125}},\ \bibinfo {pages} {050601} (\bibinfo {year} {2020})}\BibitemShut {NoStop}%
\bibitem [{\citenamefont {Miller}\ \emph {et~al.}(2021)\citenamefont {Miller}, \citenamefont {Mohammady}, \citenamefont {Perarnau-Llobet},\ and\ \citenamefont {Guarnieri}}]{Miller_PhysRevLett_2021}%
 \BibitemOpen
 \bibfield {author} {\bibinfo {author} {\bibfnamefont {H.~J.~D.}\ \bibnamefont {Miller}}, \bibinfo {author} {\bibfnamefont {M.~H.}\ \bibnamefont {Mohammady}}, \bibinfo {author} {\bibfnamefont {M.}~\bibnamefont {Perarnau-Llobet}},\ and\ \bibinfo {author} {\bibfnamefont {G.}~\bibnamefont {Guarnieri}},\ }\bibfield {title} {\bibinfo {title} {Thermodynamic uncertainty relation in slowly driven quantum heat engines},\ }\href {https://doi.org/10.1103/PhysRevLett.126.210603} {\bibfield {journal} {\bibinfo {journal} {Phys. Rev. Lett.}\ }\textbf {\bibinfo {volume} {126}},\ \bibinfo {pages} {210603} (\bibinfo {year} {2021})}\BibitemShut {NoStop}%
\bibitem [{\citenamefont {Van~Vu}\ and\ \citenamefont {Saito}(2022)}]{Tan_PhysRevLett_2022}%
 \BibitemOpen
 \bibfield {author} {\bibinfo {author} {\bibfnamefont {T.}~\bibnamefont {Van~Vu}}\ and\ \bibinfo {author} {\bibfnamefont {K.}~\bibnamefont {Saito}},\ }\bibfield {title} {\bibinfo {title} {Thermodynamics of precision in markovian open quantum dynamics},\ }\href {https://doi.org/10.1103/PhysRevLett.128.140602} {\bibfield {journal} {\bibinfo {journal} {Phys. Rev. Lett.}\ }\textbf {\bibinfo {volume} {128}},\ \bibinfo {pages} {140602} (\bibinfo {year} {2022})}\BibitemShut {NoStop}%
\bibitem [{\citenamefont {Barato}\ \emph {et~al.}(2019)\citenamefont {Barato}, \citenamefont {Chetrite}, \citenamefont {Faggionato},\ and\ \citenamefont {Gabrielli}}]{Barato.2019.JSM}%
 \BibitemOpen
 \bibfield {author} {\bibinfo {author} {\bibfnamefont {A.~C.}\ \bibnamefont {Barato}}, \bibinfo {author} {\bibfnamefont {R.}~\bibnamefont {Chetrite}}, \bibinfo {author} {\bibfnamefont {A.}~\bibnamefont {Faggionato}},\ and\ \bibinfo {author} {\bibfnamefont {D.}~\bibnamefont {Gabrielli}},\ }\bibfield {title} {\bibinfo {title} {{A unifying picture of generalized thermodynamic uncertainty relations}},\ }\href {https://doi.org/10.1088/1742-5468/ab3457} {\bibfield {journal} {\bibinfo {journal} {J. Stat. Mech.: Theory Exp.}\ }\textbf {\bibinfo {volume} {2019}},\ \bibinfo {pages} {084017}}\BibitemShut {NoStop}%
\bibitem [{\citenamefont {Chun}\ \emph {et~al.}(2019)\citenamefont {Chun}, \citenamefont {Fischer},\ and\ \citenamefont {Seifert}}]{Chun.2019.PRE}%
 \BibitemOpen
 \bibfield {author} {\bibinfo {author} {\bibfnamefont {H.-M.}\ \bibnamefont {Chun}}, \bibinfo {author} {\bibfnamefont {L.~P.}\ \bibnamefont {Fischer}},\ and\ \bibinfo {author} {\bibfnamefont {U.}~\bibnamefont {Seifert}},\ }\bibfield {title} {\bibinfo {title} {{Effect of a magnetic field on the thermodynamic uncertainty relation}},\ }\href {https://doi.org/10.1103/PhysRevE.99.042128} {\bibfield {journal} {\bibinfo {journal} {Phys. Rev. E}\ }\textbf {\bibinfo {volume} {99}},\ \bibinfo {pages} {042128} (\bibinfo {year} {2019})}\BibitemShut {NoStop}%
\bibitem [{\citenamefont {Dechant}\ and\ \citenamefont {Sasa}(2018)}]{Dechant.2018.JSM}%
 \BibitemOpen
 \bibfield {author} {\bibinfo {author} {\bibfnamefont {A.}~\bibnamefont {Dechant}}\ and\ \bibinfo {author} {\bibfnamefont {S.-i.}\ \bibnamefont {Sasa}},\ }\bibfield {title} {\bibinfo {title} {{Current fluctuations and transport efficiency for general Langevin systems}},\ }\href {https://doi.org/10.1088/1742-5468/aac91a} {\bibfield {journal} {\bibinfo {journal} {J. Stat. Mech.: Theory Exp.}\ }\textbf {\bibinfo {volume} {2018}},\ \bibinfo {pages} {063209}}\BibitemShut {NoStop}%
\bibitem [{\citenamefont {Dechant}(2019)}]{Dechant.2019.JPA}%
 \BibitemOpen
 \bibfield {author} {\bibinfo {author} {\bibfnamefont {A.}~\bibnamefont {Dechant}},\ }\bibfield {title} {\bibinfo {title} {{Multidimensional thermodynamic uncertainty relations}},\ }\href {https://doi.org/10.1088/1751-8121/aaf3ff} {\bibfield {journal} {\bibinfo {journal} {J. Phys. A}\ }\textbf {\bibinfo {volume} {52}},\ \bibinfo {pages} {035001} (\bibinfo {year} {2019})}\BibitemShut {NoStop}%
\bibitem [{\citenamefont {Dechant}\ and\ \citenamefont {Sasa}(2020)}]{Dechant.2020.PNAS}%
 \BibitemOpen
 \bibfield {author} {\bibinfo {author} {\bibfnamefont {A.}~\bibnamefont {Dechant}}\ and\ \bibinfo {author} {\bibfnamefont {S.-i.}\ \bibnamefont {Sasa}},\ }\bibfield {title} {\bibinfo {title} {{Fluctuation-response inequality out of equilibrium}},\ }\href {https://doi.org/10.1073/pnas.1918386117} {\bibfield {journal} {\bibinfo {journal} {Proc. Natl. Acad. Sci. U.S.A.}\ }\textbf {\bibinfo {volume} {117}},\ \bibinfo {pages} {6430} (\bibinfo {year} {2020})}\BibitemShut {NoStop}%
\bibitem [{\citenamefont {Dechant}\ and\ \citenamefont {Sasa}(2021{\natexlab{b}})}]{Dechant.2021.PRR}%
 \BibitemOpen
 \bibfield {author} {\bibinfo {author} {\bibfnamefont {A.}~\bibnamefont {Dechant}}\ and\ \bibinfo {author} {\bibfnamefont {S.-i.}\ \bibnamefont {Sasa}},\ }\bibfield {title} {\bibinfo {title} {{Continuous time reversal and equality in the thermodynamic uncertainty relation}},\ }\href {https://doi.org/10.1103/PhysRevResearch.3.L042012} {\bibfield {journal} {\bibinfo {journal} {Phys. Rev. Research}\ }\textbf {\bibinfo {volume} {3}},\ \bibinfo {pages} {L042012} (\bibinfo {year} {2021}{\natexlab{b}})}\BibitemShut {NoStop}%
\bibitem [{\citenamefont {Gupta}\ and\ \citenamefont {Maritan}(2020)}]{Gupta.2020.EPJB}%
 \BibitemOpen
 \bibfield {author} {\bibinfo {author} {\bibfnamefont {D.}~\bibnamefont {Gupta}}\ and\ \bibinfo {author} {\bibfnamefont {A.}~\bibnamefont {Maritan}},\ }\bibfield {title} {\bibinfo {title} {{Thermodynamic uncertainty relations in a linear system}},\ }\href {https://doi.org/10.1140/epjb/e2020-10019-4} {\bibfield {journal} {\bibinfo {journal} {Eur. Phys. J. B}\ }\textbf {\bibinfo {volume} {93}},\ \bibinfo {pages} {28} (\bibinfo {year} {2020})}\BibitemShut {NoStop}%
\bibitem [{\citenamefont {Hiura}\ and\ \citenamefont {Sasa}(2021)}]{Hiura.2021.PRE}%
 \BibitemOpen
 \bibfield {author} {\bibinfo {author} {\bibfnamefont {K.}~\bibnamefont {Hiura}}\ and\ \bibinfo {author} {\bibfnamefont {S.-i.}\ \bibnamefont {Sasa}},\ }\bibfield {title} {\bibinfo {title} {{Kinetic uncertainty relation on first-passage time for accumulated current}},\ }\href {https://doi.org/10.1103/PhysRevE.103.L050103} {\bibfield {journal} {\bibinfo {journal} {Phys. Rev. E}\ }\textbf {\bibinfo {volume} {103}},\ \bibinfo {pages} {L050103} (\bibinfo {year} {2021})}\BibitemShut {NoStop}%
\bibitem [{\citenamefont {Koyuk}\ \emph {et~al.}(2019)\citenamefont {Koyuk}, \citenamefont {Seifert},\ and\ \citenamefont {Pietzonka}}]{Koyuk.2019.JPA}%
 \BibitemOpen
 \bibfield {author} {\bibinfo {author} {\bibfnamefont {T.}~\bibnamefont {Koyuk}}, \bibinfo {author} {\bibfnamefont {U.}~\bibnamefont {Seifert}},\ and\ \bibinfo {author} {\bibfnamefont {P.}~\bibnamefont {Pietzonka}},\ }\bibfield {title} {\bibinfo {title} {{A generalization of the thermodynamic uncertainty relation to periodically driven systems}},\ }\href {https://doi.org/10.1088/1751-8121/aaeec4} {\bibfield {journal} {\bibinfo {journal} {J. Phys. A}\ }\textbf {\bibinfo {volume} {52}},\ \bibinfo {pages} {02LT02} (\bibinfo {year} {2019})}\BibitemShut {NoStop}%
\bibitem [{\citenamefont {Manikandan}\ and\ \citenamefont {Krishnamurthy}(2018)}]{Manikandan.2018.JPA}%
 \BibitemOpen
 \bibfield {author} {\bibinfo {author} {\bibfnamefont {S.~K.}\ \bibnamefont {Manikandan}}\ and\ \bibinfo {author} {\bibfnamefont {S.}~\bibnamefont {Krishnamurthy}},\ }\bibfield {title} {\bibinfo {title} {{Exact results for the finite time thermodynamic uncertainty relation}},\ }\href {https://doi.org/10.1088/1751-8121/aaaa54} {\bibfield {journal} {\bibinfo {journal} {J. Phys. A}\ }\textbf {\bibinfo {volume} {51}},\ \bibinfo {pages} {11LT01} (\bibinfo {year} {2018})}\BibitemShut {NoStop}%
\bibitem [{\citenamefont {Park}\ and\ \citenamefont {Park}(2021)}]{Park.2021.PRR}%
 \BibitemOpen
 \bibfield {author} {\bibinfo {author} {\bibfnamefont {J.-M.}\ \bibnamefont {Park}}\ and\ \bibinfo {author} {\bibfnamefont {H.}~\bibnamefont {Park}},\ }\bibfield {title} {\bibinfo {title} {{Thermodynamic uncertainty relation in the overdamped limit with a magnetic Lorentz force}},\ }\href {https://doi.org/10.1103/PhysRevResearch.3.043005} {\bibfield {journal} {\bibinfo {journal} {Phys. Rev. Research}\ }\textbf {\bibinfo {volume} {3}},\ \bibinfo {pages} {043005} (\bibinfo {year} {2021})}\BibitemShut {NoStop}%
\bibitem [{\citenamefont {Pietzonka}\ \emph {et~al.}(2017)\citenamefont {Pietzonka}, \citenamefont {Ritort},\ and\ \citenamefont {Seifert}}]{Pietzonka.2017.PRE}%
 \BibitemOpen
 \bibfield {author} {\bibinfo {author} {\bibfnamefont {P.}~\bibnamefont {Pietzonka}}, \bibinfo {author} {\bibfnamefont {F.}~\bibnamefont {Ritort}},\ and\ \bibinfo {author} {\bibfnamefont {U.}~\bibnamefont {Seifert}},\ }\bibfield {title} {\bibinfo {title} {{Finite-time generalization of the thermodynamic uncertainty relation}},\ }\href {https://doi.org/10.1103/PhysRevE.96.012101} {\bibfield {journal} {\bibinfo {journal} {Phys. Rev. E}\ }\textbf {\bibinfo {volume} {96}},\ \bibinfo {pages} {012101} (\bibinfo {year} {2017})}\BibitemShut {NoStop}%
\bibitem [{\citenamefont {Proesmans}\ and\ \citenamefont {den Broeck}(2017)}]{Proesmans.2017.EPL}%
 \BibitemOpen
 \bibfield {author} {\bibinfo {author} {\bibfnamefont {K.}~\bibnamefont {Proesmans}}\ and\ \bibinfo {author} {\bibfnamefont {C.~V.}\ \bibnamefont {den Broeck}},\ }\bibfield {title} {\bibinfo {title} {{Discrete-time thermodynamic uncertainty relation}},\ }\href {https://doi.org/10.1209/0295-5075/119/20001} {\bibfield {journal} {\bibinfo {journal} {Europhys. Lett.}\ }\textbf {\bibinfo {volume} {119}},\ \bibinfo {pages} {20001} (\bibinfo {year} {2017})}\BibitemShut {NoStop}%
\bibitem [{\citenamefont {Wolpert}(2020)}]{Wolpert.2020.PRL}%
 \BibitemOpen
 \bibfield {author} {\bibinfo {author} {\bibfnamefont {D.~H.}\ \bibnamefont {Wolpert}},\ }\bibfield {title} {\bibinfo {title} {{Uncertainty relations and fluctuation theorems for Bayes nets}},\ }\href {https://doi.org/10.1103/PhysRevLett.125.200602} {\bibfield {journal} {\bibinfo {journal} {Phys. Rev. Lett.}\ }\textbf {\bibinfo {volume} {125}},\ \bibinfo {pages} {200602} (\bibinfo {year} {2020})}\BibitemShut {NoStop}%
\bibitem [{\citenamefont {Kalaee}\ \emph {et~al.}(2021)\citenamefont {Kalaee}, \citenamefont {Wacker},\ and\ \citenamefont {Potts}}]{Kalaee.2021.PRE}%
 \BibitemOpen
 \bibfield {author} {\bibinfo {author} {\bibfnamefont {A.~A.~S.}\ \bibnamefont {Kalaee}}, \bibinfo {author} {\bibfnamefont {A.}~\bibnamefont {Wacker}},\ and\ \bibinfo {author} {\bibfnamefont {P.~P.}\ \bibnamefont {Potts}},\ }\bibfield {title} {\bibinfo {title} {{Violating the thermodynamic uncertainty relation in the three-level maser}},\ }\href {https://doi.org/10.1103/PhysRevE.104.L012103} {\bibfield {journal} {\bibinfo {journal} {Phys. Rev. E}\ }\textbf {\bibinfo {volume} {104}},\ \bibinfo {pages} {L012103} (\bibinfo {year} {2021})}\BibitemShut {NoStop}%
\bibitem [{\citenamefont {Brandner}\ \emph {et~al.}(2018)\citenamefont {Brandner}, \citenamefont {Hanazato},\ and\ \citenamefont {Saito}}]{brandner2018thermodynamic}%
 \BibitemOpen
 \bibfield {author} {\bibinfo {author} {\bibfnamefont {K.}~\bibnamefont {Brandner}}, \bibinfo {author} {\bibfnamefont {T.}~\bibnamefont {Hanazato}},\ and\ \bibinfo {author} {\bibfnamefont {K.}~\bibnamefont {Saito}},\ }\bibfield {title} {\bibinfo {title} {Thermodynamic bounds on precision in ballistic multiterminal transport},\ }\href {https://doi.org/10.1103/PhysRevLett.120.090601} {\bibfield {journal} {\bibinfo {journal} {Phys. Rev. Lett.}\ }\textbf {\bibinfo {volume} {120}},\ \bibinfo {pages} {090601} (\bibinfo {year} {2018})}\BibitemShut {NoStop}%
\bibitem [{\citenamefont {Macieszczak}\ \emph {et~al.}(2018)\citenamefont {Macieszczak}, \citenamefont {Brandner},\ and\ \citenamefont {Garrahan}}]{Macieszczak.2018.PRL}%
 \BibitemOpen
 \bibfield {author} {\bibinfo {author} {\bibfnamefont {K.}~\bibnamefont {Macieszczak}}, \bibinfo {author} {\bibfnamefont {K.}~\bibnamefont {Brandner}},\ and\ \bibinfo {author} {\bibfnamefont {J.~P.}\ \bibnamefont {Garrahan}},\ }\bibfield {title} {\bibinfo {title} {Unified thermodynamic uncertainty relations in linear response},\ }\href {https://doi.org/10.1103/PhysRevLett.121.130601} {\bibfield {journal} {\bibinfo {journal} {Phys. Rev. Lett.}\ }\textbf {\bibinfo {volume} {121}},\ \bibinfo {pages} {130601} (\bibinfo {year} {2018})}\BibitemShut {NoStop}%
\bibitem [{\citenamefont {Agarwalla}\ and\ \citenamefont {Segal}(2018)}]{Agarwalla.2018.PRB}%
 \BibitemOpen
 \bibfield {author} {\bibinfo {author} {\bibfnamefont {B.~K.}\ \bibnamefont {Agarwalla}}\ and\ \bibinfo {author} {\bibfnamefont {D.}~\bibnamefont {Segal}},\ }\bibfield {title} {\bibinfo {title} {{Assessing the validity of the thermodynamic uncertainty relation in quantum systems}},\ }\href {https://doi.org/10.1103/PhysRevB.98.155438} {\bibfield {journal} {\bibinfo {journal} {Phys. Rev. B}\ }\textbf {\bibinfo {volume} {98}},\ \bibinfo {pages} {155438} (\bibinfo {year} {2018})}\BibitemShut {NoStop}%
\bibitem [{\citenamefont {Menczel}\ \emph {et~al.}(2021)\citenamefont {Menczel}, \citenamefont {Loisa}, \citenamefont {Brandner},\ and\ \citenamefont {Flindt}}]{Menczel.2021.JPA}%
 \BibitemOpen
 \bibfield {author} {\bibinfo {author} {\bibfnamefont {P.}~\bibnamefont {Menczel}}, \bibinfo {author} {\bibfnamefont {E.}~\bibnamefont {Loisa}}, \bibinfo {author} {\bibfnamefont {K.}~\bibnamefont {Brandner}},\ and\ \bibinfo {author} {\bibfnamefont {C.}~\bibnamefont {Flindt}},\ }\bibfield {title} {\bibinfo {title} {{Thermodynamic uncertainty relations for coherently driven open quantum systems}},\ }\href {https://doi.org/10.1088/1751-8121/ac0c8f} {\bibfield {journal} {\bibinfo {journal} {J. Phys. A}\ }\textbf {\bibinfo {volume} {54}},\ \bibinfo {pages} {314002} (\bibinfo {year} {2021})}\BibitemShut {NoStop}%
\bibitem [{\citenamefont {Potts}\ and\ \citenamefont {Samuelsson}(2019)}]{Potts.2019.PRE}%
 \BibitemOpen
 \bibfield {author} {\bibinfo {author} {\bibfnamefont {P.~P.}\ \bibnamefont {Potts}}\ and\ \bibinfo {author} {\bibfnamefont {P.}~\bibnamefont {Samuelsson}},\ }\bibfield {title} {\bibinfo {title} {{Thermodynamic uncertainty relations including measurement and feedback}},\ }\href {https://doi.org/10.1103/PhysRevE.100.052137} {\bibfield {journal} {\bibinfo {journal} {Phys. Rev. E}\ }\textbf {\bibinfo {volume} {100}},\ \bibinfo {pages} {052137} (\bibinfo {year} {2019})}\BibitemShut {NoStop}%
\bibitem [{\citenamefont {Saryal}\ \emph {et~al.}(2019)\citenamefont {Saryal}, \citenamefont {Friedman}, \citenamefont {Segal},\ and\ \citenamefont {Agarwalla}}]{Saryal.2019.PRE}%
 \BibitemOpen
 \bibfield {author} {\bibinfo {author} {\bibfnamefont {S.}~\bibnamefont {Saryal}}, \bibinfo {author} {\bibfnamefont {H.~M.}\ \bibnamefont {Friedman}}, \bibinfo {author} {\bibfnamefont {D.}~\bibnamefont {Segal}},\ and\ \bibinfo {author} {\bibfnamefont {B.~K.}\ \bibnamefont {Agarwalla}},\ }\bibfield {title} {\bibinfo {title} {{Thermodynamic uncertainty relation in thermal transport}},\ }\href {https://doi.org/10.1103/PhysRevE.100.042101} {\bibfield {journal} {\bibinfo {journal} {Phys. Rev. E}\ }\textbf {\bibinfo {volume} {100}},\ \bibinfo {pages} {042101} (\bibinfo {year} {2019})}\BibitemShut {NoStop}%
\bibitem [{\citenamefont {Polettini}\ \emph {et~al.}(2016)\citenamefont {Polettini}, \citenamefont {Lazarescu},\ and\ \citenamefont {Esposito}}]{Polettini2016PRE}%
 \BibitemOpen
 \bibfield {author} {\bibinfo {author} {\bibfnamefont {M.}~\bibnamefont {Polettini}}, \bibinfo {author} {\bibfnamefont {A.}~\bibnamefont {Lazarescu}},\ and\ \bibinfo {author} {\bibfnamefont {M.}~\bibnamefont {Esposito}},\ }\bibfield {title} {\bibinfo {title} {Tightening the uncertainty principle for stochastic currents},\ }\href {https://doi.org/10.1103/PhysRevE.94.052104} {\bibfield {journal} {\bibinfo {journal} {Phys. Rev. E}\ }\textbf {\bibinfo {volume} {94}},\ \bibinfo {pages} {052104} (\bibinfo {year} {2016})}\BibitemShut {NoStop}%
\bibitem [{\citenamefont {Gingrich}\ \emph {et~al.}(2017)\citenamefont {Gingrich}, \citenamefont {Rotskoff},\ and\ \citenamefont {Horowitz}}]{Gingrich_2017}%
 \BibitemOpen
 \bibfield {author} {\bibinfo {author} {\bibfnamefont {T.~R.}\ \bibnamefont {Gingrich}}, \bibinfo {author} {\bibfnamefont {G.~M.}\ \bibnamefont {Rotskoff}},\ and\ \bibinfo {author} {\bibfnamefont {J.~M.}\ \bibnamefont {Horowitz}},\ }\bibfield {title} {\bibinfo {title} {Inferring dissipation from current fluctuations},\ }\href {https://doi.org/10.1088/1751-8121/aa672f} {\bibfield {journal} {\bibinfo {journal} {J. Phys. A: Math.l}\ }\textbf {\bibinfo {volume} {50}},\ \bibinfo {pages} {184004} (\bibinfo {year} {2017})}\BibitemShut {NoStop}%
\bibitem [{\citenamefont {Pietzonka}\ and\ \citenamefont {Seifert}(2018)}]{Pietzonka.2018.PRL}%
 \BibitemOpen
 \bibfield {author} {\bibinfo {author} {\bibfnamefont {P.}~\bibnamefont {Pietzonka}}\ and\ \bibinfo {author} {\bibfnamefont {U.}~\bibnamefont {Seifert}},\ }\bibfield {title} {\bibinfo {title} {{Universal trade-off between power, efficiency, and constancy in steady-state heat engines}},\ }\href {https://doi.org/10.1103/PhysRevLett.120.190602} {\bibfield {journal} {\bibinfo {journal} {Phys. Rev. Lett.}\ }\textbf {\bibinfo {volume} {120}},\ \bibinfo {pages} {190602} (\bibinfo {year} {2018})}\BibitemShut {NoStop}%
\bibitem [{\citenamefont {Pietzonka}\ \emph {et~al.}(2016{\natexlab{b}})\citenamefont {Pietzonka}, \citenamefont {Barato},\ and\ \citenamefont {Seifert}}]{Pietzonka.2016.JSM}%
 \BibitemOpen
 \bibfield {author} {\bibinfo {author} {\bibfnamefont {P.}~\bibnamefont {Pietzonka}}, \bibinfo {author} {\bibfnamefont {A.~C.}\ \bibnamefont {Barato}},\ and\ \bibinfo {author} {\bibfnamefont {U.}~\bibnamefont {Seifert}},\ }\bibfield {title} {\bibinfo {title} {{Universal bound on the efficiency of molecular motors}},\ }\href {https://doi.org/10.1088/1742-5468/2016/12/124004} {\bibfield {journal} {\bibinfo {journal} {J. Stat. Mech.: Theory Exp.}\ }\textbf {\bibinfo {volume} {2016}},\ \bibinfo {pages} {124004}}\BibitemShut {NoStop}%
\bibitem [{\citenamefont {Seifert}(2019)}]{Seifert.2019.ARCMP}%
 \BibitemOpen
 \bibfield {author} {\bibinfo {author} {\bibfnamefont {U.}~\bibnamefont {Seifert}},\ }\bibfield {title} {\bibinfo {title} {{From stochastic thermodynamics to thermodynamic inference}},\ }\href {https://doi.org/10.1146/annurev-conmatphys-031218-013554} {\bibfield {journal} {\bibinfo {journal} {Annu. Rev. Condens. Matter Phys.}\ }\textbf {\bibinfo {volume} {10}},\ \bibinfo {pages} {171} (\bibinfo {year} {2019})}\BibitemShut {NoStop}%
\bibitem [{\citenamefont {Li}\ \emph {et~al.}(2019)\citenamefont {Li}, \citenamefont {Horowitz}, \citenamefont {Gingrich},\ and\ \citenamefont {Fakhri}}]{Li.2019.NC}%
 \BibitemOpen
 \bibfield {author} {\bibinfo {author} {\bibfnamefont {J.}~\bibnamefont {Li}}, \bibinfo {author} {\bibfnamefont {J.~M.}\ \bibnamefont {Horowitz}}, \bibinfo {author} {\bibfnamefont {T.~R.}\ \bibnamefont {Gingrich}},\ and\ \bibinfo {author} {\bibfnamefont {N.}~\bibnamefont {Fakhri}},\ }\bibfield {title} {\bibinfo {title} {{Quantifying dissipation using fluctuating currents}},\ }\href {https://doi.org/10.1038/s41467-019-09631-x} {\bibfield {journal} {\bibinfo {journal} {Nat. Commun.}\ }\textbf {\bibinfo {volume} {10}},\ \bibinfo {pages} {1666} (\bibinfo {year} {2019})}\BibitemShut {NoStop}%
\bibitem [{\citenamefont {Manikandan}\ \emph {et~al.}(2020)\citenamefont {Manikandan}, \citenamefont {Gupta},\ and\ \citenamefont {Krishnamurthy}}]{Manikandan.2020.PRL}%
 \BibitemOpen
 \bibfield {author} {\bibinfo {author} {\bibfnamefont {S.~K.}\ \bibnamefont {Manikandan}}, \bibinfo {author} {\bibfnamefont {D.}~\bibnamefont {Gupta}},\ and\ \bibinfo {author} {\bibfnamefont {S.}~\bibnamefont {Krishnamurthy}},\ }\bibfield {title} {\bibinfo {title} {{Inferring entropy production from short experiments}},\ }\href {https://doi.org/10.1103/PhysRevLett.124.120603} {\bibfield {journal} {\bibinfo {journal} {Phys. Rev. Lett.}\ }\textbf {\bibinfo {volume} {124}},\ \bibinfo {pages} {120603} (\bibinfo {year} {2020})}\BibitemShut {NoStop}%
\bibitem [{\citenamefont {Mart{\'\i}nez}\ \emph {et~al.}(2019)\citenamefont {Mart{\'\i}nez}, \citenamefont {Bisker}, \citenamefont {Horowitz},\ and\ \citenamefont {Parrondo}}]{Martinez.2019.NC}%
 \BibitemOpen
 \bibfield {author} {\bibinfo {author} {\bibfnamefont {I.~A.}\ \bibnamefont {Mart{\'\i}nez}}, \bibinfo {author} {\bibfnamefont {G.}~\bibnamefont {Bisker}}, \bibinfo {author} {\bibfnamefont {J.~M.}\ \bibnamefont {Horowitz}},\ and\ \bibinfo {author} {\bibfnamefont {J.~M.~R.}\ \bibnamefont {Parrondo}},\ }\bibfield {title} {\bibinfo {title} {{Inferring broken detailed balance in the absence of observable currents}},\ }\href {https://doi.org/10.1038/s41467-019-11051-w} {\bibfield {journal} {\bibinfo {journal} {Nat. Commun.}\ }\textbf {\bibinfo {volume} {10}},\ \bibinfo {pages} {3542} (\bibinfo {year} {2019})}\BibitemShut {NoStop}%
\bibitem [{\citenamefont {Otsubo}\ \emph {et~al.}(2020)\citenamefont {Otsubo}, \citenamefont {Ito}, \citenamefont {Dechant},\ and\ \citenamefont {Sagawa}}]{Otsubo.2020.PRE}%
 \BibitemOpen
 \bibfield {author} {\bibinfo {author} {\bibfnamefont {S.}~\bibnamefont {Otsubo}}, \bibinfo {author} {\bibfnamefont {S.}~\bibnamefont {Ito}}, \bibinfo {author} {\bibfnamefont {A.}~\bibnamefont {Dechant}},\ and\ \bibinfo {author} {\bibfnamefont {T.}~\bibnamefont {Sagawa}},\ }\bibfield {title} {\bibinfo {title} {{Estimating entropy production by machine learning of short-time fluctuating currents}},\ }\href {https://doi.org/10.1103/PhysRevE.101.062106} {\bibfield {journal} {\bibinfo {journal} {Phys. Rev. E}\ }\textbf {\bibinfo {volume} {101}},\ \bibinfo {pages} {062106} (\bibinfo {year} {2020})}\BibitemShut {NoStop}%
\bibitem [{\citenamefont {Van~der Meer}\ \emph {et~al.}(2022)\citenamefont {Van~der Meer}, \citenamefont {Ertel},\ and\ \citenamefont {Seifert}}]{van2022thermodynamic}%
 \BibitemOpen
 \bibfield {author} {\bibinfo {author} {\bibfnamefont {J.}~\bibnamefont {Van~der Meer}}, \bibinfo {author} {\bibfnamefont {B.}~\bibnamefont {Ertel}},\ and\ \bibinfo {author} {\bibfnamefont {U.}~\bibnamefont {Seifert}},\ }\bibfield {title} {\bibinfo {title} {Thermodynamic inference in partially accessible markov networks: A unifying perspective from transition-based waiting time distributions},\ }\href {https://doi.org/10.1103/PhysRevX.12.031025} {\bibfield {journal} {\bibinfo {journal} {Phys. Rev. X}\ }\textbf {\bibinfo {volume} {12}},\ \bibinfo {pages} {031025} (\bibinfo {year} {2022})}\BibitemShut {NoStop}%
\bibitem [{\citenamefont {Bakewell-Smith}\ \emph {et~al.}(2023)\citenamefont {Bakewell-Smith}, \citenamefont {Girotti}, \citenamefont {M\u{a}d\u{a}lin},\ and\ \citenamefont {Garrahan}}]{Garrahan_PRL_2023}%
 \BibitemOpen
 \bibfield {author} {\bibinfo {author} {\bibfnamefont {G.}~\bibnamefont {Bakewell-Smith}}, \bibinfo {author} {\bibfnamefont {F.}~\bibnamefont {Girotti}}, \bibinfo {author} {\bibfnamefont {G.}~\bibnamefont {M\u{a}d\u{a}lin}},\ and\ \bibinfo {author} {\bibfnamefont {J.~P.}\ \bibnamefont {Garrahan}},\ }\bibfield {title} {\bibinfo {title} {General upper bounds on fluctuations of trajectory observables},\ }\href {https://doi.org/10.1103/PhysRevLett.131.197101} {\bibfield {journal} {\bibinfo {journal} {Phys. Rev. Lett.}\ }\textbf {\bibinfo {volume} {131}},\ \bibinfo {pages} {197101} (\bibinfo {year} {2023})}\BibitemShut {NoStop}%
\bibitem [{\citenamefont {Van~Vu}(2025)}]{vanvu2024}%
 \BibitemOpen
 \bibfield {author} {\bibinfo {author} {\bibfnamefont {T.}~\bibnamefont {Van~Vu}},\ }\bibfield {title} {\bibinfo {title} {Fundamental bounds on precision and response for quantum trajectory observables},\ }\href {https://doi.org/10.1103/PRXQuantum.6.010343} {\bibfield {journal} {\bibinfo {journal} {PRX Quantum}\ }\textbf {\bibinfo {volume} {6}},\ \bibinfo {pages} {010343} (\bibinfo {year} {2025})}\BibitemShut {NoStop}%
\bibitem [{\citenamefont {Bakewell-Smith}\ \emph {et~al.}(2024)\citenamefont {Bakewell-Smith}, \citenamefont {Girotti}, \citenamefont {Guţ\u{a}},\ and\ \citenamefont {Garrahan}}]{bakewellsmith2024}%
 \BibitemOpen
 \bibfield {author} {\bibinfo {author} {\bibfnamefont {G.}~\bibnamefont {Bakewell-Smith}}, \bibinfo {author} {\bibfnamefont {F.}~\bibnamefont {Girotti}}, \bibinfo {author} {\bibfnamefont {M.}~\bibnamefont {Guţ\u{a}}},\ and\ \bibinfo {author} {\bibfnamefont {J.~P.}\ \bibnamefont {Garrahan}},\ }\href {https://arxiv.org/abs/2405.09669} {\bibinfo {title} {Bounds on fluctuations of first passage times for counting observables in classical and quantum markov processes}} (\bibinfo {year} {2024}),\ \Eprint {https://arxiv.org/abs/2405.09669} {arXiv:2405.09669} \BibitemShut {NoStop}%
\bibitem [{\citenamefont {Reimann}\ \emph {et~al.}(2001)\citenamefont {Reimann}, \citenamefont {Van~den Broeck}, \citenamefont {Linke}, \citenamefont {H\"anggi}, \citenamefont {Rubi},\ and\ \citenamefont {P\'erez-Madrid}}]{Reimann_PRL_2001}%
 \BibitemOpen
 \bibfield {author} {\bibinfo {author} {\bibfnamefont {P.}~\bibnamefont {Reimann}}, \bibinfo {author} {\bibfnamefont {C.}~\bibnamefont {Van~den Broeck}}, \bibinfo {author} {\bibfnamefont {H.}~\bibnamefont {Linke}}, \bibinfo {author} {\bibfnamefont {P.}~\bibnamefont {H\"anggi}}, \bibinfo {author} {\bibfnamefont {J.~M.}\ \bibnamefont {Rubi}},\ and\ \bibinfo {author} {\bibfnamefont {A.}~\bibnamefont {P\'erez-Madrid}},\ }\bibfield {title} {\bibinfo {title} {Giant acceleration of free diffusion by use of tilted periodic potentials},\ }\href {https://doi.org/10.1103/PhysRevLett.87.010602} {\bibfield {journal} {\bibinfo {journal} {Phys. Rev. Lett.}\ }\textbf {\bibinfo {volume} {87}},\ \bibinfo {pages} {010602} (\bibinfo {year} {2001})}\BibitemShut {NoStop}%
\bibitem [{\citenamefont {Reimann}\ \emph {et~al.}(2002)\citenamefont {Reimann}, \citenamefont {Van~den Broeck}, \citenamefont {Linke}, \citenamefont {H\"anggi}, \citenamefont {Rubi},\ and\ \citenamefont {P\'erez-Madrid}}]{Reimann_PRE_2002}%
 \BibitemOpen
 \bibfield {author} {\bibinfo {author} {\bibfnamefont {P.}~\bibnamefont {Reimann}}, \bibinfo {author} {\bibfnamefont {C.}~\bibnamefont {Van~den Broeck}}, \bibinfo {author} {\bibfnamefont {H.}~\bibnamefont {Linke}}, \bibinfo {author} {\bibfnamefont {P.}~\bibnamefont {H\"anggi}}, \bibinfo {author} {\bibfnamefont {J.~M.}\ \bibnamefont {Rubi}},\ and\ \bibinfo {author} {\bibfnamefont {A.}~\bibnamefont {P\'erez-Madrid}},\ }\bibfield {title} {\bibinfo {title} {Diffusion in tilted periodic potentials: Enhancement, universality, and scaling},\ }\href {https://doi.org/10.1103/PhysRevE.65.031104} {\bibfield {journal} {\bibinfo {journal} {Phys. Rev. E}\ }\textbf {\bibinfo {volume} {65}},\ \bibinfo {pages} {031104} (\bibinfo {year} {2002})}\BibitemShut {NoStop}%
\bibitem [{\citenamefont {Marchenko}\ \emph {et~al.}(2023)\citenamefont {Marchenko}, \citenamefont {Aksenova}, \citenamefont {Marchenko}, \citenamefont {\L{}uczka},\ and\ \citenamefont {Spiechowicz}}]{Marchenko_PRE_2023}%
 \BibitemOpen
 \bibfield {author} {\bibinfo {author} {\bibfnamefont {I.~G.}\ \bibnamefont {Marchenko}}, \bibinfo {author} {\bibfnamefont {V.}~\bibnamefont {Aksenova}}, \bibinfo {author} {\bibfnamefont {I.~I.}\ \bibnamefont {Marchenko}}, \bibinfo {author} {\bibfnamefont {J.}~\bibnamefont {\L{}uczka}},\ and\ \bibinfo {author} {\bibfnamefont {J.}~\bibnamefont {Spiechowicz}},\ }\bibfield {title} {\bibinfo {title} {Temperature anomalies of oscillating diffusion in ac-driven periodic systems},\ }\href {https://doi.org/10.1103/PhysRevE.107.064116} {\bibfield {journal} {\bibinfo {journal} {Phys. Rev. E}\ }\textbf {\bibinfo {volume} {107}},\ \bibinfo {pages} {064116} (\bibinfo {year} {2023})}\BibitemShut {NoStop}%
\bibitem [{\citenamefont {Hayashi}\ \emph {et~al.}(2015)\citenamefont {Hayashi}, \citenamefont {Sasaki}, \citenamefont {Nakamura}, \citenamefont {Kudo}, \citenamefont {Inoue}, \citenamefont {Noji},\ and\ \citenamefont {Hayashi}}]{Hayashi_PRL_2015}%
 \BibitemOpen
 \bibfield {author} {\bibinfo {author} {\bibfnamefont {R.}~\bibnamefont {Hayashi}}, \bibinfo {author} {\bibfnamefont {K.}~\bibnamefont {Sasaki}}, \bibinfo {author} {\bibfnamefont {S.}~\bibnamefont {Nakamura}}, \bibinfo {author} {\bibfnamefont {S.}~\bibnamefont {Kudo}}, \bibinfo {author} {\bibfnamefont {Y.}~\bibnamefont {Inoue}}, \bibinfo {author} {\bibfnamefont {H.}~\bibnamefont {Noji}},\ and\ \bibinfo {author} {\bibfnamefont {K.}~\bibnamefont {Hayashi}},\ }\bibfield {title} {\bibinfo {title} {Giant acceleration of diffusion observed in a single-molecule experiment on ${\mathrm{f}}_{1}\text{\ensuremath{-}}\mathrm{ATPase}$},\ }\href {https://doi.org/10.1103/PhysRevLett.114.248101} {\bibfield {journal} {\bibinfo {journal} {Phys. Rev. Lett.}\ }\textbf {\bibinfo {volume} {114}},\ \bibinfo {pages} {248101} (\bibinfo {year} {2015})}\BibitemShut {NoStop}%
\bibitem [{\citenamefont {Shinagawa}\ and\ \citenamefont {Sasaki}(2016)}]{Shinagawa2016}%
 \BibitemOpen
 \bibfield {author} {\bibinfo {author} {\bibfnamefont {R.}~\bibnamefont {Shinagawa}}\ and\ \bibinfo {author} {\bibfnamefont {K.}~\bibnamefont {Sasaki}},\ }\bibfield {title} {\bibinfo {title} {Enhanced diffusion of molecular motors in the presence of adenosine triphosphate and external force},\ }\href {https://doi.org/10.7566/JPSJ.85.064004} {\bibfield {journal} {\bibinfo {journal} {Journal of the Physical Society of Japan}\ }\textbf {\bibinfo {volume} {85}},\ \bibinfo {pages} {064004} (\bibinfo {year} {2016})},\ \Eprint {https://arxiv.org/abs/https://doi.org/10.7566/JPSJ.85.064004} {https://doi.org/10.7566/JPSJ.85.064004} \BibitemShut {NoStop}%
\bibitem [{\citenamefont {Lee}\ and\ \citenamefont {Grier}(2006)}]{Lee_PRL_2006}%
 \BibitemOpen
 \bibfield {author} {\bibinfo {author} {\bibfnamefont {S.-H.}\ \bibnamefont {Lee}}\ and\ \bibinfo {author} {\bibfnamefont {D.~G.}\ \bibnamefont {Grier}},\ }\bibfield {title} {\bibinfo {title} {Giant colloidal diffusivity on corrugated optical vortices},\ }\href {https://doi.org/10.1103/PhysRevLett.96.190601} {\bibfield {journal} {\bibinfo {journal} {Phys. Rev. Lett.}\ }\textbf {\bibinfo {volume} {96}},\ \bibinfo {pages} {190601} (\bibinfo {year} {2006})}\BibitemShut {NoStop}%
\bibitem [{\citenamefont {Kim}\ \emph {et~al.}(2017)\citenamefont {Kim}, \citenamefont {Bowman}, \citenamefont {Del Bonis-O'Donnell}, \citenamefont {Matzavinos},\ and\ \citenamefont {Stein}}]{Kim_PRL_2017}%
 \BibitemOpen
 \bibfield {author} {\bibinfo {author} {\bibfnamefont {D.}~\bibnamefont {Kim}}, \bibinfo {author} {\bibfnamefont {C.}~\bibnamefont {Bowman}}, \bibinfo {author} {\bibfnamefont {J.~T.}\ \bibnamefont {Del Bonis-O'Donnell}}, \bibinfo {author} {\bibfnamefont {A.}~\bibnamefont {Matzavinos}},\ and\ \bibinfo {author} {\bibfnamefont {D.}~\bibnamefont {Stein}},\ }\bibfield {title} {\bibinfo {title} {Giant acceleration of dna diffusion in an array of entropic barriers},\ }\href {https://doi.org/10.1103/PhysRevLett.118.048002} {\bibfield {journal} {\bibinfo {journal} {Phys. Rev. Lett.}\ }\textbf {\bibinfo {volume} {118}},\ \bibinfo {pages} {048002} (\bibinfo {year} {2017})}\BibitemShut {NoStop}%
\bibitem [{\citenamefont {Suma}\ \emph {et~al.}(2023)\citenamefont {Suma}, \citenamefont {Carnevale},\ and\ \citenamefont {Micheletti}}]{Suma_PRL_2023}%
 \BibitemOpen
 \bibfield {author} {\bibinfo {author} {\bibfnamefont {A.}~\bibnamefont {Suma}}, \bibinfo {author} {\bibfnamefont {V.}~\bibnamefont {Carnevale}},\ and\ \bibinfo {author} {\bibfnamefont {C.}~\bibnamefont {Micheletti}},\ }\bibfield {title} {\bibinfo {title} {Nonequilibrium thermodynamics of dna nanopore unzipping},\ }\href {https://doi.org/10.1103/PhysRevLett.130.048101} {\bibfield {journal} {\bibinfo {journal} {Phys. Rev. Lett.}\ }\textbf {\bibinfo {volume} {130}},\ \bibinfo {pages} {048101} (\bibinfo {year} {2023})}\BibitemShut {NoStop}%
\bibitem [{\citenamefont {Risken}(1996)}]{risken1996fokker}%
 \BibitemOpen
 \bibfield {author} {\bibinfo {author} {\bibfnamefont {H.}~\bibnamefont {Risken}},\ }\href@noop {} {\emph {\bibinfo {title} {The {Fokker}--{Planck} Equation: Methods of Solution and Applications}}},\ \bibinfo {edition} {2nd}\ ed.\ (\bibinfo {publisher} {Springer},\ \bibinfo {address} {New York},\ \bibinfo {year} {1996})\BibitemShut {NoStop}%
\bibitem [{\citenamefont {Gardiner}(2009)}]{gardiner2009stochastic}%
 \BibitemOpen
 \bibfield {author} {\bibinfo {author} {\bibfnamefont {C.}~\bibnamefont {Gardiner}},\ }\href@noop {} {\emph {\bibinfo {title} {Stochastic Methods: A Handbook for the Natural and Social Sciences}}},\ \bibinfo {edition} {4th}\ ed.,\ Springer Series in Synergetics\ (\bibinfo {publisher} {Springer},\ \bibinfo {address} {Berlin Heidelberg},\ \bibinfo {year} {2009})\BibitemShut {NoStop}%
\bibitem [{\citenamefont {Seifert}(2012)}]{Seifert_2012}%
 \BibitemOpen
 \bibfield {author} {\bibinfo {author} {\bibfnamefont {U.}~\bibnamefont {Seifert}},\ }\bibfield {title} {\bibinfo {title} {Stochastic thermodynamics, fluctuation theorems and molecular machines},\ }\href {https://doi.org/10.1088/0034-4885/75/12/126001} {\bibfield {journal} {\bibinfo {journal} {Rep. Prog. Phys.}\ }\textbf {\bibinfo {volume} {75}},\ \bibinfo {pages} {126001} (\bibinfo {year} {2012})}\BibitemShut {NoStop}%
\bibitem [{\citenamefont {Horowitz}\ and\ \citenamefont {Gingrich}(2017)}]{horowitz2017proof}%
 \BibitemOpen
 \bibfield {author} {\bibinfo {author} {\bibfnamefont {J.~M.}\ \bibnamefont {Horowitz}}\ and\ \bibinfo {author} {\bibfnamefont {T.~R.}\ \bibnamefont {Gingrich}},\ }\bibfield {title} {\bibinfo {title} {Proof of the finite-time thermodynamic uncertainty relation for steady-state currents},\ }\href {https://doi.org/10.1103/PhysRevE.96.020103} {\bibfield {journal} {\bibinfo {journal} {Phys. Rev. E}\ }\textbf {\bibinfo {volume} {96}},\ \bibinfo {pages} {020103} (\bibinfo {year} {2017})}\BibitemShut {NoStop}%
\bibitem [{\citenamefont {Mori}\ and\ \citenamefont {Shirai}(2023)}]{mori2023symmetrized}%
 \BibitemOpen
 \bibfield {author} {\bibinfo {author} {\bibfnamefont {T.}~\bibnamefont {Mori}}\ and\ \bibinfo {author} {\bibfnamefont {T.}~\bibnamefont {Shirai}},\ }\bibfield {title} {\bibinfo {title} {Symmetrized liouvillian gap in markovian open quantum systems},\ }\href {https://doi.org/10.1103/PhysRevLett.130.230404} {\bibfield {journal} {\bibinfo {journal} {Phys. Rev. Lett.}\ }\textbf {\bibinfo {volume} {130}},\ \bibinfo {pages} {230404} (\bibinfo {year} {2023})}\BibitemShut {NoStop}%
\bibitem [{sup()}]{supple}%
 \BibitemOpen
 \href@noop {} {\bibinfo {title} {Supplementary material}}\BibitemShut {NoStop}%
\bibitem [{\citenamefont {Hartich}\ and\ \citenamefont {Godec}(2021)}]{Hartich_PRL_2021}%
 \BibitemOpen
 \bibfield {author} {\bibinfo {author} {\bibfnamefont {D.}~\bibnamefont {Hartich}}\ and\ \bibinfo {author} {\bibfnamefont {A.~c.~v.}\ \bibnamefont {Godec}},\ }\bibfield {title} {\bibinfo {title} {Thermodynamic uncertainty relation bounds the extent of anomalous diffusion},\ }\href {https://doi.org/10.1103/PhysRevLett.127.080601} {\bibfield {journal} {\bibinfo {journal} {Phys. Rev. Lett.}\ }\textbf {\bibinfo {volume} {127}},\ \bibinfo {pages} {080601} (\bibinfo {year} {2021})}\BibitemShut {NoStop}%
\end{thebibliography}
\end{document}


\title{Supplemental Material for \texorpdfstring{\\}{ }``Inverse Thermodynamic Uncertainty Relation and Entropy Production''}
\author{Van Tuan Vo}

\affiliation{Department of Physics, Kyoto University, Kyoto 606-8502, Japan}
\author{Andreas Dechant}

\affiliation{Department of Physics, Kyoto University, Kyoto 606-8502, Japan}
\author{Keiji Saito}

\affiliation{Department of Physics, Kyoto University, Kyoto 606-8502, Japan}

\maketitle
\appendix
\tableofcontents
\onecolumngrid
\newpage
\part{iTUR for Overdamped Systems}
\section{Overdamped Langevin Systems: iTUR Derivation Overview}

This section presents a brief derivation of the long-time and finite-time inverse thermodynamic uncertainty relations (iTURs) for overdamped Langevin systems, employing a variational approach.

\subsection{System Setup and Definitions}

We consider a system described by the overdamped Langevin equation:
\begin{equation}
\dot{\bm{x}}_t = \bm{A}(\bm{x}_t) + \sqrt{2\bm{B}}\bm{\xi}(t),
\label{langevin_intro}
\end{equation}
where $\bm{x}_t \in \mathbb{R}^N$ is the system's state, $\bm{A}(\bm{x}) \in \mathbb{R}^N$ is the drift, $\bm{B} \in \mathbb{R}^{N \times N}$ is a symmetric, positive-definite diffusion matrix, and $\bm{\xi}(t)$ is a vector of independent white noise processes with $\langle \xi_i(t) \rangle = 0$ and $\langle \xi_i(t) \xi_j(t') \rangle = \delta_{ij}\delta(t-t')$.  The system is assumed to reach a steady state $p_{\text{st}}(\bm{x})$, satisfying the steady-state Fokker--Planck equation:
\begin{equation}
0 = -\nabla \cdot \Bigl( \bm{\nu}_{\text{st}}(\bm{x})p_{\text{st}}(\bm{x}) \Bigr),
\end{equation}
where $\bm{\nu}_{\text{st}}(\bm{x}) := \bm{A}(\bm{x}) - \bm{B}\nabla\ln p_{\text{st}}(\bm{x})$ is the local mean velocity.

We investigate fluctuations via a general stochastic current, $\mathcal{J}$, a time-integrated quantity along a trajectory $\Gamma_\tau = \{\bm{x}_t, t \in [0,\tau]\}$:
\begin{equation}
\mathcal{J}(\Gamma_\tau) = \int_0^\tau dt  \bm{\theta}_t(\bm{x}_t) \circ \dot{\bm{x}}_t,
\label{current_intro}
\end{equation}
where $\bm{\theta}_t(\bm{x})$ is a vector-valued function, and $\circ$ denotes the Stratonovich product.  Fluctuations are characterized by the cumulant generating function (CGF):
\begin{equation}
K_{\mathcal{J}}(h,\tau) = \ln \Bigl\langle e^{\Re{h^* \mathcal{J}}} \Bigr\rangle_{P_\tau(\Gamma_\tau)},
\label{cgf_intro}
\end{equation}
where $h \in \mathbb{C}$, $*$ is complex conjugation, and $\langle \diamond \rangle_{P}$ averages with probability measure $P$.

\subsection{Variational Approach and Modified Dynamics}

The Donsker-Varadhan principle (see Section~\ref{KL_Overdamped}) expresses the CGF as:
\begin{equation*}
K_{\mathcal{J}}(h,\tau) = \sup_{Q_\tau} \left[ \Re{h^* \langle \mathcal{J} \rangle_{Q_\tau}} - D_{\text{KL}}(Q_\tau \| P_\tau) \right],
\label{kl_intro}
\end{equation*}
where the supremum is over all path probability measures, $Q_\tau$. 
We demonstrate (see Section~\ref{sec:evaluating_variational_expression}) that the optimal measure $Q_\tau$ belongs to the family of measures generated by modified Langevin dynamics of the form:
\begin{equation*}
\dot{\bm{x}}_t = \bm{A}(\bm{x}_t) + \bm{y}_t(\bm{x}_t) + \sqrt{2\bm{B}}\bm{\xi}(t),
\label{modified_dynamics_intro}
\end{equation*}
with a biased drift,  \(\bm{y}_t(\bm{x}_t)\). The modified probability density, \(q_t(\bm{x})\), and the biased drift, \(\bm{y}_t(\bm{x})\), are coupled through the modified Fokker--Planck equation. A Lagrangian formulation  yields the optimal bias:
\begin{equation*}
\hat{\bm{y}}_t(\bm{x}) = 2\bm{B} \left( \Re\{h^* \bm{\theta}_t(\bm{x})\} - \nabla \hat{\phi}_t(\bm{x}) \right),
\label{optimal_bias_intro}
\end{equation*}
where $\hat{\phi}_t(\bm{x})$ is a Lagrange multiplier function that emerges from enforcing the modified Fokker--Planck equation. 
Substituting the optimal bias back into the variational expression for the CGF and using the relationship involving $\hat{\phi}_t(\bm{x})$ to eliminate $\bm{y}_t(\bm{x}_t)$, we obtain (see Section \ref{subsec:Deriving_Max_Min_CGF})
\begin{equation*}
\begin{aligned}
K_{\mathcal{J}}(h,\tau)
=\sup_{q_t}
\inf_{\phi_t}\Biggl\{ \int_0^\tau dt \Bigl[
&\Bigl\langle \Bigl(\Re\{h^*\bm{\theta}_t\}-\nabla\phi_t\Bigr)\cdot\left( \bm{\nu}_{\text{st}} - \bm{B} \nabla \ln \frac{q_t}{p_{\text{st}}} \right)\Bigr\rangle_{q_t}\\
&+\Bigl\langle \Bigl(\Re\{h^*\bm{\theta}_t\}-\nabla\phi_t\Bigr)\cdot\bm{B}\Bigl(\Re\{h^*\bm{\theta}_t\}-\nabla\phi_t\Bigr)\Bigr\rangle_{q_t}\\
&+\Bigl\langle \phi_t \partial_t\ln q_t\Bigr\rangle_{q_t}
\Bigr]-D_{\text{KL}}(q_0\|p_\text{st})\Biggr\}.
\end{aligned}
\end{equation*}
Minimization over $\phi_t$ ensures that  $q_t$ satisfies the modified Fokker--Planck equation.
Expanding $q_t$ and $\phi_t$ for small $h$:
\begin{equation*}
\begin{aligned}
\phi_t(\bm{x}) &= -\Re\{h^* \eta_t(\bm{x})\} - \Re\bigl\{h (\chi_t(\bm{x})-\langle\chi_t\rangle_{p_\text{st}} )\bigr\}+ O(|h|^2), \\[1mm]
q_t(\bm{x}) &= p_\text{st}(\bm{x}) \biggl (1 + 2\Re\bigl\{h (\chi_t(\bm{x})-\langle\chi_t\rangle_{p_\text{st}} )\bigr\} \biggr)+ O(|h |^2),
\end{aligned}
\end{equation*}
and substituting into the representation of $K_{\mathcal{J}}(h,\tau)$, we extract a variational representation for the variance (see Section \ref{Variance_O}):
\begin{equation*}
\begin{aligned}
\text{Var}(\mathcal{J}) = 2 \sup_{\chi_t} \inf_{\eta_t}\Biggl[ \int_0^\tau dt \Bigl( & 2 \Re\Bigl\{ \text{Cov}_\text{st}\Bigl( (\bm{\theta}_t + \nabla \eta_t)\cdot\bm{\nu}_\text{st}, \chi_t^* \Bigr) - \text{Cov}_\text{st}\Bigl( \eta_t, \partial_t\chi_t^* \Bigr) \Bigr\} \\
&+ \Bigl\langle (\bm{\theta}_t + \nabla \eta_t)^*\cdot\bm{B} (\bm{\theta}_t + \nabla \eta_t) \Bigr\rangle_{p_\text{st}}
- \Bigl\langle \nabla\chi_t^*\cdot\bm{B} \nabla\chi_t \Bigr\rangle_{p_\text{st}} \Bigr)
- \Bigl( \text{Var}(\chi_\tau) + \text{Var}(\chi_0) \Bigr) \Biggr].
\end{aligned}
\label{variance_variational_intro}
\end{equation*}
\subsection{Long-Time iTUR, and Finite-Time iTUR}

$D_\infty $ has the variational representation:
\begin{equation*}
\begin{aligned}
    D_\infty := \lim_{\tau\to\infty}\frac{\text{Var}(\mathcal{J})}{2\tau} &=   \sup_{\chi} \inf_{\eta} \Biggl[ 2 \text{Cov}_{\text{st}} \Bigl( (\bm{\theta} + \nabla \eta) \cdot \bm{\nu}_{\text{st}},\chi \Bigr) \\
    &\quad\quad + \Bigl\langle (\bm{\theta} + \nabla \eta) \cdot \bm{B} (\bm{\theta} + \nabla \eta) \Bigr\rangle_{p_{\text{st}}} - \Bigl\langle \nabla\chi \cdot \bm{B} \nabla\chi \Bigr\rangle_{p_{\text{st}}} \Biggr].
\end{aligned}
\label{variational_expression_longtime_orig_notation}
\end{equation*}
Choosing the specific function $\eta = \bm{0}$ yields an upper bound. Defining  $D_0 := \langle \bm{\theta} \cdot \bm{B} \bm{\theta} \rangle_{p_{\text{st}}}$, this gives:
\begin{equation*}
    D_\infty \leq D_0 + \sup_{\chi} \left[ 2 \text{Cov}_{\text{st}} \bigl( \bm{\theta} \cdot \bm{\nu}_{\text{st}},\chi \bigr) - \Bigl\langle \nabla\chi \cdot \bm{B} \nabla\chi \Bigr\rangle_{p_{\text{st}}} \right].
\end{equation*}
Using the Cauchy-Schwarz inequality, $\text{Cov}_{\text{st}}(X, Y)^2 \leq \text{Var}_{\text{st}}(X) \text{Var}_{\text{st}}(Y)$ and optimizing, we get:
\begin{equation*}
    D_\infty \leq D_0 + \sup_{\chi} \left[ \frac{\text{Var}_{\text{st}}(\bm{\theta} \cdot \bm{\nu}_{\text{st}}) \text{Var}_{\text{st}}(\chi)}{\langle \nabla\chi \cdot \bm{B} \nabla\chi \rangle_{p_{\text{st}}}} \right].
\label{upper_bound_step2_longtime_orig_notation}
\end{equation*}
The spectral gap $\lambda$ of the symmetrized Fokker-Planck operator is variationally defined (see Eq.~\eqref{O_spectral_gap}):
\begin{equation*}
    \lambda = \inf_{\chi, \langle\chi\rangle_{p_\text{st}}=0} \left[ \frac{\langle \nabla\chi \cdot \bm{B} \nabla\chi \rangle_{p_{\text{st}}}}{\text{Var}_{\text{st}}(\chi)} \right].
\label{O_spectral_gap_orig_notation}
\end{equation*}
This leads directly to the upper bound:
\begin{equation*}
    D_\infty \leq D_0 + \frac{1}{\lambda} \text{Var}_{\text{st}} \bigl( \bm{\theta} \cdot \bm{\nu}_{\text{st}} \bigr).
\label{upper_bound_final_longtime_orig_notation}
\end{equation*}
Bounding $\text{Var}_{\text{st}} \bigl( \bm{\theta} \cdot \bm{\nu}_{\text{st}} \bigr)$ using the Cauchy-Schwarz inequality yields a bound in terms of the steady-state entropy production rate, $\sigma_{\text{st}}$, and the average current, $j_{\text{st}} = \langle \bm{\theta} \cdot \bm{\nu}_{\text{st}} \rangle_{p_{\text{st}}}$:
\begin{equation*}
 \operatorname{Var}_{\mathrm{st}}(\bm{\theta}\cdot\bm{\nu}_{\mathrm{st}}) \leq  D_{\text{max}} \sigma_{\mathrm{st}} - j_{\mathrm{st}}^2,
\end{equation*}
where  $D_{\text{max}} = \sup_{\bm{x}} \{\bm{\theta}(\bm{x}) \cdot \bm{B} \bm{\theta}(\bm{x})\}$.
Combining these, we arrive at the long-time iTUR:

\begin{equation*}
D_\infty \leq D_0 + \frac{1}{\lambda} \left( D_{\text{max}} \sigma_{\text{st}} - j_{\text{st}}^2 \right).
\end{equation*}

Deriving the finite-time iTUR requires the analysis of frequency-dependent fluctuations via the spectral density, $S(\omega)$.  With $\bm{\theta}_t(\bm{x}) = e^{i\omega t}\bm{\theta}(\bm{x})$ and $\mathcal{J}_\omega = \int_0^\tau dt  e^{i\omega t} \bm{\theta}(\bm{x}_t) \circ \dot{\bm{x}}_t$, the spectral density is given by: 
\begin{equation*}
S(\omega) = \lim_{\tau \to \infty} \frac{\text{Var}(\mathcal{J}_\omega)}{\tau}.
\end{equation*}
Using variational representation of the variance, with time-dependent auxiliary functions $\chi_t(\bm{x}) = e^{i\omega t}(\chi'(\bm{x}) + i\chi''(\bm{x}))$ and $\eta_t(\bm{x}) = -ie^{i\omega t}\phi(\bm{x})$, we obtain a variational representation for $S(\omega)$.  We then find an upper bound on $S(\omega)$ (Section~\ref{spectral_gap_upperbound_overdamped}):
\begin{equation*}
S(\omega) \leq 2D_0 + 2 \operatorname{Var}_{\mathrm{st}}\Bigl(\bm{\theta}\cdot\bm{\nu}_{\mathrm{st}}\Bigr)
\begin{cases}
\displaystyle \frac{1}{\lambda}\frac{1}{1+(\omega/\bar{\lambda})^2}, & \text{for } \omega < \bar{\lambda}, \\[1mm]
\displaystyle \frac{\bar{\lambda}}{2\lambda\omega}, & \text{for }  \omega \geq \bar{\lambda},
\end{cases}
\end{equation*}
where $\bar{\lambda} = \sqrt{\lambda (\lambda + \sigma_{\text{max}})}$, and $\sigma_{\text{max}} = \sup_{\bm{x}} \{ \bm{\nu}_{\text{st}}(\bm{x}) \cdot \bm{B}^{-1} \bm{\nu}_{\text{st}}(\bm{x}) \}$ is the maximal local entropy production rate.
$D_\tau$ is related to $S(\omega)$ by the Wiener--Khinchin theorem:
\begin{equation*}
D_\tau = \frac{\operatorname{Var}(\mathcal{J})}{2\tau} =  \frac{1}{\pi} \int_0^\infty d\omega   S(\omega) \frac{1 - \cos(\omega \tau)}{\omega^2 \tau}.
\end{equation*}
Substituting the bound on $S(\omega)$ and integrating (Section~\ref{Finite_Time_iTUR_Overdamped_derivation}) yields the finite-time iTUR:
\begin{equation*}
D_\tau \leq D_0 + \frac{   g(\bar{\lambda}\tau)}{\lambda} \left( D_{\text{max}} \sigma_{\text{st}} - j_{\text{st}}^2 \right),
\end{equation*}
where
\begin{equation*}
 g(x) := \frac{2}{\pi x} \left( \int_0^1 dz  \frac{1-\cos(zx)}{z^2(1+z^2)}
+\int_1^\infty dz  \frac{1-\cos(zx)}{2z^3} \right).
\end{equation*}
As $\tau \to \infty$, $ g(\bar{\lambda} \tau) \to 1$, recovering the long-time iTUR.

\section{Variational Methods for Overdamped Systems: CGF and Variance}
\label{VR_O}
This section develops the variational framework for the CGF and variance of currents in overdamped Langevin systems. Applying the Donsker--Varadhan principle, we link the CGF to a Kullback--Leibler divergence optimization. We demonstrate the optimal path measure arises from modified Langevin dynamics with a biasing drift, yielding a max-min CGF representation. A second-order expansion of this representation then provides the corresponding max-min variational formula for the variance. 

\subsection{Donsker--Varadhan Principle for the Cumulant Generating Function}
\label{KL_Overdamped}
We derive variational representations for the cumulant generating function and the variance of a stochastic current.
The stochastic current is defined as
\begin{equation}
\mathcal{J}(\Gamma_\tau):=\int_0^\tau dt \bm{\theta}_t(\bm{x}_t)\circ\dot{\bm{x}}_t,
\end{equation}
where $\bm{\theta}_t(\bm{x}_t)\in\mathbb{C}$ is a complex-vector-valued function, $\Gamma_\tau=\{\bm{x}_t,  t\in[0,\tau]\}$ is the trajectory, and $\circ$ denotes the Stratonovich interpretation.
The cumulant generating function is given by
\begin{equation}
K_{\mathcal{J}}(h,\tau):=\ln\Bigl\langle e^{\Re{h^*\mathcal{J}}}\Bigr\rangle_{P_\tau},
\end{equation}
with $h\in\mathbb{C}$ and $\langle\cdot\rangle_{P_\tau}$ denoting the average with respect to the path probability:
\begin{equation}
P_\tau(\Gamma_\tau)=Z^{-1}\exp\Bigl[-\int_0^\tau dt S_t(\bm{x}_t)\Bigr] p_\text{st}(\bm{x}_0).
\end{equation}
where $Z$ is a partition function, representing the equilibrium properties of the system that are not directly influenced by the drift coefficient. 
The action functional is defined as
\begin{equation}
S_t(\bm{x}_t):=\frac{1}{4}\Bigl(\dot{\bm{x}}_t-\bm{A}(\bm{x}_t)\Bigr)\cdot\Bigl(\bm{B}^{-1}\bigl[\dot{\bm{x}}_t-\bm{A}(\bm{x}_t)\bigr]\Bigr),
\end{equation}
where the dot $\cdot$ denotes the Euclidean inner product. 

The Donsker--Varadhan variational representation expresses the CGF as \cite{Donsker1983}:
\begin{equation}
K_{\mathcal{J}}(h,\tau)=\sup_{Q_\tau}\Bigl[\Re{h^*\langle\mathcal{J}\rangle_{Q_\tau}}-D_{\text{KL}}(Q_\tau\|P_\tau)\Bigr],
\label{eq:donsker_varadhan}
\end{equation}
where $Q_\tau(\Gamma_\tau)$ is any path probability measure, and $D_{\text{KL}}(Q_\tau\|P_\tau)$ is the Kullback-Leibler divergence.

The inequality in Eq.~\eqref{eq:donsker_varadhan} follows from Jensen's inequality applied to the logarithm. Starting with the definition of $K_{\mathcal{J}}(h,\tau)$ and introducing an arbitrary path measure $Q_\tau$:
\begin{equation}
\begin{aligned}
K_{\mathcal{J}}(h,\tau)&=\ln\left(\int\mathcal{D}\Gamma_\tau e^{\Re{h^*\mathcal{J}(\Gamma_\tau)}}P_\tau(\Gamma_\tau)\right)\\
&=\ln\left(\int\mathcal{D}\Gamma_\tau e^{\Re{h^*\mathcal{J}(\Gamma_\tau)}+\ln\frac{P_\tau(\Gamma_\tau)}{Q_\tau(\Gamma_\tau)}} Q_\tau(\Gamma_\tau)\right)\\
&\geq \Re{h^*\langle\mathcal{J}\rangle_{Q_\tau}} + \Bigl\langle \ln \frac{Q_\tau(\Gamma_\tau)}{P_\tau(\Gamma_\tau)} \Bigr\rangle_{Q_\tau} \quad \text{(Jensen's Inequality)} \\
&= \Re{h^*\langle\mathcal{J}\rangle_{Q_\tau}}-D_{\text{KL}}(Q_\tau\|P_\tau).
\end{aligned}
\end{equation}
The supremum is achieved because equality in Jensen's inequality holds when the argument of the logarithm is constant. This occurs for the specific tilted measure:
\begin{equation}
\hat{Q}_\tau(\Gamma_\tau):=\frac{e^{\Re{h^*\mathcal{J}(\Gamma_\tau)}}P_\tau(\Gamma_\tau)}
{\Bigl\langle e^{\Re{h^*\mathcal{J}}}\Bigr\rangle_{P_\tau}}.
\end{equation}
which yields the maximum value, thus proving Eq.~\eqref{eq:donsker_varadhan}.

\subsection{Modified Dynamics and the CGF Variational Lower Bound:}
\label{Biased_Overdamped}
We restrict our optimization to path probabilities generated by dynamics with a biased drift \cite{Dechant.2018.JSM}:
\begin{equation}
\dot{\bm{x}}_t=\bm{A}(\bm{x}_t)+\bm{y}_t(\bm{x}_t)+\sqrt{2\bm{B}} \bm{\xi}(t),
\end{equation}
where $\bm{y}_t(\bm{x}_t)$ is the bias added to the original drift $\bm{A}(\bm{x}_t)$.
The probability density of this modified process, \(q_t(\bm{x})\), satisfies a corresponding Fokker-Planck equation:
\begin{equation}
\partial_t q_t = -\nabla \cdot \bm{J}^{\bm{y}}_t, \quad \bm{J}^{\bm{y}}_t = (\bm{A}(\bm{x}) + \bm{y}_t(\bm{x})) q_t - \bm{B} \nabla q_t.
\label{modified_fokker-planck}
\end{equation}
Here, \(\bm{J}^{\bm{y}}_t\) is the probability current for the modified dynamics.
The action for the modified dynamics is given by:
\begin{equation}
\begin{aligned}
S^{\bm{y}}_t(\bm{x}_t)
&:=\frac{1}{4}\Bigl(\dot{\bm{x}}_t-\bm{A}(\bm{x}_t)-\bm{y}_t(\bm{x}_t)\Bigr)\cdot\bm{B}^{-1}\Bigl(\dot{\bm{x}}_t-\bm{A}(\bm{x}_t)-\bm{y}_t(\bm{x}_t)\Bigr)\\
&=S_t(\bm{x}_t)+\frac{1}{4}\Bigl(\bm{y}_t(\bm{x}_t)\cdot\bm{B}^{-1}\bm{y}_t(\bm{x}_t)
-2 \bm{y}_t(\bm{x}_t)\cdot\bm{B}^{-1}\bigl(\dot{\bm{x}}_t-\bm{A}(\bm{x}_t)-\bm{y}_t(\bm{x}_t)\bigr)\Bigr).
\end{aligned}
\end{equation}
Consequently, the ratio of path probabilities between the original and modified dynamics becomes
\begin{equation}
\ln\frac{P_\tau(\Gamma_\tau)}{Q_\tau(\Gamma_\tau)}
=\int_0^\tau dt \left[-\frac{1}{4}\Bigl(\bm{y}_t\cdot\bm{B}^{-1}\bm{y}_t
-2 \bm{y}_t\cdot\bm{B}^{-1}\bigl(\dot{\bm{x}}_t-\bm{A}-\bm{y}_t\bigr)\Bigr)\right]
+\ln\frac{p_\text{st}(\bm{x}_0)}{q_0(\bm{x}_0)}.
\end{equation}
The term with $\dot{\bm{x}}_t-\bm{A}-\bm{y}_t$  vanishes upon taking the expectation $\langle \dots \rangle_{Q_\tau}$ because it corresponds to the expectation of the noise $\sqrt{2\bm{B}}\bm{\xi}(t)$, which is zero.  The lower bound on the CGF is:
\begin{equation}
K_{\mathcal{J}}(h, \tau) \geq  \int_0^\tau dt \int d\bm{x} \left[ \Re\{h^* \bm{\theta}_t\} \cdot \bm{J}^{\bm{y}}_t - \frac{1}{4} \bm{y}_t \cdot \bm{B}^{-1} \bm{y}_t q_t \right] - D_{\text{KL}}(q_0 \| p_{\text{st}}),
\label{lower_bound_explicit}
\end{equation}
The expression on the right-hand side of Eq.~\eqref{lower_bound_explicit} provides a lower bound on the CGF. To find the best such bound, the optimization is performed over pairs $(q_t, \bm{y}_t)$ subject to the constraint of dynamical consistency; specifically, $q_t$ must evolve according to the modified Fokker--Planck equation \eqref{modified_fokker-planck} determined by the bias $\bm{y}_t$.

To incorporate the constraint imposed by the modified Fokker--Planck equation \eqref{modified_fokker-planck}, we introduce a Lagrange multiplier field $\phi_t(\bm{x})$. The constrained optimization in Eq.~\eqref{lower_bound_explicit} can then be solved by finding the saddle point of the following Lagrangian functional $\mathcal{L}(q_t, \dot{q}_t, \phi_t, \bm{y}_t)$:\begin{equation}
\begin{aligned}
\mathcal{L}(q_t, \dot{q}_t, \phi_t, \bm{y}_t) = \int_0^\tau dt \int d\bm{x} \Biggl\{ &\Re\{h^* \bm{\theta}_t(\bm{x})\} \cdot \bm{J}^{\bm{y}}_t(\bm{x}, t) - \frac{1}{4} \bm{y}_t(\bm{x}, t) \cdot \bm{B}^{-1} \bm{y}_t(\bm{x}, t) q_t(\bm{x}, t) \\
&+ \phi_t(\bm{x}, t) \Bigl(\partial_t q_t(\bm{x}, t) + \nabla \cdot \bm{J}^{\bm{y}}_t(\bm{x}, t)\Bigr) \Biggr\} - D_{\text{KL}}(q_0 \| p_{\text{st}}).
\end{aligned}
\label{lagrangian_alpha} 
\end{equation}
The corresponding Lagrangian density, obtained after integration by parts, is:
\begin{equation}
\ell(q_t, \dot{q}_t, \phi_t, \bm{y}_t) =  (\Re\{h^* \bm{\theta}_t\} - \nabla \phi_t) \cdot (\bm{A} + \bm{y}_t - \bm{B} \nabla) q_t - \frac{1}{4} \bm{y}_t \cdot \bm{B}^{-1} \bm{y}_t q_t + \phi_t \partial_t q_t,
\label{density_L}
\end{equation}
where we have used integration by parts on the $\phi_t \nabla \cdot \bm{J}^{\bm{y}}_t$ term, discarding the boundary term.
We are interested in the quantity $G_{\mathcal{J}}(h,\tau)$, which is determined by the following optimization:
\begin{equation}
G_{\mathcal{J}}(h,\tau) := \sup_{q_t} \inf_{\phi_t} \sup_{ \bm{y}_t} \mathcal{L}(q_t, \dot{q}_t, \phi_t, \bm{y}_t).
\end{equation}
As demonstrated in Section~\ref{sec:evaluating_variational_expression}, the value $G_{\mathcal{J}}(h,\tau)$ obtained via this saddle-point calculation is exactly equal to the true CGF $K_{\mathcal{J}}(h,\tau)$. This confirms that restricting the optimization to path measures generated by biased Langevin dynamics is sufficient.

\subsection{Optimal Modified Dynamics Yield the CGF}
\label{sec:evaluating_variational_expression}
The proof of $G_{\mathcal{J}}(h,\tau) = K_{\mathcal{J}}(h,\tau)$ proceeds by evaluating the variational expression $G_{\mathcal{J}}(h,\tau)$ at its optimal point. This requires first identifying the optimal fields $(\hat{q}_t, \hat{\phi}_t, \hat{\bm{y}}_t)$ which satisfy the stationary point equations derived from the Lagrangian $\mathcal{L}(q_t, \dot{q}_t, \phi_t, \bm{y}_t)$ (Eq.~\eqref{lagrangian_alpha}). We then show that substituting these optimal fields into the definition of $G_{\mathcal{J}}(h,\tau)$ precisely recovers the value of $K_{\mathcal{J}}(h,\tau)$.
\subsubsection{Justification for Swapping Optimization Order}
The standard procedure for a constrained optimization problem, as formulated in Eq.~\eqref{lower_bound_explicit} subject to the Fokker--Planck equation \eqref{modified_fokker-planck}, leads to the formulation of the Lagrangian functional $\mathcal{L}$. The original problem is to maximize over the primal variables ($\bm{q}_t, \bm{y}_t$) and minimize over the dual (Lagrange multiplier) variable ($\bm{\phi}_t$):
\begin{equation}
 G_{\mathcal{J}}(h,\tau) = \sup_{q_t, \bm{y}_t} \inf_{\phi_t} \mathcal{L}(q_t, \dot{q}_t, \phi_t, \bm{y}_t).
 \label{eq:fundamental_saddle_overdamped}
\end{equation}
The key insight is that for this specific Lagrangian, the order of the optimization operations over the fields $\bm{y}_t$ and $\phi_t$ can be exchanged, leading to the expression in Eq.~\eqref{K_with_L_final}:
\begin{equation}
 G_{\mathcal{J}}(h,\tau) = \sup_{q_t} \inf_{\phi_t} \sup_{ \bm{y}_t} \mathcal{L}(q_t, \dot{q}_t, \phi_t, \bm{y}_t).
 \label{eq:K_with_L_final_overdamped}
\end{equation}
This reordering is permitted by a Minimax Theorem \cite{SionMinMax}, which requires the function to be convex in the minimization variable and concave in the maximization variable for a fixed value of the other variables. We verify these conditions for the Lagrangian density $\ell(q_t, \dot{q}_t, \phi_t, \bm{y}_t)$ from Eq.~\eqref{density_L}, holding $q_t$ and its derivatives fixed.

\begin{itemize}
	\item {Convexity in the minimization variable} $\bm{\phi}_t$: The Lagrangian density $\ell$ is \emph{linear} in $\phi_t$ and its spatial derivatives $\nabla\phi_t$. A linear function is a special case of a convex function. Thus, $\mathcal{L}$ is convex in $\bm{\phi}_t$.

	\item {Concavity in the maximization variable} $\bm{y}_t$: The terms in $\ell$ that depend on $\bm{y}_t$ are:
	$$ \ell_{\bm{y}} = (\Re\{h^* \bm{\theta}_t\} - \nabla \phi_t) \cdot \bm{y}_t q_t - \frac{1}{4} \bm{y}_t \cdot \bm{B}^{-1} \bm{y}_t q_t $$
	This is a quadratic form in the vector $\bm{y}_t$. To check for concavity, we can examine the Hessian matrix of second partial derivatives with respect to the components of $\bm{y}_t$. The Hessian is given by $-\frac{1}{2} q_t \bm{B}^{-1}$. Since the diffusion matrix $\bm{B}$ is positive-definite, its inverse $\bm{B}^{-1}$ is also positive-definite. As the probability density $q_t$ is non-negative, the Hessian is negative-semidefinite. This confirms that the Lagrangian density is concave in $\bm{y}_t$.
\end{itemize}
Since both conditions are met, strong duality holds, and the swap of the infimum and supremum is justified. We can therefore proceed by first performing the unconstrained maximization over the biasing field $\bm{y}_t$, which yields the optimal bias as a function of the other fields, and then substituting this result back into the Lagrangian.

\subsubsection{Optimal Conditions from Stationary Point Equations}
\label{subsubsec:optimal_conditions_el}

To find the optimal $(\hat{q}_t, \hat{\phi}_t, \hat{\bm{y}}_t)$, we analyze the stationary points of the action associated with the Lagrangian $\mathcal{L}(q_t, \dot{q}_t, \phi_t, \bm{y}_t)$. 
We now derive the stationary point equations.  First, variation with respect to $q_t$ yields:
\begin{equation}
\frac{\delta \mathcal{L}(q_t, \dot{q}_t, \phi_t, \bm{y}_t)}{\delta q_t} := \frac{\partial \ell}{\partial q_t} - \frac{\partial}{\partial t} \frac{\partial \ell}{\partial (\partial_t q_t)} - \nabla \cdot \frac{\partial \ell}{\partial (\nabla q_t)} = 0,
\end{equation}
which yields:
\begin{equation}
	(\Re\{h^* \bm{\theta}_t\} - \nabla \phi_t) \cdot (\bm{A} + \bm{y}_t) - \frac{1}{4} \bm{y}_t \cdot \bm{B}^{-1} \bm{y}_t - \partial_t \phi_t + \nabla \cdot ( \bm{B}(\Re\{h^* \bm{\theta}_t\} - \nabla \phi_t)) = 0.
	\label{EL_qt}
\end{equation}
The stationary point equation $\frac{\delta \mathcal{L}(q_t, \dot{q}_t, \phi_t, \bm{y}_t)}{\delta \phi_t} = 0$ simply recovers the modified Fokker-Planck equation \eqref{modified_fokker-planck}.
For each component $y_{t,i}$ of $\bm{y}_t$, the stationary point equation is:
\begin{align}
	 \frac{\delta \mathcal{L}(q_t, \dot{q}_t, \phi_t, \bm{y}_t)}{\delta y_{t,i}} &= (\Re\{h^* \theta_{t,i}\} - \partial_i \phi_t) q_t - \frac{1}{2} (\bm{B}^{-1} \bm{y}_t)_i q_t = 0.
	  \label{EL_yt}
\end{align}
Additionally, we impose the transversality condition
\begin{equation}
\left.\frac{\partial \ell}{\partial (\partial_t q_t(\bm{x}_t))}\right|_{t=\tau} = \phi_\tau(\bm{x}_\tau)=0.
\end{equation}
This implies that the optimal Lagrange multiplier $\hat{\phi}_t(\bm{x})$ vanishes at the final time.

The optimal fields $(\hat{q}_t, \hat{\phi}_t, \hat{\bm{y}}_t)$ satisfy the stationary point equations derived from the Lagrangian $\mathcal{L}(q_t, \dot{q}_t, \phi_t, \bm{y}_t)$ (Eq.~\eqref{lagrangian_alpha}). Key relations emerging from these equations are the expression for the optimal bias $\hat{\bm{y}}_t$ in terms of the Lagrange multiplier $\hat{\phi}_t$:
\begin{equation}
\hat{\bm{y}}_t(\bm{x})=2\bm{B}\Bigl(\Re\{h^*\bm{\theta}_t(\bm{x})\}-\nabla\hat{\phi}_t(\bm{x})\Bigr),
\label{optimal_bias_explicit}  
\end{equation}
and the evolution equation for the partial time derivative of $\hat{\phi}_t$:
\begin{align}
\partial_t \hat{\phi}_t&=\frac{1}{2} \hat{\bm{y}}_t\cdot\bigl(\bm{B}^{-1}(\bm{A}+\hat{\bm{y}}_t)\bigr)- \frac{1}{4} \hat{\bm{y}}_t\cdot\bm{B}^{-1}\hat{\bm{y}}_t +\frac{1}{2} \nabla\cdot\hat{\bm{y}}_t.
\label{phi_evolution_eq_subsubsec}  
\end{align}

We examine total time derivative of $\hat{\phi}_t$  along a trajectory $\bm{x}_t$ governed by the optimal modified Langevin equation, $\dot{\bm{x}}_t = \bm{A} + \hat{\bm{y}}_t + \sqrt{2\bm{B}}\bm{\xi}(t)$. Since $\hat{\phi}_t(\bm{x}_t)$ depends on time both explicitly and implicitly through the stochastic process $\bm{x}_t$, we employ It\^{o}'s lemma:
\begin{equation*}
\frac{d}{dt}\hat{\phi}_t = \partial_t \hat{\phi}_t + \nabla \hat{\phi}_t \bullet \dot{\bm{x}}_t + (\bm{B}\nabla)\cdot\nabla \hat{\phi}_t,
\end{equation*}
where $\bullet$ signifies the It\^{o} product and the last term is the It\^{o} correction. Substituting the expressions for $\partial_t \hat{\phi}_t$ (Eq.~\eqref{phi_evolution_eq_subsubsec}) and $\nabla \hat{\phi}_t$ (obtained by rearranging Eq.~\eqref{optimal_bias_explicit}) into the It\^{o} formula leads to:
\begin{equation*}
\frac{d}{dt}\hat{\phi}_t = \Bigl( \Re\{h^*\bm{\theta}_t\} \bullet \dot{\bm{x}}_t + (\bm{B}\nabla)\cdot\Re\{h^*\bm{\theta}_t\} \Bigr)
- \frac{1}{4}\hat{\bm{y}}_t\cdot\bm{B}^{-1}\hat{\bm{y}}_t - \frac{1}{2}\hat{\bm{y}}_t\cdot\bm{B}^{-1}\bigl(\dot{\bm{x}}_t-\bm{A}-\hat{\bm{y}}_t\bigr).
\end{equation*}
Applying the conversion $\alpha \circ \dot{\bm{x}}_t = \alpha \bullet \dot{\bm{x}}_t + (\bm{B}\nabla)\cdot\alpha$ results in:
\begin{equation}
\frac{d}{dt}\hat{\phi}_t = \Re\{h^*\bm{\theta}_t\}\circ\dot{\bm{x}}_t - \frac{1}{4}\hat{\bm{y}}_t\cdot\bm{B}^{-1}\hat{\bm{y}}_t - \frac{1}{2}\hat{\bm{y}}_t\cdot\bm{B}^{-1}\bigl(\dot{\bm{x}}_t-\bm{A}-\hat{\bm{y}}_t\bigr).
\label{phi_ito_result_subsubsec}  
\end{equation}
The final term, involving $\dot{\bm{x}}_t-\bm{A}-\hat{\bm{y}}_t$, is proportional to the noise $\sqrt{2\bm{B}}\bm{\xi}(t)$ driving the modified dynamics.

\subsubsection{Equivalence of the Optimized Variational Bound and the CGF}
\label{subsubsec:equivalence_G_K_evaluated}
Having established the conditions satisfied by the optimal fields $(\hat{q}_t, \hat{\phi}_t, \hat{\bm{y}}_t)$, we now evaluate the variational expression $G_{\mathcal{J}}(h,\tau)$ at this optimum and demonstrate its identity with the CGF $K_{\mathcal{J}}(h,\tau)$.
We can rewrite $G_{\mathcal{J}}(h,\tau)$ as follows:
\begin{equation}
\begin{aligned}
	G_{\mathcal{J}}(h,\tau) &= \int_0^\tau dt \Bigl\langle\Re\{h^*\bm{\theta}_t\}\circ\dot{\bm{x}}_t
-\frac{1}{4}\hat{\bm{y}}_t\cdot\bm{B}^{-1}\hat{\bm{y}}_t\Bigr\rangle_{\hat{Q}_\tau}
-D_{\text{KL}}(\hat{q}_0\|p_\text{st})\\
&= \int_0^\tau dt \left\langle \frac{d}{dt}\hat{\phi}_t + \frac{1}{2}\hat{\bm{y}}_t\cdot\bm{B}^{-1}\bigl(\dot{\bm{x}}_t-\bm{A}-\hat{\bm{y}}_t\bigr) \right\rangle_{\hat{Q}_\tau}
-D_{\text{KL}}(\hat{q}_0\|p_\text{st}).
\end{aligned}
\end{equation}
The expectation of the term involving $(\dot{\bm{x}}_t-\bm{A}-\hat{\bm{y}}_t)$ vanishes under $\hat{Q}_\tau$. Integrating the remaining term and using the transversality condition $\hat{\phi}_\tau=0$:
\begin{equation}
\begin{aligned}
G_{\mathcal{J}}(h,\tau) &= \langle \hat{\phi}_\tau - \hat{\phi}_0 \rangle_{\hat{Q}_\tau} - D_{\text{KL}}(\hat{q}_0\|p_\text{st}) \\
&= \langle -\hat{\phi}_0(\bm{x}_0) \rangle_{\hat{Q}_\tau} - \int d\bm{x}_0 \hat{q}_0(\bm{x}_0)\ln\frac{\hat{q}_0(\bm{x}_0)}{p_\text{st}(\bm{x}_0)} \\
&=\int d\bm{x}_0\Bigl[-\hat{\phi}_0(\bm{x}_0)\hat{q}_0(\bm{x}_0)
-\hat{q}_0(\bm{x}_0)\ln\frac{\hat{q}_0(\bm{x}_0)}{p_\text{st}(\bm{x}_0)}\Bigr].
\end{aligned}
\end{equation}
The optimal initial distribution $\hat{q}_0(\bm{x})$ is found by maximizing the expression for $G_{\mathcal{J}}(h,\tau)$ derived above with respect to $\hat{q}_0$, enforcing the normalization condition $\int \hat{q}_0(\bm{x}) d\bm{x} = 1$:
\begin{equation}
\hat{q}_0(\bm{x})
=\frac{e^{-\hat{\phi}_0(\bm{x})}p_\text{st}(\bm{x})}{\int d\bm{x}' e^{-\hat{\phi}_0(\bm{x}')}p_\text{st}(\bm{x}')}.
\label{optimal_q0_subsubsec} 
\end{equation}
Inserting the optimal initial distribution $\hat{q}_0(\bm{x})$  simplifies the expression for the optimized bound $G_{\mathcal{J}}(h,\tau)$ to:
\begin{equation}
G_{\mathcal{J}}(h,\tau)
=\ln\Biggl(\int d\bm{x} e^{-\hat{\phi}_0(\bm{x})}p_\text{st}(\bm{x})\Biggr).
\label{G_f}
\end{equation}

Next, we evaluate the cumulant generating function using the optimal bias $\hat{\bm{y}}_t(\bm{x})$. By substituting Eq.~\eqref{phi_ito_result_subsubsec} into the definition of the path probability ratio, we obtain:
\begin{equation}
\begin{aligned}
\Re\{h^* \mathcal{J}\}+\ln\frac{P_\tau(\Gamma_\tau)}{\hat{Q}_\tau(\Gamma_\tau)}
-\ln\frac{p_\text{st}(\bm{x}_0)}{\hat{q}_0(\bm{x}_0)}
&=\Re\{h^* \mathcal{J}\}-\int_0^\tau dt \frac{1}{4}\Bigl(\hat{\bm{y}}_t\cdot\bm{B}^{-1}\hat{\bm{y}}_t
-2 \hat{\bm{y}}_t\cdot\bigl(\bm{B}^{-1}(\dot{\bm{x}}_t-\bm{A}-\hat{\bm{y}}_t)\bigr)\Bigr)\\
&=\int_0^\tau dt  \frac{d}{dt}\hat{\phi}_t\\
&=-\hat{\phi}_0(\bm{x}_0).
\label{path_prob_ratio_result_subsubsec}
\end{aligned}
\end{equation}
As a result, the cumulant generating function can be written as:
\begin{equation}
\begin{aligned}
K_{\mathcal{J}}(h,\tau)
&=\ln\Biggl(\int \mathcal{D}\Gamma_\tau e^{\Re\{h^*\mathcal{J}(\Gamma_\tau)\}+\ln\frac{P_\tau(\Gamma_\tau)}{\hat{Q}_\tau(\Gamma_\tau)}}\hat{Q}_\tau(\Gamma_\tau)\Biggr)\\
&=\ln\Biggl(\int \mathcal{D}\Gamma_\tau e^{-\hat{\phi}_0(\bm{x}_0) + \ln\frac{p_\text{st}(\bm{x}_0)}{\hat{q}_0(\bm{x}_0)}} \hat{Q}_\tau(\Gamma_\tau)\Biggr) \\
&=\ln\Biggl(\int d\bm{x}_0 e^{-\hat{\phi}_0(\bm{x}_0)} \frac{p_\text{st}(\bm{x}_0)}{\hat{q}_0(\bm{x}_0)} \hat{q}_0(\bm{x}_0) \Biggr)\\
&=\ln\Biggl(\int d\bm{x} e^{-\hat{\phi}_0(\bm{x})} p_\text{st}(\bm{x})\Biggr).
\label{K_f}
\end{aligned}
\end{equation}
Comparing Eq.~\eqref{G_f} and Eq.~\eqref{K_f}, we can verify  when optimized over the family of modified dynamics, $G_{\mathcal{J}}(h,\tau)$ is equivalent to the cumulant generating function:
\begin{equation}
	K_{\mathcal{J}}(h,\tau) = G_{\mathcal{J}}(h,\tau).
\end{equation}
This confirms that the variational approach limited to biased Langevin dynamics is sufficient to calculate the CGF and justifies the subsequent derivation of the maximum-minimum representation of the CGF.

\subsection{Max-Min Representation for the CGF}
\label{subsec:Deriving_Max_Min_CGF}
We simplify the variational expression by eliminating the explicit dependence on the biasing field $\bm{y}_t$. 
Substituting  $\hat{\bm{y}}_t$ back into the Lagrangian density $\ell(q_t, \dot{q}_t, \phi_t, \bm{y}_t)$ (Eq.~\eqref{optimal_bias_explicit}) yields,
\begin{align}
\ell(q_t, \dot{q}_t, \phi_t, \hat{\bm{y}}_t) 
&= (\Re\{h^* \bm{\theta}_t\} - \nabla \phi_t) \cdot (\bm{A} + 2\bm{B}(\Re\{h^*\bm{\theta}_t\} - \nabla \phi_t) - \bm{B} \nabla) q_t \nonumber \\
&\quad - \frac{1}{4} [2\bm{B}(\Re\{h^*\bm{\theta}_t\} - \nabla \phi_t)] \cdot \bm{B}^{-1} [2\bm{B}(\Re\{h^*\bm{\theta}_t\} - \nabla \phi_t)] q_t + \phi_t \partial_t q_t \nonumber \\
&= (\Re\{h^* \bm{\theta}_t\} - \nabla \phi_t) \cdot (\bm{A} - \bm{B} \nabla \ln p_{\text{st}} - \bm{B} \nabla + \bm{B} \nabla \ln p_{\text{st}}) q_t \nonumber \\
&\quad + (\Re\{h^* \bm{\theta}_t\} - \nabla \phi_t) \cdot \bm{B} (\Re\{h^* \bm{\theta}_t\} - \nabla \phi_t) q_t + \phi_t \partial_t q_t \nonumber \\
&= (\Re\{h^* \bm{\theta}_t\} - \nabla \phi_t) \cdot \left(\bm{\nu}_{\text{st}} - \bm{B} \nabla \ln \frac{q_t}{p_{\text{st}}}\right) q_t \nonumber \\
&\quad + (\Re\{h^* \bm{\theta}_t\} - \nabla \phi_t) \cdot \bm{B} (\Re\{h^* \bm{\theta}_t\} - \nabla \phi_t) q_t + \phi_t \partial_t q_t.
\label{lagrangian_density_optimized_y}
\end{align}
where $\bm{\nu}_{\text{st}} = \bm{A} - \bm{B}\nabla\ln p_{\text{st}}$.
Integrating this optimized Lagrangian density gives the final max-min representation for the CGF, where the optimization is now only over $q_t$ and $\phi_t$:
\begin{equation}
\begin{aligned}
K_{\mathcal{J}}(h,\tau)
=\sup_{q_t} \inf_{\phi_t} 
\Biggl\{ \int_0^\tau dt \Bigl[
&\Bigl\langle \Bigl(\Re\{h^*\bm{\theta}_t\}-\nabla\phi_t\Bigr)\cdot\left( \bm{\nu}_{\text{st}} - \bm{B} \nabla \ln \frac{q_t}{p_{\text{st}}} \right)\Bigr\rangle_{q_t}\\
&+\Bigl\langle \Bigl(\Re\{h^*\bm{\theta}_t\}-\nabla\phi_t\Bigr)\cdot\bm{B}\Bigl(\Re\{h^*\bm{\theta}_t\}-\nabla\phi_t\Bigr)\Bigr\rangle_{q_t}\\
&+\Bigl\langle \phi_t \partial_t\ln q_t\Bigr\rangle_{q_t}
\Bigr]-D_{\text{KL}}(q_0\|p_\text{st})\Biggr\}.
\label{K_final}
\end{aligned}
\end{equation}
Here, the maximization is over arbitrary probability densities $q_t(\bm{x})$, and the minimization over the auxiliary fields $\phi_t(\bm{x})$ implicitly enforces the modified dynamics (Eq.~\eqref{modified_fokker-planck}) associated with the optimal bias $\hat{\bm{y}}_t$.

\subsection{Variational Representation of the Current Variance}
\label{Variance_O}
We seek a variational representation for the variance of the general complex-valued stochastic current $\mathcal{J}$. We write the current and the biasing parameter $h$ in terms of their real (subscript $r$) and imaginary (subscript $i$) parts:
\begin{equation}
\mathcal{J} = \mathcal{J}_r + i \mathcal{J}_i \quad \text{and} \quad h = h_r + i h_i.
\end{equation}
The total variance of the complex current is defined as the sum of the variances of its real and imaginary parts:
\begin{equation}
\text{Var}(\mathcal{J}) := \text{Var}(\mathcal{J}_r) + \text{Var}(\mathcal{J}_i).
\end{equation}

This variance is related to the second-order expansion of the cumulant generating function for small $|h|$. The argument of the exponential in the CGF definition involves the real part of $h^* \mathcal{J}$:
\begin{equation}
\Re\{h^* \mathcal{J}\} = \Re\{(h_r - i h_i)(\mathcal{J}_r + i \mathcal{J}_i)\} = h_r \mathcal{J}_r + h_i \mathcal{J}_i.
\end{equation}
Expanding the CGF $K_{\mathcal{J}}(h,\tau) = \ln \langle e^{\Re\{h^* \mathcal{J}\}} \rangle_{P_\tau}$ yields:
\begin{equation}
K_{\mathcal{J}}(h,\tau) = h_r \langle \mathcal{J}_r \rangle + h_i \langle \mathcal{J}_i \rangle + \frac{1}{2} \left[h_r^2 \text{Var}(\mathcal{J}_r) + h_i^2 \text{Var}(\mathcal{J}_i) + 2 h_r h_i \text{Cov}(\mathcal{J}_r, \mathcal{J}_i) \right] +  O(|h|^3).
\label{eq:CGF_expansion_detailed}
\end{equation}
From Eq.~\eqref{eq:CGF_expansion_detailed}, we see that the total variance $\text{Var}(\mathcal{J})$ corresponds precisely to the sum of the coefficients multiplying $h_r^2$ and $h_i^2$ in the expansion of $2 K_{\mathcal{J}}(h,\tau)$.

Our strategy is to derive $\text{Var}(\mathcal{J})$ by expanding the max-min variational representation of $K_{\mathcal{J}}(h,\tau)$ (Eq.~\eqref{K_final}) to second order in $h$.  At $h=0$, the system follows the original dynamics, so the optimal $q_t(\bm{x})$ is the steady state $p_\text{st}(\bm{x})$, and the optimal $\phi_t(\bm{x})$ is zero (since the bias $\hat{\bm{y}}_t$ is zero). For small $h$, we introduce first-order corrections parameterized by complex auxiliary fields $\rho_t(\bm{x})$ and $\psi_t(\bm{x})$:
\begin{align}
\phi_t(\bm{x}) &= -\Re\{h^* \rho_t(\bm{x})\} + O(|h|^2), \\[1mm]
q_t(\bm{x}) &= p_\text{st}(\bm{x}) \biggl (1 + \Re\bigl\{h (\psi_t^*(\bm{x})-\langle\psi_t^*\rangle_{p_\text{st}} )\bigr\} \biggr)+ O(|h |^2).
\label{eq:field_expansions}
\end{align}
The term $-\langle\psi_t^*\rangle_{p_\text{st}}$ within the expansion for $q_t$ ensures that the normalization condition $\int q_t(\bm{x}) d\bm{x} = 1$ is satisfied up to first order in $h$.
Substituting these expansions into the max-min representation of $K_{\mathcal{J}}(h,\tau)$ (Eq. \eqref{K_final}) and keeping only the terms proportional to $h_r^2$ and $h_i^2$, we obtain a variational representation for the variance:
\begin{equation}
\begin{aligned}
\text{Var}(\mathcal{J}) = \sup_{\psi_t} \inf_{\rho_t} \Biggl[ 2\int_0^\tau dt \Bigl( & \Re\Bigl\{ \text{Cov}_\text{st}\Bigl( (\bm{\theta}_t + \nabla \rho_t)\cdot\bm{\nu}_\text{st}, \psi_t^* \Bigr) - \Bigl\langle (\bm{\theta}_t + \nabla \rho_t)\cdot\bm{B} \nabla \psi_t^* \Bigr\rangle_{p_\text{st}} - \text{Cov}_\text{st}\Bigl( \rho_t, \partial_t\psi_t^* \Bigr) \Bigr\} \\
&+ \Bigl\langle (\bm{\theta}_t + \nabla \rho_t)^*\cdot\bm{B} (\bm{\theta}_t + \nabla \rho_t) \Bigr\rangle_{p_\text{st}}  \Bigr) - \text{Var}(\psi_0) \Biggr].
\label{var_rep_initial}
\end{aligned}
\end{equation}
Here, $\text{Cov}_\text{st}(\alpha,\beta) = \langle \alpha \beta \rangle_{p_\text{st}} - \langle \alpha \rangle_{p_\text{st}} \langle \beta \rangle_{p_\text{st}}$ denotes the covariance with respect to the steady state.
To simplify this expression, we introduce a change of variables:
\begin{align}
	\eta_t(\bm{x})= \rho_t(\bm{x}) - \frac{1}{2} \psi_t(\bm{x}).
	\label{change_of_variables}
\end{align}
This change of variables helps to eliminate cross-terms involving both $\rho_t$ and $\psi_t$.
We also use the following identity, which holds for any real-valued function $\alpha(\bm{x})$ due to the stationarity condition $\nabla \cdot (\bm{\nu}_\text{st} p_\text{st}) = 0$ and the assumption that boundary terms vanish upon integration:
\begin{equation}
\bigl\langle \nabla \alpha \cdot \bm{\nu}_\text{st} \bigr\rangle_{p_\text{st}} = 0.
\end{equation}
These identities are used to simplify terms involving the gradient of the auxiliary fields:
\begin{equation}
\begin{aligned}
\text{Var}(\mathcal{J}) = \sup_{\psi_t} \inf_{\eta_t}\Biggl[ 2\int_0^\tau dt \Bigl( & \Re\Bigl\{ \text{Cov}_\text{st}\Bigl( (\bm{\theta}_t + \nabla \eta_t)\cdot\bm{\nu}_\text{st}, \psi_t^* \Bigr)  - \text{Cov}_\text{st}\Bigl( \eta_t, \partial_t\psi_t^* \Bigr) \Bigr\} \\
&+ \Bigl\langle (\bm{\theta}_t + \nabla \eta_t)^*\cdot\bm{B} (\bm{\theta}_t + \nabla \eta_t) \Bigr\rangle_{p_\text{st}} - \frac{1}{4} \Bigl\langle  \nabla \psi_t^*\cdot\bm{B} \nabla \psi_t \Bigr\rangle_{p_\text{st}}\Bigr) \\
&- \frac{1}{2}\Bigl( \text{Var}(\psi_\tau) + \text{Var}(\psi_0) \Bigr) \Biggr].
\label{var_rep_intermediate}
\end{aligned}
\end{equation}
We perform a final change of variables $\chi_t(\bm{x}) = \frac{1}{2}\psi_t(\bm{x})$ to obtain the variational representation for the variance:
\begin{equation}
\begin{aligned}
\text{Var}(\mathcal{J}) = 2 \sup_{\psi_t} \inf_{\eta_t}\Biggl[ \int_0^\tau dt \Bigl( & 2 \Re\Bigl\{ \text{Cov}_\text{st}\Bigl( (\bm{\theta}_t + \nabla \eta_t)\cdot\bm{\nu}_\text{st}, \chi_t^* \Bigr) - \text{Cov}_\text{st}\Bigl( \eta_t, \partial_t\chi_t^* \Bigr) \Bigr\} \\
&+ \Bigl\langle (\bm{\theta}_t + \nabla \eta_t)^*\cdot\bm{B} (\bm{\theta}_t + \nabla \eta_t) \Bigr\rangle_{p_\text{st}} - \Bigl\langle \nabla\chi_t^*\cdot\bm{B} \nabla\chi_t \Bigr\rangle_{p_\text{st}} \Bigr) \\
&- \Bigl( \text{Var}(\chi_\tau) + \text{Var}(\chi_0) \Bigr) \Biggr].
\label{var_rep_final}
\end{aligned}
\end{equation}

\section{Long-Time iTUR for Overdamped Systems}
\label{long_time_iTUR_Overdamped}
We consider the case where $\bm{\theta}_t = \bm{\theta}$ is a time-independent and real function. Using Eq.~\eqref{var_rep_final}, the variational representation for $D_\infty$ is given by:
\begin{equation}
\begin{aligned}
	D_\infty:= \lim_{\tau\to\infty}\frac{\operatorname{Var}(\mca{J})}{2\tau} &=  \inf_{\eta} \sup_{\chi} \Biggl[ 2 \text{Cov}_{\text{st}} \Bigl( (\bm{\theta} + \nabla \eta) \cdot \bm{\nu}_{\text{st}},\chi \Bigr)  + \Bigl\langle (\bm{\theta} + \nabla \eta) \cdot \bm{B} (\bm{\theta} + \nabla \eta) \Bigr\rangle_{p_\text{st}} - \Bigl\langle \nabla\chi \cdot \bm{B} \nabla\chi \Bigr\rangle_{p_\text{st}} \Biggr].
\end{aligned}
\label{variational_expression_longtime}
\end{equation}
Setting the auxiliary function $\eta(\bm{x}) = \bm{0}$ in the variational expression \eqref{variational_expression_longtime} removes the infimum over $\eta$ and yields an upper bound on $D_\infty$:
\begin{equation}
	D_\infty \leq D_0 + \sup_{\chi} \left[ 2 \text{Cov}_{\text{st}} \bigl( \bm{\theta} \cdot \bm{\nu}_{\text{st}},\chi \bigr) - \Bigl\langle \nabla\chi \cdot \bm{B} \nabla\chi \Bigr\rangle_{p_\text{st}} \right],
\label{upper_bound_step1_longtime}
\end{equation}
with $D_0=\langle \bm{\theta} \cdot \bm{B} \bm{\theta}\rangle_{p_\text{st}}$.
We apply the Cauchy--Schwarz inequality to the covariance term:
\begin{equation}
 \text{Cov}_{\text{st}} \bigl( \bm{\theta} \cdot \bm{\nu}_{\text{st}},\chi \bigr)^2 \leq \text{Var}_{\text{st}}(\bm{\theta} \cdot \bm{\nu}_{\text{st}}) \text{Var}_{\text{st}}(\chi).
\end{equation}
This allows us to rewrite the bound as
\begin{equation}
\begin{aligned}
		D_\infty &\leq D_0 + \sup_{\chi} \left[ 2 \sqrt{\text{Var}_{\text{st}}(\bm{\theta} \cdot \bm{\nu}_{\text{st}}) \text{Var}_{\text{st}}(\chi)} - \langle \nabla\chi \cdot \bm{B} \nabla\chi \rangle_{p_\text{st}} \right]\\
		 &\leq D_0 + \sup_{\chi} \left[ \frac{ \text{Var}_{\text{st}}(\chi)\text{Var}_{\text{st}} \bigl( \bm{\theta} \cdot \bm{\nu}_{\text{st}} \bigr)  }{ \langle \nabla\chi \cdot \bm{B} \nabla\chi \rangle_{p_\text{st}} } \right].
\end{aligned}
\label{upper_bound_step2_longtime}
\end{equation}
The last step above uses that $2ab \leq a^2 + b^2$, and specifically, that the maximum value of $2ab - b^2$ with respect to $b$ is $a^2$ (which occurs when $b = a$).
Recall the definition of the symmetrized spectral gap (Eq.~\eqref{O_spectral_gap}):
\begin{equation}
	\lambda = \inf_{\chi} \left[ \frac{\langle \nabla\chi \cdot \bm{B} \nabla\chi \rangle_{p_\text{st}}}{\text{Var}_{\text{st}}(\chi)} \right].
\end{equation}
The supremum in Eq.~\eqref{upper_bound_step2_longtime} is essentially optimized by a function that nearly saturates this Rayleigh quotient.  Therefore,
\begin{equation}
	D_\infty \leq D_0 + \frac{1}{\lambda}\text{Var}_{\text{st}} \bigl( \bm{\theta} \cdot \bm{\nu}_{\text{st}} \bigr).
\label{upper_bound_final_longtime}
\end{equation}

We seek an upper bound for the stationary variance $\operatorname{Var}_{\mathrm{st}}(\bm{\theta} \cdot \bm{\nu}_{\mathrm{st}}) = \langle (\bm{\theta} \cdot \bm{\nu}_{\mathrm{st}})^2 \rangle_{p_{\mathrm{st}}} - j_{\mathrm{st}}^2$, where $j_{\mathrm{st}} := \langle \bm{\theta} \cdot \bm{\nu}_{\mathrm{st}} \rangle_{p_{\mathrm{st}}}$ is the average steady-state current.
Applying the pointwise Cauchy-Schwarz inequality $(\bm{\theta} \cdot \bm{\nu}_{\mathrm{st}})^2 \leq (\bm{\theta} \cdot \bm{B} \bm{\theta}) (\bm{\nu}_{\mathrm{st}} \cdot \bm{B}^{-1} \bm{\nu}_{\mathrm{st}})$ and taking the expectation yields:
\begin{equation}
\langle (\bm{\theta} \cdot \bm{\nu}_{\mathrm{st}})^2 \rangle_{p_{\mathrm{st}}} \leq \langle (\bm{\theta} \cdot \bm{B} \bm{\theta}) (\bm{\nu}_{\mathrm{st}} \cdot \bm{B}^{-1} \bm{\nu}_{\mathrm{st}}) \rangle_{p_{\mathrm{st}}}.
\label{eq:cs_averaged_concise}
\end{equation}
Let $D_{\text{max}} := \sup_{\bm{x}} \{ \bm{\theta}(\bm{x}) \cdot \bm{B} \bm{\theta}(\bm{x}) \}$ and recall the total steady-state entropy production rate $\sigma_{\mathrm{st}} := \langle \bm{\nu}_{\mathrm{st}} \cdot \bm{B}^{-1} \bm{\nu}_{\mathrm{st}} \rangle_{p_{\mathrm{st}}}$. We can bound the right-hand side of Eq.~\eqref{eq:cs_averaged_concise}:
\begin{equation}
\langle (\bm{\theta} \cdot \bm{B} \bm{\theta}) (\bm{\nu}_{\mathrm{st}} \cdot \bm{B}^{-1} \bm{\nu}_{\mathrm{st}}) \rangle_{p_{\mathrm{st}}} \leq D_{\text{max}} \langle \bm{\nu}_{\mathrm{st}} \cdot \bm{B}^{-1} \bm{\nu}_{\mathrm{st}} \rangle_{p_{\mathrm{st}}} = D_{\text{max}} \sigma_{\mathrm{st}}.
\label{eq:mean_square_bound_concise}
\end{equation}
Combining these results, we arrive at the bound:
\begin{equation}
\operatorname{Var}_{\mathrm{st}}(\bm{\theta} \cdot \bm{\nu}_{\mathrm{st}}) \leq D_{\text{max}} \sigma_{\mathrm{st}} - j_{\mathrm{st}}^2.
\label{eq:var_bound_final_Dmax_concise}
\end{equation}
The long-time limit of the iTUR is given by:
\begin{equation}
	D_\infty \leq D_0 + \frac{1}{\lambda}\Bigl( D_{\text{max}}\sigma_{\text{st}} - j^2_{\text{st}} \Bigr).
\label{final_iTUR_longtime}
\end{equation}
If $\bm{\theta}$ is an all-ones vector, then $D_0 = \bm{\theta} \cdot \bm{B} \bm{\theta} = D_{\text{max}}$, yielding: 
\begin{equation}
D_\infty \leq D_0 + \frac{1}{\lambda} \left( D_0 \sigma_{\mathrm{st}} - j_{\mathrm{st}}^2 \right).
\end{equation}
 
\section{Finite Time iTUR for Overdamped Systems}
\label{Finite_Time_iTUR_Overdamped}

\subsection{Upper Bound on the Spectral Density}
\label{spectral_gap_upperbound_overdamped}
To derive a finite-time iTUR, we analyze the spectral density of the system.  We consider a complex observable:
\begin{equation}
	\bm{\theta}_t(\bm{x}) = e^{i\omega t} \bm{\theta}(\bm{x}),
\end{equation}
with $\bm{\theta}(\bm{x})$ a time-independent, real-valued vector function. The corresponding complex integrated current is:
\begin{equation}
	\mca{J}_{\omega} := \int_0^\tau dt  \bm{\theta}_t(\bm{x}_t)\circ\dot{\bm{x}}_t \, .
\end{equation}
The spectral density, $S(\omega)$, characterizes the fluctuations of this current in the long-time limit and is defined as:
\begin{equation}
	S(\omega):=\lim_{\tau\to\infty}\frac{\operatorname{Var}(\mca{J}_{\omega})}{\tau}.
\end{equation}
In the long-time limit ($\tau \to \infty$), we can neglect non-extensive boundary terms. Using Eq.~\eqref{var_rep_final}, we obtain  a variational representation for the spectral density:
\begin{equation}
\begin{aligned}
S(\omega)=\lim_{\tau\to\infty}\frac{2}{\tau}\sup_{\psi_t} \inf_{\eta_t}\Biggl[ \int_0^\tau dt  \Bigl( &2 \Re\Bigl\{ \operatorname{Cov}_{\text{st}}\Bigl( \bigl(\bm{\theta}_t + \nabla\eta_t\bigr)\cdot\bm{\nu}_{\text{st}}, \chi_t^* \Bigr)  - \operatorname{Cov}_{\text{st}}\Bigl( \eta_t, \partial_t\chi_t^* \Bigr)\Bigr\} \\
&+ \Bigl\langle \bigl(\bm{\theta}_t + \nabla\eta_t\bigr)^*\cdot\bm{B} \bigl(\bm{\theta}_t + \nabla\eta_t\bigr) \Bigr\rangle_{p_{\text{st}}} \\
&- \Bigl\langle \nabla\chi_t^*\cdot\bm{B} \nabla\chi_t \Bigr\rangle_{p_{\text{st}}} \Bigr) \Biggr].
\end{aligned}
\label{S1}
\end{equation}
To simplify this expression, we make the following choice for the auxiliary fields $\chi_t(\bm{x})$ and $\eta_t(\bm{x})$:
\begin{align}
	\chi_t(\bm{x}) &= e^{i\omega t} \bar{\chi}_t(\bm{x}),\\
	\eta_t(\bm{x}) &= e^{i\omega t} \eta(\bm{x}),
	\end{align}
where $\eta(\bm{x})$ is time-independent. Then the spectral density is upper bounded by
\begin{equation}
\begin{aligned}
S(\omega)\leq\lim_{\tau\to\infty}\frac{2}{\tau} \sup_{\bar{\chi}_t} \inf_{\eta} \Biggl[ \int_0^\tau dt  \Bigl( &2 \Re\Bigl\{ \operatorname{Cov}_{\text{st}}\Bigl( \bigl(\bm{\theta} + \nabla\eta\bigr)\cdot\bm{\nu}_{\text{st}}+ i\omega\eta, \bar{\chi}_t^* \Bigr)  - \operatorname{Cov}_{\text{st}}\Bigl( \eta, \partial_t\bar{\chi}_t^* \Bigr)\Bigr\} \\
&+ \Bigl\langle \bigl(\bm{\theta} + \nabla\eta\bigr)^*\cdot\bm{B} \bigl(\bm{\theta} + \nabla\eta\bigr) \Bigr\rangle_{p_{\text{st}}} \\
&- \Bigl\langle \nabla\bar{\chi}_t^*\cdot\bm{B} \nabla\bar{\chi}_t \Bigr\rangle_{p_{\text{st}}} \Bigr) \Biggr].
\end{aligned}
\label{S_omega_simplified_Overdamped}
\end{equation}
The stationary point equation yields a time-independent solution.  This allows us to choose a simplified, time-independent form for $\bar{\chi}_t$ and $\eta$ to make the upper bound more manageable.  We select:
\begin{align}
	\bar{\chi}_t(\bm{x}) &= \chi'(\bm{x}) + i \chi''(\bm{x}),\\
	\eta(\bm{x}) &= -i \phi(\bm{x}),
\end{align}
where $\chi'(\bm{x})$, $\chi''(\bm{x})$, and $\phi(\bm{x})$ are real-valued functions that do not depend on time.  Substituting these expressions into Eq.~\eqref{S_omega_simplified_Overdamped} gives:
\begin{equation}
\begin{aligned}
S(\omega) \leq 2  \sup_{\chi', \chi''} \inf_{\phi} \Biggl[ 
& 2 \operatorname{Cov}_{\mathrm{st}}\Bigl(\bm{\theta}\cdot\bm{\nu}_{\mathrm{st}}+\omega \phi,  \chi'\Bigr)
+2 \operatorname{Cov}_{\mathrm{st}}\Bigl(\nabla\phi\cdot\bm{\nu}_{\mathrm{st}},  \chi''\Bigr) \\[1mm]
&+\left\langle \bm{\theta}\cdot\bm{B} \bm{\theta}\right\rangle_{p_{\mathrm{st}}}
+\left\langle \nabla\phi\cdot\bm{B} \nabla\phi\right\rangle_{p_{\mathrm{st}}} \\[1mm]
&-\left\langle \nabla\chi'\cdot\bm{B} \nabla\chi'\right\rangle_{p_{\mathrm{st}}}
-\left\langle \nabla\chi''\cdot\bm{B} \nabla\chi''\right\rangle_{p_{\mathrm{st}}}
\Biggr] .
\label{eq:S_omega_before_abc}
\end{aligned}
\end{equation}
In this expression, we can identify  $D_0 = \left\langle \bm{\theta}\cdot\bm{B} \bm{\theta}\right\rangle_{p_{\mathrm{st}}}$.

To further tighten this bound, we optimize not only over the functional shapes of $\chi'$, $\chi''$, and $\phi$, but also over their overall magnitudes. We introduce real scaling parameters $a, b, c $ and replace the auxiliary functions as follows:
\begin{align}
\chi'(\bm{x}) &\longrightarrow a\chi'(\bm{x}), \label{scale1}\\[1mm]
\chi''(\bm{x}) &\longrightarrow b\chi''(\bm{x}), \label{scale2}\\[1mm]
\phi(\bm{x}) &\longrightarrow c\phi(\bm{x}). \label{scale3}
\end{align}
Substituting these into the inequality \eqref{eq:S_omega_before_abc} results in an expression that is quadratic in $a, b, c$ (for fixed functional forms $\chi', \chi'', \phi$)
After performing this optimization over the scaling parameters, we obtain:
\begin{equation}
\begin{aligned}
S(\omega) &\leq 2D_0 + 2  \sup_{\chi', \chi''} \inf_{\phi}   \left[ \frac{\operatorname{Cov}_{\mathrm{st}}(\bm{\theta}\cdot\bm{\nu}_{\mathrm{st}},  \chi')^2}{\left\langle \nabla\chi'\cdot\bm{B} \nabla\chi'\right\rangle_{p_{\mathrm{st}}}}\left(
\frac{ \omega^2 \frac{\operatorname{Cov}_{\mathrm{st}}(\phi,  \chi')^2}{\left\langle \nabla\chi'\cdot\bm{B} \nabla\chi'\right\rangle_{p_{\mathrm{st}}}} }{ \left\langle \nabla\phi\cdot\bm{B} \nabla\phi\right\rangle_{p_{\mathrm{st}}} + \frac{\operatorname{Cov}_{\mathrm{st}}(\nabla\phi\cdot\bm{\nu}_{\mathrm{st}},  \chi'')^2}{\left\langle \nabla\chi''\cdot\bm{B} \nabla\chi''\right\rangle_{p_{\mathrm{st}}}} } +1\right)^{-1} \right].
\label{eq:S_omega_after_abc_opt}
\end{aligned}
\end{equation}
We observe that the fraction inside the square brackets in Eq.~\eqref{eq:S_omega_after_abc_opt} is a monotonically increasing function of:
\begin{equation}
 Q = \frac{\text{Cov}_{\text{st}}\Bigl(\nabla\phi\cdot\bm{\nu}_{\text{st}}, \chi''\Bigr)^2}{\langle \nabla\chi''\cdot\bm{B} \nabla\chi'' \rangle_{p_{\text{st}}}}.
 \label{eq:Q_def_direct}
\end{equation}
First, we bound the numerator. Using the Cauchy--Schwarz inequality for covariance, the fact that variance is less than or equal to the mean square ($\text{Var}_{\text{st}}(X) \leq \langle X^2 \rangle_{p_{\text{st}}}$), the pointwise Cauchy--Schwarz inequality $(\bm{u} \cdot \bm{v})^2 \leq (\bm{u} \cdot \bm{B} \bm{u}) (\bm{v} \cdot \bm{B}^{-1} \bm{v})$, and the definition of the maximum local entropy production rate $\sigma_{\text{max}} := \sup_{\bm{x}} (\bm{\nu}_{\text{st}} \cdot \bm{B}^{-1} \bm{\nu}_{\text{st}})$, we chain the inequalities:
\begin{align}
 \text{Cov}_{\text{st}}\Bigl(\nabla\phi\cdot\bm{\nu}_{\text{st}}, \chi''\Bigr)^2 &\leq \text{Var}_{\text{st}}(\nabla\phi \cdot \bm{\nu}_{\text{st}}) \text{Var}_{\text{st}}(\chi'') \nonumber \\
 &\leq \langle (\nabla\phi \cdot \bm{\nu}_{\text{st}})^2 \rangle_{p_{\text{st}}} \text{Var}_{\text{st}}(\chi'') \nonumber \\
 &\leq \langle (\nabla\phi \cdot \bm{B} \nabla\phi) (\bm{\nu}_{\text{st}} \cdot \bm{B}^{-1} \bm{\nu}_{\text{st}}) \rangle_{p_{\text{st}}} \text{Var}_{\text{st}}(\chi'') \nonumber \\
 &\leq \sigma_{\text{max}} \langle \nabla\phi \cdot \bm{B} \nabla\phi \rangle_{p_{\text{st}}} \text{Var}_{\text{st}}(\chi'').
 \label{eq:numerator_bound_direct}
\end{align}
Substituting this bound into Eq.~\eqref{eq:Q_def_direct}:
\begin{equation}
Q \leq \sigma_{\text{max}} \langle \nabla\phi \cdot \bm{B} \nabla\phi \rangle_{p_{\text{st}}} \frac{\text{Var}_{\text{st}}(\chi'')}{\langle \nabla\chi''\cdot\bm{B} \nabla\chi'' \rangle_{p_{\text{st}}}}.
 \label{eq:Q_intermediate_bound_direct}
\end{equation}
Finally, using the variational definition of the spectral gap $\lambda$, which implies $\text{Var}_{\text{st}}(\chi'') / \langle \nabla\chi''\cdot\bm{B} \nabla\chi'' \rangle_{p_{\text{st}}} \leq 1/\lambda$ (see Eq.~\eqref{O_spectral_gap}), we arrive at the upper bound for $Q$:
\begin{equation}
Q \leq \frac{\sigma_{\text{max}}}{\lambda} \langle \nabla\phi \cdot \bm{B} \nabla\phi \rangle_{p_{\text{st}}}.
 \label{eq:Q_final_bound_direct}
\end{equation}
Substituting the bound for $Q$ into the expression for $S(\omega)$ (Eq.~\eqref{eq:S_omega_after_abc_opt}) yields:
\begin{equation}
S(\omega) \leq 2D_0 + 2  \sup_{\chi', \chi''} \inf_{\phi}   \Biggl[  \frac{ \operatorname{Cov}_{\mathrm{st}}(\bm{\theta}\cdot\bm{\nu}_{\mathrm{st}},  \chi')^2  }{  \left\langle \nabla\chi'\cdot\bm{B} \nabla\chi'\right\rangle_{p_{\mathrm{st}}} +  \frac{\omega^2}{1+\sigma_{\text{max}}/\lambda} \frac{\operatorname{Cov}_{\mathrm{st}}(\phi,  \chi')^2} {\left\langle \nabla\phi\cdot\bm{B} \nabla\phi\right\rangle_{p_{\mathrm{st}}}} }   \Biggr].
\end{equation}
To obtain a simpler bound, we make the specific choice $\phi = \chi'$. This eliminates the infimum and the auxiliary function $\chi''$. Using $\operatorname{Cov}_{\mathrm{st}}(\chi', \chi') = \operatorname{Var}_{\mathrm{st}}(\chi')$, the inequality becomes:
\begin{equation}
 S(\omega) \leq 2D_0 + 2 \sup_{\chi'}\frac{ \operatorname{Cov}_{\mathrm{st}}(\bm{\theta}\cdot\bm{\nu}_{\mathrm{st}},  \chi')^2  }{  \left\langle \nabla\chi'\cdot\bm{B} \nabla\chi'\right\rangle_{p_{\mathrm{st}}} +  \frac{\omega^2}{1+\sigma_{\text{max}}/\lambda} \frac{(\operatorname{Var}_{\mathrm{st}}( \chi') )^2}{\left\langle \nabla\chi'\cdot\bm{B} \nabla\chi'\right\rangle_{p_{\mathrm{st}}} }  }.
\end{equation}
Applying the Cauchy--Schwarz inequality to the numerator, $\operatorname{Cov}_{\mathrm{st}}(\bm{\theta}\cdot\bm{\nu}_{\mathrm{st}}, \chi')^2 \leq \operatorname{Var}_{\mathrm{st}}(\bm{\theta}\cdot\bm{\nu}_{\mathrm{st}}) \operatorname{Var}_{\mathrm{st}}(\chi')$, we get:
\begin{equation}
S(\omega) \leq 2D_0 + 2 \operatorname{Var}_{\mathrm{st}}(\bm{\theta}\cdot\bm{\nu}_{\mathrm{st}}) \sup_{\chi'}  \frac{  \operatorname{Var}_{\mathrm{st}}(\chi') }{  \left\langle \nabla\chi'\cdot\bm{B} \nabla\chi'\right\rangle_{p_{\mathrm{st}}} +  \frac{\omega^2}{1+\sigma_{\text{max}}/\lambda} \frac{(\operatorname{Var}_{\mathrm{st}}( \chi') )^2}{\left\langle \nabla\chi'\cdot\bm{B} \nabla\chi'\right\rangle_{p_{\mathrm{st}}} }  }.
\end{equation}
We introduce a change of variable:
\begin{equation}
z = \frac{\operatorname{Var}_{\mathrm{st}}(\chi')}{\left\langle \nabla\chi'\cdot\bm{B} \nabla\chi'\right\rangle_{p_{\mathrm{st}}}}.
\end{equation}
From the variational definition of the spectral gap $\lambda$ (Eq.~\eqref{O_spectral_gap}), we know that $z \leq 1/\lambda$. Substituting $z$ into the bound for $S(\omega)$:
\begin{equation}
S(\omega) \leq 2D_0 + 2 \operatorname{Var}_{\mathrm{st}}(\bm{\theta}\cdot\bm{\nu}_{\mathrm{st}}) \sup_{0 \leq z \leq 1/\lambda}  \left[ \frac{z}{1 + \frac{\omega^2}{1+\sigma_{\text{max}}/\lambda} z^2} \right].
\end{equation}
Define $\bar{\lambda} := \sqrt{\lambda(\lambda + \sigma_{\text{max}})}$. The supremum of the function $f(z) = z / (1 + (\omega^2 \lambda / \bar{\lambda}^2) z^2)$ over the interval $z \in [0, 1/\lambda]$ is evaluated by comparing the function's value at the unconstrained maximum ($z = \bar{\lambda}/(\omega\sqrt{\lambda})$) with its value at the boundary $z=1/\lambda$. This leads to the final bound:
\begin{equation}
S(\omega) \leq 2D_0 + 2 \operatorname{Var}_{\mathrm{st}}(\bm{\theta}\cdot\bm{\nu}_{\mathrm{st}})
\begin{cases}
\displaystyle \frac{1}{\lambda}\frac{1}{1+(\omega/\bar{\lambda})^2}, & \text{for } \omega < \bar{\lambda}, \\[2mm] \displaystyle \frac{\bar{\lambda}}{2\lambda\omega}, & \text{for }  \omega \geq \bar{\lambda}.
\end{cases}
\label{O_S_final}
\end{equation}

\subsection{From Spectral Density to Finite-Time iTUR}
\label{Finite_Time_iTUR_Overdamped_derivation}
The Wiener--Khinchin theorem relates the spectral density, $S(\omega)$, of a stationary process to its autocovariance function.  For the current observable $j(t) = \bm{\theta}(\bm{x}_t) \circ \dot{\bm{x}}_t$, the theorem is
\begin{equation}
S(\omega) = \int_{-\infty}^\infty dt  e^{i\omega t} \operatorname{Cov}_{\mathrm{st}}\bigl(j(t),j(0)\bigr)
= 2 \int_0^\infty dt  \cos(\omega t) \operatorname{Cov}_{\mathrm{st}}\bigl(j(t),j(0)\bigr).
\label{wiener_khinchin}
\end{equation}
The second equality follows from the even symmetry of the autocovariance for stationary processes.
Expanding the variance, we obtain:
\begin{equation}
D_\tau = \frac{1}{2\tau}\operatorname{Var}(\mathcal J_\tau) =  \frac{1}{\tau} \int_0^\tau \int_0^\tau \operatorname{Cov}_{\text{st}}(j(t),j(s)) dt ds =  \frac{2}{\tau} \int_0^\tau dt  \int_0^t  ds\operatorname{Cov}_{\text{st}}(j(t),j(s)).
\end{equation}
Using the inverse Fourier transform of Eq.~\eqref{wiener_khinchin} and inserting this into the expression for $D_\tau$, we obtain:
\begin{equation}
D_\tau = \frac{1}{\pi} \int_0^\infty d\omega \, S(\omega) \frac{1 - \cos(\omega \tau)}{\omega^2 \tau}.
\label{eq:D_tau_S_omega_relation}
\end{equation}
Substituting the upper bound for $S(\omega)$ (Eq.~\eqref{O_S_final}) into Eq.~\eqref{eq:D_tau_S_omega_relation} gives:
\begin{align}
D_\tau &\leq \frac{1}{\pi} \int_0^\infty d\omega \, (2D_0) \frac{1 - \cos(\omega \tau)}{\omega^2 \tau} \nonumber  + \frac{2 \operatorname{Var}_{\mathrm{st}}(\bm{\theta}\cdot\bm{\nu}_{\mathrm{st}})}{\pi \lambda} \left[ \int_0^{\bar{\lambda}} d\omega \, \frac{1}{1+(\omega/\bar{\lambda})^2} \frac{1 - \cos(\omega \tau)}{\omega^2 \tau} + \int_{\bar{\lambda}}^\infty d\omega \, \frac{\bar{\lambda}}{2\omega} \frac{1 - \cos(\omega \tau)}{\omega^2 \tau} \right].
\end{align}
The first integral evaluates exactly to $D_0$, using the standard identity $\int_0^\infty d\omega \, (1 - \cos(\omega \tau)) / (\omega^2 \tau) = \pi/2$.
For the remaining integrals, we perform a change of variables $z = \omega/\bar{\lambda}$:
\begin{equation}
D_\tau \leq D_0 + \frac{\operatorname{Var}_{\mathrm{st}}(\bm{\theta}\cdot\bm{\nu}_{\mathrm{st}})}{\lambda} \underbrace{\frac{2}{\pi (\bar{\lambda} \tau)} \left[ \int_0^1 dz \frac{1 - \cos(z\bar{\lambda} \tau)}{z^2 (1+z^2)} + \int_1^\infty dz \frac{1 - \cos(z\bar{\lambda} \tau)}{2z^3} \right]}_{=: g(\bar{\lambda} \tau)}.
\end{equation}
The function $g(x)$ indicated above captures the time dependence.
Finally, using the bound $\operatorname{Var}_{\mathrm{st}}(\bm{\theta}\cdot\bm{\nu}_{\mathrm{st}}) \leq D_{\text{max}} \sigma_{\mathrm{st}} - j_{\mathrm{st}}^2$ (derived in Eq.~\eqref{eq:var_bound_final_Dmax_concise}), we arrive at the finite-time iTUR:
\begin{equation}
D_\tau \leq D_0 + \frac{g(\bar{\lambda} \tau)}{\lambda} \left( D_{\text{max}} \sigma_{\mathrm{st}} - j_{\mathrm{st}}^2 \right).
\label{eq:finite_time_itur_result}
\end{equation}

\subsection{Accessible Bounds on Current Fluctuations}
We now develop a hierarchy of more practical bounds on the variance. Our approach relies on a key inequality relating the dynamics to two fundamental scalar fields: the local current fluctuations $D(\bm{x}) := \boldsymbol{\theta}(\bm{x}) \cdot \mathbf{B} \boldsymbol{\theta}(\bm{x})$ and the local entropy production rate $\sigma(\bm{x}) := \boldsymbol{\nu}_{\text{st}}(\bm{x}) \cdot \mathbf{B}^{-1} \boldsymbol{\nu}_{\text{st}}(\bm{x})$. Note that $g(z) \leq 1$ for any $z$, so we can write:
\begin{equation*}
\label{eq:supremum_bound}
D_\tau \leq D_0 + \frac{1}{\lambda} \left( D_{\text{max}} \sigma_{\text{st}} - j_{\text{st}}^2 \right).
\end{equation*}
This bound has limited utility because estimating global suprema, such as $D_{\text{max}}$, is challenging when $D_{\text{max}}$ is unbounded.
A more robust approach is to use statistical moments, which are averages over the NESS. By definition, the correlation term is exactly $\langle D\sigma \rangle_{\text{st}} = \operatorname{Cov}_{\text{st}}(D, \sigma) + \langle D \rangle_{\text{st}}\langle \sigma \rangle_{\text{st}} = \operatorname{Cov}_{\text{st}}(D, \sigma) + D_0 \sigma_{\text{st}}$. Substituting this identity into $D_\tau \leq D_0 + \frac{g(\bar{\lambda}\tau)}{\lambda} \operatorname{Var}_{\text{st}}(\boldsymbol{\theta} \cdot \boldsymbol{\nu}_{\text{st}})$ gives the tighter bound based on second-order moments:
\begin{equation*}
\label{eq:cov_bound}
D_\tau \leq D_0 + \frac{1}{\lambda} \left( D_0 \sigma_{\text{st}} + \operatorname{Cov}_{\text{st}}(D, \sigma) - j_{\text{st}}^2 \right).
\end{equation*}
This bound can be loosened slightly by applying the standard Cauchy-Schwarz inequality for random variables, $\operatorname{Cov}_{\text{st}}(D, \sigma) \le \sqrt{\operatorname{Var}_{\text{st}}(D) \operatorname{Var}_{\text{st}}(\sigma)}$, to obtain a relation involving only variances:
\begin{equation*}
\label{eq:var_prod_bound}
D_\tau \leq D_0 + \frac{1}{\lambda} \left( D_0 \sigma_{\text{st}} + \sqrt{\operatorname{Var}_{\text{st}}(D) \operatorname{Var}_{\text{st}}(\sigma)} - j_{\text{st}}^2 \right).
\end{equation*}

Finally, the suprema can be replaced with probabilistic estimates based on Cantelli's inequality. For a random variable $X$ with mean $\mu_X$ and standard deviation $s_X$, Cantelli's inequality provides an upper bound $X_{1-\epsilon} = \mu_X + s_X \sqrt{(1-\epsilon)/\epsilon}$ that holds with probability at least $1-\epsilon$. Applying this method to the two symmetric forms of the supremum bound (one using $D_{\text{max}}$, the other using $\sigma_{\text{max}} = \sup_{\bm{x}} \sigma(\bm{x})$) yields practical bounds for experimental or numerical applications. With a confidence level of $1-\epsilon$, the following bounds can be used:

\begin{enumerate}
    \item Bound based on diffusion field moments:
    \begin{equation*}
    D_\tau \leq D_0 + \frac{1}{\lambda} \left[ \left( D_0 + \sqrt{\operatorname{Var}_{\text{st}}(D)} \sqrt{\frac{1-\epsilon}{\epsilon}} \right) \sigma_{\text{st}} - j_{\text{st}}^2 \right]
    \end{equation*}
    \item Bound based on dissipation field moments:
    \begin{equation*}
    D_\tau \leq D_0 + \frac{1}{\lambda} \left[ D_0 \left( \sigma_{\text{st}} + \sqrt{\operatorname{Var}_{\text{st}}(\sigma)} \sqrt{\frac{1-\epsilon}{\epsilon}} \right) - j_{\text{st}}^2 \right]
    \end{equation*}
\end{enumerate}

These bounds allow the selection of the one that depends on the moments of the more well-behaved or easily measurable field ($D$ or $\sigma$) for a specific system. By adjusting the confidence level $1-\epsilon$, the bounds can be tailored to balance precision and robustness in practical scenarios.

\section{Asymptotic Analysis for Large Driving Force}
Here we provide a detailed asymptotic analysis of the diffusion coefficients and thermodynamic bounds in the limit of large driving force $F \to \infty$. We consider the system of a particle in a tilted sinusoidal potential, where the periodic component is $V(x) = V_0 \cos(kx)$ with period $L=2\pi/k$. The dynamics are governed by the overdamped Langevin equation:
\[ \dot x_t = -V'(x_t) + F + \sqrt{2D_0}\,\xi(t) = k V_0 \sin(kx_t) + F + \sqrt{2D_0}\,\xi(t). \]
We adopt the conventions of the main document: mobility $\mu=1$ and thermal energy scale given by the bare diffusion coefficient $D_0=k_B T$.
As $F \to \infty$, the potential becomes a small perturbation on a constant drift. The asymptotic behaviors of the relevant quantities are derived as follows:
\begin{align*}
D_\infty &= D_0 + D_0 \left\langle (V')^2 \right\rangle_x F^{-2} + \mathcal{O}(F^{-4}) = D_0 \left[ 1 + \frac{k^2 V_0^2}{2 F^2} + \mathcal{O}(F^{-4}) \right]. \\
j_{\text{st}} &= F - \left\langle (V')^2 \right\rangle_x F^{-1} + \mathcal{O}(F^{-3}) = F \left[ 1 - \frac{k^2 V_0^2}{2 F^2} + \mathcal{O}(F^{-4}) \right]. \\
\sigma_{\text{st}} &= \frac{F}{D_0} \left( F \left[ 1 - \frac{k^2 V_0^2}{2 F^2} + \dots \right] \right) = \frac{F^2}{D_0} \left[ 1 - \frac{k^2 V_0^2}{2 F^2} + \mathcal{O}(F^{-4}) \right].
\end{align*}
In the high-force limit, the particle is washed over the potential. The relaxation is governed by diffusion over one period, so the gap converges to that of free diffusion on a ring of length $L=2\pi/k$. The operator for free diffusion has eigenvalues $-(n 2\pi/L)^2 D_0 = -(nk)^2 D_0$. The first non-zero eigenvalue ($n=1$) gives the gap:
    \[ \lim_{F\to\infty} \lambda = D_0 k^2. \]
The standard TUR provides a lower bound $D_\infty \ge D_\infty^{\text{TUR}} = j_{\text{st}}^2/\sigma_{\text{st}}$. Using the expressions above, this is equivalent to $D_0 j_{\text{st}}/F$:
\[ D_\infty^{\text{TUR}} = \frac{D_0 j_{\text{st}}}{F} = \frac{D_0}{F} \left( F \left[ 1 - \frac{k^2 V_0^2}{2 F^2} + \dots \right] \right) = D_0 - \frac{D_0 k^2 V_0^2}{2} F^{-2} + \mathcal{O}(F^{-4}). \]
In the limit, the TUR bound becomes asymptotically tight: $\lim_{F\to\infty} D_\infty^{\text{TUR}} = D_0$.

The iTUR provides an upper bound $D_\infty^{\text{iTUR}} = D_0 + (D_0 \sigma_{\text{st}} - j_{\text{st}}^2)/\lambda$. We calculate the numerator $N(F) = D_0 \sigma_{\text{st}} - j_{\text{st}}^2$, also given by $j_{\text{st}}(F - j_{\text{st}})$, using the asymptotic expansions:
\begin{align*}
N(F) = D_0 \sigma_{\text{st}} - j_{\text{st}}^2 &= D_0 \left( \frac{F^2}{D_0} \left[ 1 - \frac{k^2 V_0^2}{2 F^2} + \dots \right] \right) - \left( F \left[ 1 - \frac{k^2 V_0^2}{2 F^2} + \dots \right] \right)^2 \\
&= F^2 \left( 1 - \frac{k^2 V_0^2}{2 F^2} \right) - F^2 \left( 1 - \frac{k^2 V_0^2}{F^2} + \dots \right) \\
&\approx F^2 \left( \frac{k^2 V_0^2}{2 F^2} \right) = \frac{k^2 V_0^2}{2}.
\end{align*}
The numerator approaches the finite constant $\left\langle (V')^2 \right\rangle_x$. The iTUR bound becomes:
\[
\lim_{F\to\infty} D_\infty^{\text{iTUR}} = D_{0} + \frac{\lim_{F\to\infty} N(F)}{\lim_{F\to\infty} \lambda} = D_{0} + \frac{k^2 V_0^2 / 2}{D_{0} k^2} = D_{0} + \frac{V_0^2}{2D_0}.
\]
This is a finite value strictly greater than $D_0$ (for $V_0 \ne 0$), confirming that the iTUR remains a valid upper bound. The bound is not asymptotically tight in this limit, as $\lim D_\infty = D_0$ while $\lim D_\infty^{\text{iTUR}} > D_0$.

\newpage
\part{iTUR for Continuous-Time Markov Jump Processes}
\section{Continuous-Time Markov Jump Processes: iTUR Derivation Overview}
\subsection{System Setup and Definitions}
We consider a system evolving via a continuous-time Markov jump process (CTMJP) on a discrete state space.  The probability of being in state $k$ at time $t$ is $p_k(t)$. The dynamics are governed by the master equation:
\begin{equation}
\frac{d}{dt} p_k(t) = \sum_{m} W_{km}(t) p_m(t) - W_{mk}(t) p_k(t),
\label{master_intro}
\end{equation}
where $W_{km}(t)$ is the transition rate from state $m$ to state $k$ at time $t$.  We primarily consider time-independent rates, $W_{km}$, leading to a steady state $\pi_k$:
\begin{equation}
0 = \sum_{m} \Bigl[W_{km} \pi_m - W_{mk} \pi_k\Bigr].
\end{equation}

We consider a general stochastic current, $\mathcal{J}$, representing a sum over jumps in a trajectory $\Gamma_\tau = (k_0, k_1, \dots, k_N)$ with $N-1$ jumps occurring at times $0 < t_1 < \cdots < t_N < \tau$:
\begin{equation}
\mathcal{J}(\Gamma_\tau) = \sum_{i=1}^{N} d_{k_i, k_{i-1}}(t_i).
\label{current_intro_M}
\end{equation}
$d_{k_i, k_{i-1}}(t_i)$ is the observable increment associated with the jump from state $k_{i-1}$ to $k_i$ at time $t_i$.  Fluctuations are characterized by the cumulant generating function (CGF):
\begin{equation}
K_{\mathcal{J}}(h,\tau) = \ln \Bigl\langle e^{\Re{h^* \mathcal{J}}} \Bigr\rangle_{P_\tau},
\label{cgf_intro_M}
\end{equation}
where $h \in \mathbb{C}$, and $\langle \diamond \rangle_{P_\tau}$ averages over trajectories $\Gamma_\tau$ with probability $P_\tau$.

\subsection{Variational Approach and Modified Dynamics}
The Donsker-Varadhan principle (see Section~\ref{KL_MK}) expresses the CGF as:
\begin{equation*}
K_{\mathcal{J}}(h,\tau) = \sup_{Q_\tau} \left[ \Re{h^* \langle \mathcal{J} \rangle_{Q_\tau}} - D_{\text{KL}}(Q_\tau \| P_\tau) \right],
\label{kl_intro_M}
\end{equation*}
where the supremum is over all path probability measures, $Q_\tau$.  We show that the optimal path probability measure generated by a modified master equation with tilted transition rates (see Section~\ref{Bias_MK}):
\begin{equation*}
\tilde{W}_{km}(t) = W_{km} e^{\zeta_{km}(t)},
\end{equation*}
The modified dynamics are defined by tilted parameters $\zeta_{km}(t)$, resulting in modified probabilities $q_k(t)$ that satisfy the modified master equation. The optimal tilted parameters $\zeta_{km}(t)$ are determined via a Lagrangian formulation:
\begin{equation*}
\hat{\zeta}_{km}(t) = \Re\{h^* d_{km}(t)\} - {\phi}_k(t) + {\phi}_m(t),
\label{optimal_bias_intro_M}
\end{equation*}
where ${\phi}_k(t)$ are Lagrange multipliers. This leads to a max-min representation of the CGF:
\begin{equation*}
K_{\mathcal{J}}(h,\tau) =  \sup_{\bm{q}} \inf_{\bm{\phi}}\left\{ \int_{0}^{\tau} dt \left[
\sum_{k \neq m} W_{km}q_m(t)\biggl( e^{\Re\{h^*d_{km}(t)\}-\phi_k(t)+\phi_m(t)} -1 \biggr)
+\sum_k \phi_k(t)\dot{q}_k(t)
\right]  - \sum_k q_k(0)\ln \frac{q_k(0)}{p_k(0)} \right\},
\label{max_min_intro_M}
\end{equation*}
where $\bm{q} = \{q_k(t)\}$ is an arbitrary probability distribution, and minimizing over $\bm{\phi} = \{\phi_k(t)\}$ implicitly enforces the constraint that $\bm{q}$ satisfies the modified  master equation.
Expanding $q_k(t)$ and $\phi_k(t)$ for small $h$ (see Section~\ref{Variance_MK}):
\begin{equation*}
\begin{aligned}
q_k(t) &= p_k(t) \Bigl( 1 + \Re\bigl\{h^* \psi_k(t)\bigr\} + O(|h|^2) \Bigr), \\
\phi_k(t) &= \Re\bigl\{h^* \rho_k(t)\bigr\} + O(|h|^2).
\end{aligned}
\end{equation*}
Substituting into the representation of $K_{\mathcal{J}}(h,\tau)$, we obtain a variational representation for the variance:
\begin{equation*}
\begin{aligned}
\text{Var}(\mathcal{J}) =  \sup_{\bm{\psi}, \langle \bm{\psi} \rangle_p = 0}  \inf_{\bm{\rho}} \Biggl\{ \int_{0}^{\tau} dt  \Biggl[ &\sum_{k \neq m} W_{km}(t)p_m(t) \Bigl( \bigl|d_{km}(t) - \rho_k(t) + \rho_m(t)\bigr|^2  \\
&+ 2\Re\bigl\{ \psi_m(t) \bigl(d_{km}^*(t) - \rho_k^*(t) + \rho_m^*(t)\bigr) \bigr\} \Bigr) \\
&+ 2\sum_k \Re\bigl\{ \rho_k(t) \partial_t (\psi_k^*(t) p_k(t)) \bigr\} \Biggr] - \sum_k \abs{\psi_k(0)}^2 p_k(0) \Biggr\}.
\end{aligned}
\label{variance_variational_intro_M}
\end{equation*}

\subsection{Long-Time iTUR and Finite-Time iTUR}
$D_\infty$ has the variational representation:
\begin{equation*}
\begin{aligned}
	D_\infty := \lim_{\tau \to \infty} \frac{\text{Var}(\mathcal{J})}{2\tau}= \frac{1}{2}  \sup_{\chi, \langle\chi \rangle_{\bm{\pi}} = 0} \inf_{\eta} \Biggl\{ \sum_{k \neq m} W_{km}\pi_m \Bigl( &(d_{km} - \eta_k + \eta_m)^2 - (\chi_k - \chi_m)^2 \\
&+ 2 (\chi_k + \chi_m)(d_{km} - \eta_k + \eta_m) \Bigr) \Biggr\}.
\end{aligned}
\end{equation*}
 By making the specific choices for the auxiliary functions $\psi_k(t) = \chi_k/2$ and $\rho_k(t) = -\chi_k/2$ (where $\chi_k$ are time-independent due to the long-time limit and stationarity), we obtain the following inequality:
\begin{equation*}
D_\infty \leq D_0 + \sup_{\chi,\langle\chi\rangle_{\bm{\pi}}=0} \Biggl[ \frac{\left( \sum_{k \neq m} d_{km}(\chi_k + \chi_m)W_{km}\pi_m \right)^2}{ \sum_{k \neq m} (\chi_k - \chi_m)^2 W_{km}\pi_m } \Biggr].
\label{eq:D_infty_bound_pre_lambda}
\end{equation*}
where  
$
D_0 = \frac{1}{2}\sum_{k \neq m} W_{km}\pi_m d_{km}^2.
$
The denominator in the supremum term is related to the spectral gap $\lambda$ of the symmetrized generator. Recall the variational definition of the spectral gap (from Section~\ref{Spectral_Gap_M}):
\begin{equation*}
\lambda = \inf_{\substack{\chi \\ \langle \chi \rangle_{\bm{\pi}} = 0}} \frac{\frac{1}{2} \sum_{k \ne m} W_{km} \pi_m ( \chi_k - \chi_m)^2}{\sum_k \pi_k \chi_k^2}.
\end{equation*}
Using this, we can simplify the above upper bound on $D_\infty$:
\begin{equation*}
D_\infty \leq D_0 + \sup_{\chi,\langle\chi\rangle_{\bm{\pi}}=0} \Biggl[ \frac{\left( \sum_{k \neq m} d_{km}(\chi_k + \chi_m)W_{km}\pi_m \right)^2}{2 \lambda \sum_k \chi_k^2 \pi_k } \Biggr].
\label{eq:D_infty_bound_with_lambda}
\end{equation*}
Further bounding the numerator using the Cauchy-Schwarz inequality and a bound involving the concave function  $f(x,y)= (x^2/4y)\mathfrak{h}(x/(2y))^{-2}$ where $\mathfrak{h}$ is the inverse of the function $x \tanh(x)$ as detailed in Section~\ref{L_iTUR_C}, we arrive at the long-time iTUR for current-like observables:
\begin{equation*}
D_\infty \leq D_0  + \frac{ \kappa  f(d_{\text{max}}^2\sigma_\text{st},  2D_0 ) - j_\text{st}^2}{\lambda} .
\end{equation*}
where   $\kappa = \max_k \sum_{m(\neq k)} W_{mk}$ is the maximum escape rate, $j_\text{st} = \sum_{k \neq m} d_{km}   W_{km} \pi_m$ is the average current, $\sigma_{\text{st}} = \frac{1}{2}\sum_{k, m} (W_{km}\pi_m - W_{mk}\pi_k) \ln \frac{W_{km}\pi_m}{W_{mk}\pi_k}$ is the total entropy production rate. 

The finite-time iTUR uses the spectral density, $S(\omega)$, describing frequency-dependent fluctuations.  With $d_{km}(t) = e^{i\omega t}d_{km}$ and $\mathcal{J}_\omega = \sum_{i=1}^{N}  e^{i\omega t_i}d_{k_i, k_{i-1}}$, the spectral density is:
\begin{equation*}
S(\omega) = \lim_{\tau \to \infty} \frac{\text{Var}(\mathcal{J}_\omega)}{\tau}.
\end{equation*}
Using variational representation of the variance, we find an upper bound on $S(\omega)$ (see Section~\ref{Upper_bound_SP_MK}):
\begin{equation*}
S(\omega) \leq 2D_0 + 2 \Bigl( \kappa f(d_{\text{max}}^2\sigma_\text{st},  2D_0) - j_{\text{st}}^2 \Bigr)
\begin{cases}
\displaystyle \frac{1}{\lambda}\frac{1}{1+(\omega/\tilde{\lambda})^2}, & \text{for } \omega < \tilde{\lambda}, \\[1mm]
\displaystyle \frac{\tilde{\lambda}}{2\lambda\omega}, & \text{for }  \omega \geq \tilde{\lambda}.
\end{cases}
\end{equation*}
where  $\tilde{\lambda} = \sqrt{\lambda(\lambda + 2\kappa)}$.
$D_\tau$ is related to $S(\omega)$ by the Wiener--Khinchin theorem:
\begin{equation*}
D_\tau =  \frac{\operatorname{Var}(\mathcal{J})}{2\tau} = \frac{1}{\pi} \int_0^\infty d\omega   S(\omega) \frac{1 - \cos(\omega \tau)}{\omega^2 \tau}.
\end{equation*}
Substituting the bound on $S(\omega)$ and integrating (Section~\ref{Finite_time_iTUR_MK}) yields the finite-time iTUR:
\begin{equation*}
D_\tau \leq D_0 + \frac{ g(\tilde{\lambda}\tau)
}{\lambda} \Bigl( \kappa f(d_{\text{max}}^2\sigma_\text{st},  2D_0) - j_{\text{st}}^2 \Bigr),
\end{equation*}
where
\begin{equation*}
   g(x) = \frac{2}{\pi x} \left( \int_0^1 dz  \frac{1-\cos(zx)}{z^2(1+z^2)}
+\int_1^\infty dz  \frac{1-\cos(zx)}{2z^3} \right).
\end{equation*}
$   g(x)$ captures the time-dependence, and as $\tau \to \infty$, $   g(\tilde{\lambda} \tau) \to 1$, recovering the long-time iTUR. Importantly, $ f(x,y)\leq \min\{x/2, y\}$. As a consequence, our iTUR provides a stricter upper bound on $D_\tau$  than the one in Ref.~\cite{Garrahan_PRL_2023}, which is $D_\tau \leq D_0 (1 + \frac{ 2\kappa}{\lambda})$.

\section{Variational Methods for CTMJPs: CGF and Variance}
\label{VR_ML}
This section derives variational representations for the CGF and variance of CTMJP currents using the Donsker--Varadhan principle. We show the optimizing path measure arises from a modified CTMJP with tilted transition rates, leading to a  max-min CGF representation. Expanding this yields the variational expression for the variance, forming the basis for the iTUR derivations.

\subsection{Donsker--Varadhan Principle for CTMJPs}
\label{KL_MK}
Let $\mathcal{J}(\Gamma_\tau)$ be a current observable defined along a trajectory as:
\begin{equation}
\mathcal{J}(\Gamma_\tau) = \sum_{i=1}^{N} d_{k_i, k_{i-1}}(t_i),
\end{equation}
where $d_{mn}(t)$ is the time-dependent increment associated with the transition from state $n$ to state $m$ at time $t$. This represents the change in the observable $\mathcal{J}$ when the system jumps from state $n$ to state $m$. 
The cumulant generating function is given by:
\begin{equation}
K_{\mathcal{J}}(h,\tau) = \ln \Bigl\langle \exp\Bigl( \Re\bigl\{h^* \mathcal{J}(\Gamma_\tau)\bigr\} \Bigr) \Bigr\rangle,
\end{equation}
where $h$ is a  complex number, and $h^*$ is its complex conjugate. 

The Donsker--Varadhan variational representation expresses the CGF in terms of the Kullback--Leibler divergence:
\begin{equation}
K_{\mathcal{J}}(h,\tau) = \sup_{Q_\tau} \Bigl[ \Re\bigl\{h^* \langle \mathcal{J} \rangle_{Q_\tau} \bigr\} - D_{\text{KL}}(Q_\tau \| P_\tau) \Bigr],
\label{general_KL_final}
\end{equation}
where the supremum is over all possible path probability measures, $Q_\tau$.  

\subsection{Modified Dynamics and the CGF Variational Lower Bound:}
\label{Bias_MK}
Consider a family of modified dynamics, where the transition rates are modified: $\tilde{W}_{km} =  W_{km}(t) e^{\zeta_{km}(t)}$.
Under these modified dynamics, the probability of the system being in state $k$ at time $t$, denoted by $q_k(t)$, evolves according to a modified master equation:
\begin{equation}
\frac{d}{dt} q_k(t)  = \sum_m  W_{km}(t) e^{\zeta_{km}(t)} q_m(t) - W_{mk}(t) e^{\zeta_{mk}(t)} q_k(t) .
\label{tilde_master_final}
\end{equation}
We restrict this optimization to the family of path probabilities $Q_\tau$ generated by the modified dynamics defined by Eq.~\eqref{tilde_master_final}. This restriction gives us a lower bound:
\begin{equation}
K_{\mathcal{J}}(h,\tau) \geq \sup_{Q_\tau} \Bigl[ \Re\bigl\{h^* \langle \mathcal{J} \rangle_{Q_\tau} \bigr\} - D_{\text{KL}}(Q_\tau \| P_\tau) \Bigr].
\label{restricted_KL_final}
\end{equation}

Next, we evaluate the terms inside the supremum. The average of the current observable $\mathcal{J}$ under the modified dynamics is:
\begin{equation}
\langle \mathcal{J} \rangle_{Q_\tau} = \int_0^\tau dt \sum_{k \neq m} d_{km}(t) \tilde{W}_{km}(t) q_m(t) = \int_0^\tau dt \sum_{k \neq m} d_{km}(t) W_{km}(t) e^{\zeta_{km}(t)} q_m(t).
\label{average_J_modified_final}
\end{equation}

The KL divergence between the original path ensemble $P_\tau(\Gamma_\tau)$ and the modified ensemble $Q_\tau(\Gamma_\tau)$ can be split into two contributions:
\begin{equation}
D_{\text{KL}}\bigl[Q_\tau(\Gamma_\tau)\| P_\tau(\Gamma_\tau) \bigr] = D_{\text{KL}}^{\mathrm{j}}\bigl[Q_\tau(\Gamma_\tau)\| P_\tau(\Gamma_\tau)\bigr] + \sum_k q_k(0) \ln \frac{q_k(0)}{p_k(0)} ,
\end{equation}
where $D_{\text{KL}}^{\mathrm{j}}$ is the contribution from the jump events. Considering a small time interval $[t, t+\Delta t]$, the contribution of jumps to the KL divergence is expressed as follows:
\begin{equation}
dD_{\text{KL}}^{\mathrm{j}}\bigl[Q_\tau(\Gamma_\tau)\| P_\tau(\Gamma_\tau)\bigr] = \sum_{m,k} q_m(t)  q(k,t+\Delta t \mid m,t) \ln \frac{q(k,t+\Delta t \mid m,t)}{p(k,t+\Delta t \mid m,t)},
\end{equation}
where $p(k,t+\Delta t \mid m,t)$ describes the probability of transitioning from state $m$ at time $t$ to state $k$ at time $t+\Delta t$.
This conditional probability can be expanded to first order in the time increment:
\begin{align} p(k,t+\Delta t \mid m,t) = \delta_{mk} + \Delta t\left(W_{mk}(t) - \delta_{mk}\sum_{l(\neq m)}W_{ml}(t)\right) + {\cal O}(\Delta t^2).
\end{align} 
Separating the terms corresponding to no jump ($k = m$) and jump events ($k \neq m$), and expanding the logarithm to first order in $\Delta t$, we obtain
\begin{align}
d D_{\text{KL}}^{\mathrm{j}}[Q_\tau(\Gamma_\tau)\| P_\tau(\Gamma_\tau)] &= \sum_m q_m(t) \left( 1 - \Delta t \sum_{k (\neq m)} W_{km}(t) e^{\zeta_{km}(t)} \right) \ln \left( \frac{1 - \Delta t \sum_{k (\neq m)} W_{km}(t) e^{\zeta_{km}(t)}}{1 - \Delta t \sum_{k (\neq m)} W_{km}(t)} \right) \nonumber \\
&\quad \quad \quad  + \Delta t \sum_{k \neq m} q_m(t) W_{km}(t) e^{\zeta_{km}(t)} \zeta_{km}(t) + \mathcal{O}(\Delta t^2).
\end{align}
Expanding the logarithm to first order in $ \Delta t $ and neglecting higher-order terms, we simplify the expression:
\begin{align}
d D_{\text{KL}}^{\mathrm{j}}[Q_\tau(\Gamma_\tau) \| P_\tau(\Gamma_\tau)] &= \Delta t \sum_{k \neq m} W_{km}(t) q_m(t) \left[ (\zeta_{km}(t) - 1) e^{\zeta_{km}(t)} + 1 \right] + \mathcal{O}(\Delta t^2).
\end{align}
Integrating in time and including the initial condition, the full KL divergence is
\begin{equation}
D_{\text{KL}}\bigl[Q_\tau (\Gamma_\tau) \| P_\tau(\Gamma_\tau)\bigr] = \int_0^{\tau} dt \sum_{k \neq m} W_{km}(t)q_m(t) \left[ \bigl(\zeta_{km}(t)-1\bigr)e^{\zeta_{km}(t)} + 1 \right] + \sum_k q_k(0) \ln \frac{q_k(0)}{p_k(0)} .
\label{KL_modified_final}
\end{equation}
Substituting Eq.~\eqref{average_J_modified_final} and Eq.~\eqref{KL_modified_final} into Eq.~\eqref{restricted_KL_final}, we obtain a lower bound for $K_{\mathcal{J}}(h,\tau)$:
\begin{equation}
\begin{aligned}
K_{\mathcal{J}}(h,\tau) &\geq  \sup_{\bm{q}, \bm{\zeta}} \Biggl\{ \int_0^{\tau} dt \Biggl[ \sum_{k \neq m} W_{km}(t)q_m(t)e^{\zeta_{km}(t)} \Bigl(\Re\bigl\{h^* d_{km}(t)\bigr\} + 1 - \zeta_{km}(t)\Bigr) \\
&\qquad - \sum_{k \neq m} W_{km}(t)q_m(t)  \Biggr] - \sum_k q_k(0)\ln \frac{q_k(0)}{p_k(0)} \Biggr\} 
\end{aligned}
\label{G_definition_final}
\end{equation}
where the supremum is taken over all time-dependent functions $\bm{q} = \{q_k(t)\}$ and $\bm \zeta = \{\zeta_{km}(t)\}$.
However, the modified probabilities $q_k(t)$  are constrained by the modified master equation (Eq.~\eqref{tilde_master_final}). 
To enforce this constraint, we introduce the Lagrangian:
\begin{equation}
\begin{aligned}
\mathcal{L}(\bm{q}, \dot{\bm{q}}, \bm{\phi},  \bm{\zeta}) = \int_0^\tau dt \Biggl\{ &\sum_{k \neq m} W_{km}(t)q_m(t)e^{\zeta_{km}(t)} \Bigl(\Re\bigl\{h^* d_{km}(t)\bigr\} + 1 - \zeta_{km}(t)\Bigr) - \sum_{k \neq m} W_{km}(t)q_m(t)\\
&+ \sum_k \phi_k(t) \Bigl[ \dot{q}_k(t) - \sum_{k \neq m} \Bigl( W_{km}(t)q_m(t)e^{\zeta_{km}(t)} - W_{mk}(t)q_k(t)e^{\zeta_{mk}(t)} \Bigr) \Bigr] \Biggr\} - \sum_k q_k(0)\ln \frac{q_k(0)}{p_k(0)}.
\end{aligned}
\label{lagrangian_full_final}
\end{equation}
where $\phi_k(t)$ are time-dependent Lagrange multipliers.
The corresponding Lagrangian density (the integrand in Eq.~\eqref{lagrangian_full_final}) is:
\begin{equation}
\begin{aligned}
\ell(\bm{q}, \dot{\bm{q}}, \bm{\phi},  \bm{\zeta}) = &\sum_{k \neq m} W_{km}(t)q_m(t)e^{\zeta_{km}(t)} \Bigl(\Re\bigl\{h^* d_{km}(t)\bigr\} + 1 - \zeta_{km}(t)\Bigr) - \sum_{k \neq m} W_{km}(t)q_m(t)\\
&+ \sum_k \phi_k(t) \Bigl[ \dot{q}_k(t) - \sum_{m (\neq k)} \Bigl( W_{km}(t)q_m(t)e^{\zeta_{km}(t)} - W_{mk}(t)q_k(t)e^{\zeta_{mk}(t)} \Bigr) \Bigr].
\end{aligned}
\label{lagrangian_density_mk}
\end{equation}
We are interested in the quantity $G_{\mathcal{J}}(h,\tau)$, which is determined by the following optimization:
 \begin{equation}
 G_{\mathcal{J}}(h,\tau) = \sup_{\bm{q}}  \inf_{\bm{\phi}} \sup_{\bm{\zeta}}  \mathcal{L}(\bm{q}, \dot{\bm{q}}, \bm{\phi},  \bm{\zeta}).
 \label{K_with_L_final} 
\end{equation}

As shown in Section~\ref{sec:optimal_dynamics_cgf_mk}, the value $G_{\mathcal{J}}(h,\tau)$, derived from this saddle-point calculation, precisely matches the true CGF $K_{\mathcal{J}}(h,\tau)$. This verifies that limiting the optimization to path measures produced by the modified dynamics is adequate. 

\subsection{Optimal Modified Dynamics Yield the CGF}
\label{sec:optimal_dynamics_cgf_mk}

We now demonstrate that $G_{\mathcal{J}}(h,\tau)$ is in fact equal to the true cumulant generating function $K_{\mathcal{J}}(h,\tau)$. This confirms that the optimal path probability measure is indeed generated by such a modified process. The structure mirrors the derivation for overdamped Langevin systems (Sec.~\ref{sec:evaluating_variational_expression}).

\subsubsection{Justification for Swapping Optimization Order}
The constrained optimization problem is equivalent to finding the saddle point of the Lagrangian $\mathcal{L}$, where we maximize over the primal variables ($\bm{q}, \bm{\zeta}$) and minimize over the dual variable ($\bm{\phi}$):
\begin{equation}
	G_{\mathcal{J}}(h,\tau) = \sup_{\bm{q}, \bm{\zeta}} \inf_{\bm{\phi}} \mathcal{L}(\bm{q}, \dot{\bm{q}}, \bm{\phi}, \bm{\zeta}).
	\label{eq:fundamental_saddle}
\end{equation}
For this specific Lagrangian, the order of the supremum and infimum operations can be exchanged:
\begin{equation}
	G_{\mathcal{J}}(h,\tau) = \sup_{\bm{q}} \inf_{\bm{\phi}} \sup_{\bm{\zeta}} \mathcal{L}(\bm{q}, \dot{\bm{q}}, \bm{\phi}, \bm{\zeta}).
	\label{eq:K_with_L_final}
\end{equation}
This reordering is justified because, for any function $F(x,y)$, it holds that $\sup_x \inf_y F(x,y) \leq \inf_y \sup_x F(x,y)$. Equality holds under the conditions of the Minimax Theorem \cite{SionMinMax}, which requires $F$ to be convex in the minimization variable $y$ and concave in the maximization variable $x$. We need to check convexity in $\bm{\phi}$ (the minimization variable) and concavity in $\bm{\zeta}$ (the maximization variable) when $\bm{q}$ is fixed.
\begin{itemize}
	\item {Convexity in the minimization variable} $\bm{\phi}$: The Lagrangian density $\ell$ is \emph{linear} in each component $\phi_k(t)$. A linear function is a special case of a convex function. Thus, $\mathcal{L}$ is convex in $\bm{\phi}$.

	\item {Concavity in the maximization variable} $\bm{\zeta}$: The optimization over $\bm{\zeta}$ can be performed for each pair $(k,m)$ and time $t$ independently. Let's examine the terms in $\ell$ containing a single component $\zeta_{km}(t)$:
	$$ \ell_{\zeta_{km}} = W_{km}q_m e^{\zeta_{km}} \left[ \Re\{h^* d_{km}\} + 1 - \phi_k + \phi_m - \zeta_{km} \right] $$
	To check for concavity, we compute the second derivative:
	\begin{align*}
	\frac{\partial \ell_{\zeta_{km}}}{\partial \zeta_{km}} &= W_{km}q_m e^{\zeta_{km}} \left[ \Re\{h^* d_{km}\} + 1 - \phi_k + \phi_m - \zeta_{km} - 1 \right] \\
	\frac{\partial^2 \ell_{\zeta_{km}}}{\partial \zeta_{km}^2} &= W_{km}q_m e^{\zeta_{km}} \left[ \Re\{h^* d_{km}\} - \phi_k + \phi_m - \zeta_{km} - 1 \right]
	\end{align*}
	The stationary point $\hat{\zeta}_{km}$ is found by setting the first derivative to zero, which requires the term in brackets to be zero: $\Re\{h^* d_{km}\} - \phi_k + \phi_m - \hat{\zeta}_{km} = 0$. Substituting this back into the second derivative gives:
	$$ \left.\frac{\partial^2 \ell_{\zeta_{km}}}{\partial \zeta_{km}^2}\right|_{\hat{\zeta}_{km}} = W_{km}q_m e^{\hat{\zeta}_{km}} [0 - 1] = -W_{km}q_m e^{\hat{\zeta}_{km}} < 0 $$
	Since $W_{km}q_m > 0$ for all relevant transitions, the second derivative is strictly negative, confirming that $\ell_{\zeta_{km}}$ is strictly concave. As the Hessian of the Lagrangian $\mathcal{L}$ has zero off-diagonal elements and is negative definite, $\mathcal{L}$ is concave with respect to $\bm{\zeta}$.
\end{itemize}
Since both conditions are met, strong duality holds, and the swap of inf and sup is justified. We can now proceed by solving the saddle-point problem in the convenient order of Eq.~\eqref{eq:K_with_L_final}.

\subsubsection{Optimal Conditions from Stationary Point Equations}
\label{subsubsec:optimal_conditions_el_mk}
We find the optimal modified dynamics $(\hat{\bm{q}}(t), \hat{\bm{\zeta}}(t))$ and the corresponding optimal Lagrange multipliers $\hat{\bm{\phi}}(t)$ by solving the stationary point equations derived from the Lagrangian function. 
The stationary point equation for $\zeta_{km}$ is $\frac{\partial \ell}{\partial \zeta_{km}} = 0$.  Calculating the partial derivative with respect to $\zeta_{km}(t)$:
\begin{align*}
\frac{\partial \ell}{\partial \zeta_{km}} 
&= W_{km} q_m \left[ e^{\zeta_{km}} (\Re\{h^* d_{km}\} + 1 - \phi_k + \phi_m) - (1 \cdot e^{\zeta_{km}} + \zeta_{km} e^{\zeta_{km}}) \right] \\
&= W_{km} q_m e^{\zeta_{km}} \left[ \Re\{h^* d_{km}\} - \phi_k + \phi_m - \zeta_{km} \right].
\end{align*}
Setting this to zero, we obtain  the optimal tilted parameters $\hat{\zeta}_{km}(t)$:
\begin{equation}
\hat{\zeta}_{km}(t) = \Re\bigl\{h^* d_{km}(t)\bigr\} - {\phi}_k(t) + {\phi}_m(t).
\label{optimal_bias_mk_explicit}
\end{equation}
The stationary point equation for $\phi_k$ is $ \frac{\partial \ell}{\partial \phi_k} = 0$, recovering the modified master equation:
\begin{equation}
\dot{q}_k(t) = \sum_{m (\neq k)} \Bigl( W_{km}(t)q_m(t)e^{\zeta_{km}(t)} - W_{mk}(t)q_k(t)e^{\zeta_{mk}(t)} \Bigr).
\label{EL_phi_mk}
\end{equation}
The stationary point equation for the $k$th component of $q$ is $\frac{\partial \mathcal{L}}{\partial q_k} - \frac{d}{dt} \frac{\partial \mathcal{L}}{\partial \dot{q}_k} = 0$.
This equation can be rewritten as follows:
\begin{align}
\dot{\phi}_k(t)
  = &\sum_{j (\neq k)} W_{jk}(t) \left[ e^{\zeta_{jk}(t)} \Bigl(\Re\bigl\{h^* d_{jk}(t)\bigr\} + 1 - \zeta_{jk}(t)\Bigr) - 1 \right] - \sum_{j (\neq k)} \phi_j(t) W_{jk}(t)e^{\zeta_{jk}(t)} + \phi_k(t) \sum_{m (\neq k)} W_{mk}(t)e^{\zeta_{mk}(t)}.
\end{align}
 Now, we substitute the optimal tilted parameters $\hat{\zeta}_{jk}(t)$ into the expression for $\frac{\partial \ell}{\partial q_k}$:
\begin{equation}
	\begin{aligned}
\dot{\hat{\phi}}_k(t) &= \sum_{j (\neq k)} W_{jk}(t) \left[ e^{\Re\{h^* d_{jk}\} - \hat{\phi}_j + \hat{\phi}_k} \Bigl(\Re\bigl\{h^* d_{jk}(t)\bigr\} + 1 - (\Re\bigl\{h^* d_{jk}(t)\bigr\} - \hat{\phi}_j(t) + \hat{\phi}_k(t))\Bigr) - 1 \right] \\
&\quad - \sum_{j(\neq k)} \hat{\phi}_j(t) W_{jk}(t) e^{\Re\{h^* d_{jk}\} - \hat{\phi}_j + \hat{\phi}_k} + \hat{\phi}_k(t) \sum_{m (\neq k)} W_{mk}(t)e^{\Re\{h^* d_{mk}\} - \hat{\phi}_m + \hat{\phi}_k} \\
&= \sum_{j (\neq k)} W_{jk}(t) (e^{\hat{\zeta}_{jk}(t)} - 1).
\label{phi_evolution_eq_mk}
\end{aligned}
\end{equation}

Furthermore, the variational principle requires the transversality condition at the final time:
\begin{equation}
\hat{\phi}_k(\tau) = 0 \quad \text{for all } k.
\label{transversality_mk_again}
\end{equation}

\subsubsection{Equivalence of the Optimized Variational Bound and the CGF}
\label{subsubsec:equivalence_G_K_mk_evaluated}

We now show that the true CGF $K_{\mathcal{J}}(h,\tau)$ equals the value $G_{\mathcal{J}}(h,\tau)$ found in Eq.~\eqref{G_f_mk}. Substituting the optimal tilted parameters $\hat{\zeta}_{km}(t)$ (Eq.~\eqref{optimal_bias_mk_explicit}) into  $\int_0^\tau dt \, \ell(\bm{q}, \dot{\bm{q}}, \bm{\phi}, \bm{\zeta})$  yields:
\begin{equation}
\int_0^\tau dt \, \ell[\hat{\bm{q}}, \dot{\hat{\bm{q}}}, \hat{\bm{\phi}}, \hat{\bm{\zeta}}] = \int_0^\tau dt \sum_k \Biggl( \hat{q}_k(t) \sum_{m(\neq k)} W_{mk}(t) \bigl(e^{\hat{\zeta}_{mk}(t)}-1\bigr) + \hat{\phi}_k(t) \dot{\hat{q}}_k(t) \Biggr).
\end{equation}
Using the evolution equation for $\hat{\phi}_k(t)$ (Eq.~\eqref{phi_evolution_eq_mk}), the first term becomes $\sum_k \hat{q}_k(t) \dot{\hat{\phi}}_k(t)$. Thus, the integrand is a total time derivative:
\begin{equation}
\int_0^\tau dt \, \ell[\hat{\bm{q}}, \dot{\hat{\bm{q}}}, \hat{\bm{\phi}}, \hat{\bm{\zeta}}] = \int_0^\tau dt \frac{d}{dt} \sum_k \hat{\phi}_k(t) \hat{q}_k(t) = \Biggl[ \sum_k \hat{\phi}_k(t) \hat{q}_k(t) \Biggr]_0^\tau.
\end{equation}
Inserting this into Eq.~\eqref{K_with_L_final} and applying the transversality condition $\hat{\phi}_k(\tau)=0$:
\begin{equation}
 G_{\mathcal{J}}(h,\tau) = - \sum_k \hat{\phi}_k(0)\hat{q}_k(0) - \sum_k \hat{q}_k(0)\ln \frac{\hat{q}_k(0)}{p_k(0)} .
 \label{G_before_q0_opt_mk}
\end{equation}
This must be maximized with respect to the initial distribution $\hat{\bm{q}}(0)$ (subject to normalization $\sum_k \hat{q}_k(0) = 1$). This optimization yields the optimal initial distribution:
\begin{equation}
\hat{q}_k(0) = p_k(0) \frac{e^{-\hat{\phi}_k(0)}}{\sum_j p_j(0) e^{-\hat{\phi}_j(0)}}.
\label{optimal_q0_mk_explicit}
\end{equation}
Substituting this optimal $\hat{\bm{q}}(0)$ back into Eq.~\eqref{G_before_q0_opt_mk} gives the final result for the evaluated optimal bound:
\begin{equation}
G_{\mathcal{J}}(h,\tau) = \ln \left( \sum_k p_k(0) e^{-\hat{\phi}_k(0)} \right).
\label{G_f_mk} 
\end{equation}

Next, we will evaluate the CGF.
The CGF can be written as an expectation under the optimally modified path measure $\hat{Q}_\tau$:
\begin{equation}
K_{\mathcal{J}}(h,\tau) = \ln \Biggl\langle \exp\Bigl( \Re\{h^* \mathcal{J}(\Gamma_\tau) \}+\ln\frac{P_\tau(\Gamma_\tau)}{\hat{Q}_\tau(\Gamma_\tau)}\Bigr) \Biggr\rangle_{\hat{Q}_\tau}.
\end{equation}
For a trajectory $\Gamma_\tau = (k_0, t_0=0; \dots; k_N, t_N)$, the ratio is given by:
\begin{equation}
\begin{aligned}
\ln \frac{P_\tau(\Gamma_\tau)}{\hat{Q}_\tau(\Gamma_\tau)}
&= \ln \frac{p_{k_0}(0)}{\hat{q}_{k_0}(0)} - \sum_{i=1}^N \hat{\zeta}_{k_i k_{i-1}}(t_{i-1})+  \sum_{i=1}^{N+1} \int_{t_{i-1}}^{t_i} \sum_{l\neq k_{i-1}} W_{l k_{i-1}}(t)\Bigl( e^{\hat{\zeta}_{l k_{i-1}}(t)}   - 1\Bigr)
 d t ,
\end{aligned}
\end{equation}
where $t_{N+1} = \tau$.
Substitute the optimal tilted parameters $\hat{\zeta}_{km}(t) = \Re\{h^* d_{km}(t)\} - \hat{\phi}_k(t) + \hat{\phi}_m(t)$ (Eq.~\eqref{optimal_bias_mk_explicit}) and use the relation $\sum_{l \neq k(t)} W_{l k(t)}(t) ( e^{\hat{\zeta}_{l k(t)}(t)} - 1 ) = \dot{\hat{\phi}}_{k(t)}(t)$ (Eq.~\eqref{phi_evolution_eq_mk}), the log-ratio becomes:
\begin{equation}
\ln \frac{P_\tau(\Gamma_\tau)}{\hat{Q}_\tau(\Gamma_\tau)} = \ln \frac{p_{k_0}(0)}{\hat{q}_{k_0}(0)} - \Re\{h^* \mathcal{J}(\Gamma_\tau)\} + \hat{\phi}_{k_N}(\tau) - \hat{\phi}_{k_0}(0).
\end{equation}
The exponent inside the expectation for $K_{\mathcal{J}}(h,\tau)$ is then:
\begin{equation}
\Re\{h^* \mathcal{J}(\Gamma_\tau) \}+\ln\frac{P_\tau(\Gamma_\tau)}{\hat{Q}_\tau(\Gamma_\tau)} = \ln \frac{p_{k_0}(0)}{\hat{q}_{k_0}(0)} + \hat{\phi}_{k_N}(\tau) - \hat{\phi}_{k_0}(0).
\end{equation}
Using the transversality condition $\hat{\phi}_{k_N}(\tau)=0$, this simplifies to $\ln (p_{k_0}(0)/\hat{q}_{k_0}(0)) - \hat{\phi}_{k_0}(0)$. This depends only on the initial state $k_0$. The expectation averages over the initial distribution $\hat{q}_{k_0}(0)$:
\begin{equation}
\begin{aligned}
K_{\mathcal{J}}(h,\tau) &= \ln \Biggl( \sum_{k_0} \hat{q}_{k_0}(0) \exp\Bigl( \ln \frac{p_{k_0}(0)}{\hat{q}_{k_0}(0)} - \hat{\phi}_{k_0}(0) \Bigr) \Biggr) \\
&= \ln \Biggl( \sum_{k_0} p_{k_0}(0) e^{-\hat{\phi}_{k_0}(0)} \Biggr).
\label{K_f_mk} 
\end{aligned}
\end{equation}
Comparing Eq.~\eqref{G_f_mk} and Eq.~\eqref{K_f_mk}, we observe that the optimization is equivalent to the cumulant generating function:
\begin{equation}
K_{\mathcal{J}}(h,\tau) = G_{\mathcal{J}}(h,\tau).
\end{equation}
This confirms that restricting the Donsker--Varadhan optimization to the family of modified Markov jump processes is sufficient to calculate the CGF.

\subsection{Max-Min Representation for the CGF}
\label{subsec:max_min_cgf_mkc} 
Having established $K_{\mathcal{J}}(h,\tau) = G_{\mathcal{J}}(h,\tau)$, we now derive the max-min representation by eliminating the explicit dependence on the tilted parameters $\bm{\zeta}$ from the saddle-point expression (Eq.~\eqref{K_with_L_final}).
\begin{equation}
K_{\mathcal{J}}(h,\tau) =  \sup_{\bm{q}}  \inf_{\bm{\phi}} \sup_{\bm{\zeta}}  \Biggl\{\int_0^\tau dt \, \ell(\bm{q}, \dot{\bm{q}}, \bm{\phi},  \bm{\zeta}) - \sum_k q_k(0)\ln \frac{q_k(0)}{p_k(0)} \Biggr\}.
\end{equation}
Substituting this optimal $\hat{\bm{\zeta}}$ back into the Lagrangian density $\ell(\bm{q}, \dot{\bm{q}}, \bm{\phi}, \bm{\zeta})$ (Eq.~\eqref{lagrangian_density_mk}), we obtain:
\begin{equation}
\ell(\bm{q}, \dot{\bm{q}}, \bm{\phi},  \hat{\bm{\zeta}}) = \sum_{k \neq m} W_{km}(t)q_m(t) \Bigl( e^{\Re\{h^*d_{km}(t)\}-\phi_k(t)+\phi_m(t)} - 1 \Bigr) + \sum_k \phi_k(t) \dot{q}_k(t).
\label{lagrangian_density_optimized_zeta} 
\end{equation}
Integrating this optimized Lagrangian density gives the final max-min representation for the CGF, where the optimization is now only over $\bm{q}$ and $\bm{\phi}$:
\begin{equation}
K_{\mathcal{J}}(h,\tau) =  \sup_{\bm{q}} \inf_{\bm{\phi}}\left\{ \int_{0}^{\tau} dt \left[
\sum_{k \neq m} W_{km}(t)q_m(t)\biggl( e^{\Re\{h^*d_{km}(t)\}-\phi_k(t)+\phi_m(t)} -1 \biggr)
+\sum_k \phi_k(t)\dot{q}_k(t)
\right]  - \sum_k q_k(0)\ln \frac{q_k(0)}{p_k(0)} \right\}.
\label{K_final_mk_concise} 
\end{equation}
Here, the maximization is over arbitrary probability vectors $\bm{q}(t)$, and the minimization over the auxiliary fields $\bm{\phi}(t)$ implicitly enforces the modified dynamics (Eq.~\eqref{EL_phi_mk}) associated with the optimal  $\hat{\bm{\zeta}}$. This is the starting point for deriving the variance expression.

\subsection{Variational Representation of the Variance}
\label{Variance_MK}
We derive a variational representation for the variance of the current observable, $\text{Var}(\mathcal{J})$, starting from the above max-min representation of the cumulant generating function,  
For small $h$, we expand the exponential:
\begin{equation}
\exp\Bigl( \Re\bigl\{h^* \mathcal{J}\bigr\} \Bigr)= 1 + \Re\bigl\{h^* \mathcal{J}\bigr\} + \frac{1}{2} \bigl(\Re\bigl\{h^* \mathcal{J}\bigr\}\bigr)^2 +  O(|h|^3).
\end{equation}
Using the definitions $h = h_r + i h_i$ and $\mathcal{J} = \mathcal{J}_r + i \mathcal{J}_i$ (where subscripts $r$ and $i$ denote real and imaginary parts), we obtain:
\begin{equation}
K_{\mathcal{J}}(h,\tau)
=  h_r \langle \mathcal{J}_r \rangle + h_i \langle \mathcal{J}_i \rangle + \frac{1}{2} \left[h_r^2 \text{Var}(\mathcal{J}_r) + h_i^2 \text{Var}(\mathcal{J}_i) + 2 h_r h_i \text{Cov}(\mathcal{J}_r, \mathcal{J}_i) \right] +  O(|h|^3).
\end{equation}
 The total variance of the complex observable is defined as $\text{Var}(\mathcal{J}) = \text{Var}(\mathcal{J}_r) + \text{Var}(\mathcal{J}_i)$. We can extract this variance by considering the quadratic terms in the expansion of the CGF for small $h$. Specifically, $\text{Var}(\mathcal{J})$ is the sum of the coefficients of $h_r^2$ and $h_i^2$ in the expansion of $2 K_{\mathcal{J}}(h,\tau)$.

We now expand the max-min representation of $K_{\mathcal{J}}(h,\tau)$ (Eq.~\eqref{K_final_mk_concise}) for small $h$. At $h = 0$, we have $q_k(t) = p_k(t)$ and $\phi_k(t) = 0$. 
For small $h$, we introduce first-order corrections using complex auxiliary functions $\psi_k(t)$ and $\rho_k(t)$:
\begin{align}
q_k(t) &:= p_k(t) \Bigl( 1 + \Re\bigl\{h^* \psi_k(t)\bigr\} + O(|h|^2) \Bigr) ,\\
\phi_k(t) &:= \Re\bigl\{h^* \rho_k(t)\bigr\} + O(|h|^2).
\end{align}
where $\psi_k(t)$ and $\rho_k(t)$ are complex functions.
We also impose the constraint $\langle \bm{\psi}(t) \rangle_p = \sum_k \psi_k(t) p_k(t) = 0$ for all $t$, because the modified probabilities must sum to unity: $\sum_k q_k(t) = 1$. We substitute these into the variational representation of $K_{\mathcal{J}}(h,\tau)$ (Eq.~\eqref{K_final_mk_concise}). Keeping terms related to $h_r^2$ and $h_i^2$, we find:
\begin{equation}
\begin{aligned}
\text{Var}(\mathcal{J}) =  \sup_{\bm{\psi}, \langle \bm{\psi} \rangle_p = 0}  \inf_{\bm{\rho}} \Biggl\{ \int_{0}^{\tau} dt  \Biggl[ &\sum_{k \neq m} W_{km}(t)p_m(t) \Bigl( \bigl|d_{km}(t) - \rho_k(t) + \rho_m(t)\bigr|^2  \\
&+ 2\Re\bigl\{ \psi_m(t) \bigl(d_{km}^*(t) - \rho_k^*(t) + \rho_m^*(t)\bigr) \bigr\} \Bigr) \\
&+ 2\sum_k \Re\bigl\{ \rho_k(t) \partial_t (\psi_k^*(t) p_k(t)) \bigr\} \Biggr] - \sum_k \abs{\psi_k(0)}^2 p_k(0) \Biggr\}.
\end{aligned}
\label{var_J_complex_1}
\end{equation}
To create a more symmetric structure and facilitate subsequent analysis, we perform a change of variables from $(\bm{\psi}, \bm{\rho})$ to $(\bm{\chi}, \bm{\eta})$ defined as:
\begin{align}
	\psi_k(t) &:= 2\chi_k(t), \\
	\rho_k(t) &:= \eta_k(t) - \chi_k(t). \label{change_of_variables_MK}
\end{align}
The constraint $\langle \bm{\psi} \rangle_p = 0$ becomes $\langle \bm{\chi} \rangle_p = \sum_k \chi_k(t) p_k(t) = 0$. Substituting these into the variational expression for the variance (Eq.~\eqref{var_J_complex_1}):
\begin{equation}
\begin{aligned}
\text{Var}(\mathcal{J}) = \sup_{\chi, \langle\chi \rangle_p = 0} \inf_{\eta}  \Biggl\{ \int_{0}^{\tau} dt  \Biggl[ &\sum_{k \neq m} W_{km}(t)p_m(t) \Bigl( \bigl|d_{km}(t) - \eta_k(t) + \eta_m(t)\bigr|^2 +  \bigl| \chi_k(t) - \chi_m(t) \bigr|^2 - 4\Re{\chi_m(\chi_k^* - \chi_m^*)}\\
&+ 2\Re\bigl\{ (\chi_k(t) + \chi_m(t)  )\bigl(d_{km}^*(t) - \eta_k^*(t) + \eta_m^*(t)\bigr) \bigr\} \Bigr) \\
&+ 4\sum_k \Re\bigl\{ \eta_k(t) \partial_t (\chi_k^*(t) p_k(t)) \bigr\} -2 \sum_k \abs{\chi_k(t)}^2\dot p_k(t) \Biggr] \\
&- 2\sum_k (\abs{\chi_k(0)}^2p_k(0) + \abs{\chi_k(\tau)p_k(\tau)}^2) \Biggr\}.
\end{aligned}
\end{equation}
Using the master equation (Eq.~\eqref{master_intro}), we can simplify the time derivative term:
\begin{equation}
\begin{aligned}
\sum_k \abs{\chi_k(t)}^2 \dot{p}_k(t) &= \sum_k  \abs{\chi_k(t)}^2 \sum_{m} \Bigl( W_{km}(t) p_m(t) - W_{mk}(t) p_k(t) \Bigr) \\
&= \sum_{k \neq m} ( \abs{\chi_k(t)}^2 -  \abs{\chi_m(t)}^2) W_{km}(t) p_m(t).
\end{aligned}
\end{equation}
We obtain the final variational representation for the variance:
\begin{equation}
\begin{aligned}
\text{Var}(\mathcal{J}) = \sup_{\chi, \langle\chi \rangle_p = 0} \inf_{\eta}   \Biggl\{ \int_{0}^{\tau} dt  \Biggl[ &\sum_{k \neq m} W_{km}(t)p_m(t) \Bigl( \bigl|d_{km}(t) - \eta_k(t) + \eta_m(t)\bigr|^2 - \bigl| \chi_k(t) - \chi_m(t) \bigr|^2\\ 
&+ 2\Re\bigl\{ (\chi_k(t) + \chi_m(t)  )\bigl(d_{km}^*(t) - \eta_k^*(t) + \eta_m^*(t)\bigr) \bigr\} \Bigr) \\
&+ 4\sum_k \Re\bigl\{ \eta_k(t) \partial_t(\chi_k^*(t) p_k(t) )\bigr\} \Biggr] \\
&- 2\sum_k (\abs{\chi_k(0)}^2p_k(0) + \abs{\chi_k(\tau)}^2p_k(\tau)) \Biggr\}.
\end{aligned}
\label{var_J_complex}
\end{equation}

\section{Long-Time iTUR for Continuous-Time Markov Jump Processes}
\label{long_time_iTUR_MK}

We focus on the long-time behavior of the current fluctuations,  $D_\infty := \lim_{\tau \to \infty} \frac{\text{Var}(\mathcal{J})}{2\tau}$. We assume the system reaches a steady state described by the distribution $\bm{\pi}$, the transition rates $W_{km}$ are time-independent, and the current increments $d_{km}$ are real and time-independent. In this limit, the auxiliary fields $\chi_k(t)$ and $\eta_k(t)$ in the variational representation of the variance (Eq.~\eqref{var_J_complex}) are expected to become time-independent, denoted $\chi_k$ and $\eta_k$. The terms involving time derivatives and boundary contributions vanish when divided by $\tau$. Taking the limit $\tau \to \infty$ and dividing by $2\tau$, Eq.~\eqref{var_J_complex} yields the variational representation for $D_\infty$:
\begin{equation}
\begin{aligned}
D_\infty = \frac{1}{2}  \sup_{\chi, \langle\chi \rangle_{\bm{\pi}} = 0} \inf_{\eta} \Biggl\{ \sum_{k \neq m} W_{km}\pi_m \Bigl( &(d_{km} - \eta_k + \eta_m)^2 - (\chi_k - \chi_m)^2 \\
&+ 2 (\chi_k + \chi_m)(d_{km} - \eta_k + \eta_m) \Bigr) \Biggr\}.
\end{aligned}
\label{D_infty_variational_start}
\end{equation}
To obtain an upper bound, we make the specific choice $\eta_k = 0$ for all $k$. This eliminates the infimum and provides an inequality:
\begin{equation}
\begin{aligned}
D_\infty &\leq \frac{1}{2} \sup_{\chi, \langle\chi \rangle_{\bm{\pi}} = 0} \Biggl\{ \sum_{k \neq m} W_{km}\pi_m \Bigl( d_{km}^2 - (\chi_k - \chi_m)^2 + 2 (\chi_k + \chi_m)d_{km} \Bigr) \Biggr\} \\
&= \frac{1}{2} \sum_{k \neq m} W_{km}\pi_m d_{km}^2 + \frac{1}{2} \sup_{\chi, \langle\chi \rangle_{\bm{\pi}} = 0} \Biggl\{ \sum_{k \neq m} W_{km}\pi_m \Bigl( 2d_{km}(\chi_k + \chi_m) - (\chi_k - \chi_m)^2 \Bigr) \Biggr\}.
\end{aligned}
\label{D_infty_eta_zero}
\end{equation}
We identify $D_0 = \frac{1}{2} \sum_{k \neq m} W_{km}\pi_m d_{km}^2$. The remaining supremum term can be bounded using the inequality $2AB - B^2 \leq A^2$, specifically:
\begin{equation}
\sum_{k \neq m} W_{km}\pi_m \Bigl(2d_{km}(\chi_k + \chi_m) - (\chi_k - \chi_m)^2 \Bigr) \leq \frac{\left( \sum_{k \neq m} d_{km}(\chi_k + \chi_m)W_{km}\pi_m \right)^2}{\sum_{k \neq m} (\chi_k - \chi_m)^2 W_{km}\pi_m }.
\label{chi_term_bound}
\end{equation}
Substituting this into Eq.~\eqref{D_infty_eta_zero} gives:
\begin{equation}
D_\infty \leq D_0 + \frac{1}{2} \sup_{\chi, \langle\chi \rangle_{\bm{\pi}} = 0} \Biggl\{ \frac{\left( \sum_{k \neq m} d_{km}(\chi_k + \chi_m)W_{km}\pi_m \right)^2}{\sum_{k \neq m} (\chi_k - \chi_m)^2 W_{km}\pi_m } \Biggr\}.
\label{D_infty_bound_ratio}
\end{equation}
Now, we introduce the spectral gap $\lambda$ of the symmetrized generator, given by its variational representation (Eq.~\eqref{MK_spectral_gap}):
\begin{equation}
\lambda = \inf_{\substack{\bm{\alpha} \\ \sum_k \alpha_k \pi_k = 0}} \frac{\frac{1}{2} \sum_{k \ne m} W_{km} \pi_k ( \alpha_k - \alpha_m)^2}{\sum_k \pi_k \alpha_k^2}.
\end{equation}
This implies that for any $\chi$ satisfying $\langle\chi\rangle_{\bm{\pi}}=0$, the denominator in Eq.~\eqref{D_infty_bound_ratio} is bounded below:
\begin{equation}
\sum_{k \neq m} (\chi_k - \chi_m)^2 W_{km}\pi_m \geq 2 \lambda \sum_k \chi_k^2 \pi_k.
\end{equation}
Using this lower bound in the denominator of Eq.~\eqref{D_infty_bound_ratio} yields:
\begin{equation}
D_\infty \leq D_0 + \sup_{\chi,\langle\chi\rangle_{\bm{\pi}}=0} \Biggl[ \frac{\left( \sum_{k \neq m} d_{km}(\chi_k + \chi_m)W_{km}\pi_m \right)^2}{4 \lambda \sum_k \chi_k^2 \pi_k } \Biggr].
\label{long_iTUR_start}
\end{equation}
This expression serves as our starting point for deriving the specific inverse thermodynamic uncertainty relations. To obtain the iTUR, we will subsequently find an upper bound for the numerator term $\left( \sum_{k \neq m} d_{km}(\chi_k + \chi_m)W_{km}\pi_m \right)^2$.

\subsection{Long-Time iTUR for Current Observables}
\label{L_iTUR_C}
We consider current-like observables ($d_{km} = -d_{mk}$).  
 In the steady state, the time average of the observable (the average current) is given by
\begin{equation}
j_{\text{st}} := \sum_{k \neq m} d_{km}W_{km}\pi_m.
\end{equation}

Due to the constraint $\langle\chi \rangle_{\bm{\pi}} = 0$, we have $\sum_k \chi_k \pi_k = 0$.  This allows us to add a term $- j_\text{st} \sum_k \chi_k \pi_k$ to the numerator of the bound without changing its value.  We rewrite the numerator of Eq.~\eqref{long_iTUR_start} as follows:
\begin{equation}
\left( \sum_{k \neq m} d_{km}(\chi_k + \chi_m)W_{km}\pi_m \right)^2 = \left(\sum_k \chi_k \left[\sum_{m\neq k} d_{km}\left( W_{km}\pi_m - W_{mk}\pi_k \right)  -2\pi_k j_\text{st}\right]\right)^2.
\label{numerator_rewrite_antisym}
\end{equation}
Applying the Cauchy--Schwarz inequality ($\left(\sum_k a_k b_k\right)^2 \le \left(\sum_k a_k^2/\pi_k\right) \left(\sum_k b_k^2 \pi_k\right)$ with $a_k = \chi_k \pi_k$ and $b_k = [\dots]/\pi_k$) to the right-hand side yields:
\begin{equation}
\begin{aligned}
\left( \dots \right)^2 &\leq \left( \sum_k \chi_k^2 \pi_k \right) \left( \sum_k \frac{1}{\pi_k} \left[ \sum_{m (\neq k)} d_{km} (W_{km}\pi_m - W_{mk}\pi_k) - 2 \pi_k j_{\text{st}} \right]^2 \right) \\
&= \left( \sum_k \chi_k^2 \pi_k \right) \left( \sum_k \frac{1}{\pi_k} \left[ \sum_{m (\neq k)} d_{km} (W_{km}\pi_m - W_{mk}\pi_k) \right]^2 - 4 j_{\text{st}}^2 \right).
\label{cauchy_schwarz_antisym}
\end{aligned}
\end{equation}
Apply a different Cauchy--Schwarz inequality to the inner sum:
\begin{equation}
\left( \sum_{m (\neq k)} d_{km} \left( W_{km}\pi_m - W_{mk}\pi_k \right) \right)^2 \leq  \left( \sum_{m (\neq k)}  \left( W_{km} \pi_m + W_{mk} \pi_k \right) \right) \left( \sum_{m (\neq k)} \frac{d_{km}^2\left( W_{km}\pi_m - W_{mk}\pi_k \right)^2}{W_{km} \pi_m + W_{mk} \pi_k} \right).
\label{cauchy_schwarz_antisym_3}
\end{equation}
Using the steady-state condition and the maximum escape rate $\kappa := \max_k \sum_{m(\neq k)} W_{mk}$, we obtain the bound for the first factor on the right:
\begin{equation}
\sum_{m (\neq k)} (W_{km} \pi_m + W_{mk} \pi_k) = 2 \pi_k \sum_{m (\neq k)} W_{mk} \leq 2 \pi_k \kappa.
\label{eq:escape_rate_bound_concise}\end{equation}
To bound the second factor in Eq. \eqref{cauchy_schwarz_antisym_3}, we use the following identity from Ref.~\cite{Vo_2022}:
\begin{equation}
(\alpha - \beta)^2 = \frac{\left[(\alpha-\beta) \ln \left(\frac{\alpha}{\beta}\right)\right]^2}{4}  \mfr{h}\left(\frac{(\alpha-\beta) \ln \left(\frac{\alpha}{\beta}\right)}{2(\alpha+\beta)}\right)^{-2}
\end{equation}
where $\mfr{h}$ is the inverse of the function $x \tanh(x)$. 
Therefore, the second factor can be rewritten in terms of a function $f(x,y) = \frac{x^2}{4y} \mfr{h}\left(\frac{x}{2y}\right)^{-2}$, which is known to be concave for $x, y > 0$:
\begin{equation}
\begin{aligned}
	\sum_{k \neq m} \frac{d_{km}^2\left( W_{km}\pi_m - W_{mk}\pi_k \right)^2}{W_{km} \pi_m + W_{mk} \pi_k} = 2  \sum_{k > m} f \bigg(d_{km}^2(W_{km}\pi_m-W_{mk}\pi_k) \ln \left(\frac{W_{km}\pi_m}{W_{mk}\pi_k}\right), d_{km}^2(W_{km}\pi_m+W_{mk}\pi_k)       \bigg)
\end{aligned}
\end{equation}
Applying Jensen's inequality:
\begin{equation}
\begin{aligned}
	\sum_{k \neq m} \frac{d_{km}^2\left( W_{km}\pi_m - W_{mk}\pi_k \right)^2}{W_{km} \pi_m + W_{mk} \pi_k} \leq  2f \bigg( \sum_{k > m}d_{km}^2(W_{km}\pi_m-W_{mk}\pi_k) \ln \left(\frac{W_{km}\pi_m}{W_{mk}\pi_k}\right),    \sum_{k > m}d_{km}^2(W_{km}\pi_m+W_{mk}\pi_k)       \bigg)
\end{aligned}
\end{equation}
We define the maximum squared increment and the entropy production rate:
\begin{align}
	d_{\text{max}}^2 &:= \max_{k, m} d_{km}^2,\\
	 \sigma_\text{st} &:=  \sum_{k > m} (W_{km} \pi_m - W_{mk} \pi_k) \ln \frac{W_{km} \pi_m}{W_{mk} \pi_k},
\end{align}
respectively.
Then the second factor in Eq. \eqref{cauchy_schwarz_antisym_3} is bounded by
\begin{equation}
\sum_{k \neq m} \frac{d_{km}^2\left( W_{km}\pi_m - W_{mk}\pi_k \right)^2}{W_{km} \pi_m + W_{mk} \pi_k}\leq   
2 f(d_{\text{max}}^2\sigma_\text{st},  2D_0).
\end{equation}
Combining these results:
\begin{equation}
\left( \sum_{k \neq m} d_{km}(\chi_k + \chi_m)W_{km}\pi_m \right)^2  \leq  4\left( \sum_k \chi_k^2 \pi_k \right) (\kappa f(d_{\text{max}}^2\sigma_\text{st},  2D_0) - j_{\text{st}}^2).
\label{before_itur_long}
\end{equation}
Substituting this bound back into the inequality for $D_\infty$ (Eq.~\eqref{long_iTUR_start}), we arrive at the long-time iTUR for current-like observables:
\begin{equation}
\label{itur_final_concave}
D_\infty \leq D_0 + \frac{ \kappa f(d_{\text{max}}^2\sigma_\text{st}, 2D_0) - j_{\text{st}}^2}{\lambda}.
\end{equation}
We note the property $f(x,y) \leq \min\{x/2, y\}$. Using $f(x,y) \le y$, this bound implies $D_\infty \leq D_0 + (2D_0 \kappa - j_{\text{st}}^2)/\lambda$, which is related to the bound $D_\infty \leq D_0 (1 + 2\kappa/\lambda)$ found in Ref.~\cite{Garrahan_PRL_2023}.

\subsection{Long-Time iTUR for Arbitrary Observables}
\label{L_iTUR_G}
We consider the increments $d_{km}$ associated with a jump from state $m$ to state $k$ to be arbitrary. 
First, we rewrite the sum within the square to group terms multiplying each auxiliary field component $\chi_k$:
\begin{equation}
\begin{aligned}
\left( \sum_{k \neq m} d_{km} (\chi_k + \chi_m) W_{km} \pi_m \right)^2 &= \left( \sum_{k \neq m} d_{km} \chi_k W_{km} \pi_m + d_{km} \chi_m W_{km} \pi_m \right)^2 \\
&= \left( \sum_k \chi_k \left( \sum_{m (\neq k)} (d_{mk} W_{mk}  + d_{km} W_{mk}^*) \pi_k \right) \right)^2\\
&= \left( \sum_k \chi_k \left( \pi_k\sum_{m (\neq k)} (d_{mk} W_{mk}  + d_{km} W_{mk}^* - 2  j_\text{st})  \right) \right)^2,
\end{aligned}
\end{equation}
where $W^*_{mk} := W_{km} \pi_m /\pi_k $.
Applying the Cauchy—Schwarz inequality, the numerator in Eq.\eqref{long_iTUR_start} can be upper bounded as:
\begin{equation}
	\left( \sum_{k \neq m} d_{km} (\chi_k + \chi_m) W_{km} \pi_m \right)^2  \leq  \left( \sum_k \left(\sum_{m (\neq k)} d_{mk} W_{mk}  + d_{km} W_{mk}^* - 2  j_\text{st} \right)^2 \pi_k \right)  \qty(\sum _{k }  \chi_k ^2\pi_k  ).
\end{equation}
Apply the Cauchy—Schwarz inequality to the inner sum over $m(\neq k)$.
\begin{equation}
\begin{aligned}
    \sum _{k} \pi_k \qty( \sum_{m(\neq k)}(d_{mk} W_{mk} + d_{km} W^*_{mk}))^2 &\leq \sum _{k} \pi_k \qty( \sum_{m(\neq k)}(W_{mk} + W^*_{mk})) \qty( \sum_{m(\neq k)}(d_{mk}^2 W_{mk} + d_{km}^2 W^*_{mk})) \\
    &\leq \sum_k \pi_k (2\kappa) \qty( \sum_{m(\neq k)}(d_{mk}^2 W_{mk} + d_{km}^2 W^*_{mk}) )\\
    &= 4 \kappa \sum_{k\neq m} d_{km}^2 W_{km} \pi_m = 8 \kappa D_0. 
    \label{eq:Yprime_bound_explained}
\end{aligned}
\end{equation}
Here, $\kappa = \max_k \sum_{m(\neq k)} W_{mk}$ is the maximum escape rate from any state, and $D_0 = \frac{1}{2}\sum_{k\neq m} d_{km}^2 W_{km} \pi_m$. 
Incorporating the average current $j_{\text{st}}$, we have 
\begin{equation}
	\left( \sum_{k \neq m} d_{km} (\chi_k + \chi_m) W_{km} \pi_m \right)^2  \leq 4\qty( 2 \kappa D_0    -  j_\text{st} ^2) \qty(\sum _{k }  \chi_k ^2\pi_k  )
\label{kappa_itur_before}
\end{equation}
We obtain the iTUR for all observables:
\begin{equation}
	D_\infty \leq D_0 \qty(1+  \frac{ 2 \kappa  }{\lambda})  - \frac{j_\text{st} ^2}{\lambda}.
\end{equation}

\section{Finite-Time iTUR for Continuous-time Markov Jump Processes}
\label{Finite_Time_TUR_MK}

\subsection{Upper Bound on the Spectral Density}
\label{Upper_bound_SP_MK}

 We are interested in the fluctuating observable  $d_{km}(t) = e^{i\omega t} d_{km}$, where $d_{km} = -d_{mk}$ are anti-symmetric, time-independent real numbers. The spectral density, $S(\omega)$, is defined as the long-time limit of its variance divided by the observation time:
\begin{equation}
S(\omega) := \lim_{\tau \to \infty} \frac{\text{Var}(\mathcal{J})}{\tau}.
\end{equation}
We assume the system reaches a stationary state. We replace time-dependent probabilities $p_k(t)$ with stationary counterparts $\pi_k$ and time-independent transition rates $W_{km}(t)$ with $W_{km}$. The variational representation for the spectral density is given by:
\begin{equation}
\begin{aligned}
S(\omega) = \lim_{\tau \to \infty} \frac{1}{\tau} \sup_{\substack{\chi \\ \langle\chi \rangle_{\bm{\pi}} = 0}} \inf_{\eta}  \Biggl\{ \int_{0}^{\tau} dt  \Biggl[ &\sum_{k \neq m} W_{km}\pi_m \Bigl( \bigl|d_{km}(t) - \eta_k(t) + \eta_m(t)\bigr|^2 -  \bigl| \chi_k(t) - \chi_m(t) \bigr|^2\\
&+ 2\Re\bigl\{ (\chi_k(t) + \chi_m(t)  )\bigl(d_{km}^*(t) - \eta_k^*(t) + \eta_m^*(t)\bigr) \bigr\} \Bigr) \\
&+ 4\sum_k \Re\bigl\{ \eta_k(t) \dot{\chi}_k^*(t) \pi_k \bigr\} \Biggr]  \Biggr\}.
\end{aligned}
\label{S_omega_long_time}
\end{equation}
We introduce specific forms for the auxiliary functions $\eta_k(t)$ and $\chi_k(t)$ to simplify the expression:\begin{align}
	\eta_k(t) &= e^{i\omega t} \eta_k,\\
	\chi_k(t) &= e^{i\omega t}\bar{\chi}_k(t),
\end{align}
where $\eta_k$ is a time-independent complex number.  
Substituting these into Eq.~ \eqref{S_omega_long_time} and using $d_{km}(t) = e^{i\omega t} d_{km}$, we obtain:
\begin{equation}
\begin{aligned}
S(\omega) \leq \lim_{\tau \to \infty} \frac{1}{\tau}  \sup_{\substack{\bar{\chi} \inf_{\eta}\\ \langle \bar{\chi}\rangle_{\bm{\pi}} = 0}}  \inf_{\eta}\Biggl\{ \int_{0}^{\tau} dt  \Biggl[ &\sum_{k \neq m} W_{km}\pi_m \Bigl( \bigl|d_{km} - \eta_k + \eta_m\bigr|^2  - \bigl| \bar{\chi}_k(t) - \bar{\chi}_m(t) \bigr|^2\\
&+ 2\Re\bigl\{\bar{\chi}_k(t) + \bar{\chi}_m(t) \bigl(d_{km} - \eta_k^* + \eta_m^*\bigr) \bigr\} \Bigr) \\
&+ 4\sum_k \Re\bigl\{ -i\omega\eta_k  \bar{\chi}_k^*(t) \pi_k \bigr\} \Biggr]  \Biggr\}.
\end{aligned}
\label{S_omega_simplified}
\end{equation}
We have used the fact that the term involving $\frac{1}{\tau} \int_0^\tau  \eta_k \dot{\bar{\chi}}_k^*(t) \pi_k dt$ vanishes as $\tau \to \infty$. 
Because the integrand in \eqref{S_omega_simplified} is now independent of time except for the possible time dependence of $\bar{\chi}_m(t)$, the optimal $\bar{\chi}$ must be time-independent.  
We can further decompose $\bar{\chi}_k$ into its real and imaginary parts: $\bar{\chi}_k = \chi'_k + i \chi''_k$, where $\chi'_k$ and $\chi''_k$ are real and time-independent. We also restrict $\eta_k$ to have only an imaginary part: $\eta_k = i \eta_k''$, where $\eta_k''$ is a real number.
We obtain an upper bound for the spectral density:
\begin{equation}
\begin{aligned}
S(\omega) \leq \sup_{\substack{\chi',\chi'' \\ \langle\chi' \rangle_{\bm{\pi}} =  \langle\chi'' \rangle_{\bm{\pi}} = 0}}  \inf_{\eta} &\Biggl\{  \sum_{k \neq m} W_{km}\pi_m \Bigl[  d_{km}^2 +  2d_{km}(\chi'_k +\chi'_m )  - (\chi'_k-\chi'_m)^2 \Bigr]\\
&\qquad + \sum_{k \neq m} W_{km}\pi_m  \Bigl[(\eta''_k - \eta''_m)^2 - 2(\eta''_k - \eta''_m)(\chi''_k + \chi''_m)   - (\chi''_k - \chi''_m)^2\Bigr] \\
&\qquad + 4\omega \sum_k \eta''_k \chi'_k  \pi_k \Biggr\}.
\end{aligned}
\label{S_omega_simplified_rewrite_v2_2}
\end{equation}

To further optimize, we introduce real-valued scaling parameters $a$, $b$, and $c$, and perform the following substitutions:
\begin{equation}
\begin{aligned}
\chi'_k &\rightarrow a \chi'_k, \\
\chi''_k &\rightarrow b \chi''_k,\\
\eta''_k &\rightarrow c \eta''_k.
\end{aligned}
\end{equation}
Substituting these scaled fields into Eq.~\eqref{S_omega_simplified_rewrite_v2_2} transforms the bound into a quadratic form in $a$, $b$, and $c$. Optimizing over these scaling parameters, we arrive at:
\begin{equation}
	\begin{aligned}
S(\omega) &\leq 2D_0 + \sup_{\substack{\chi',\chi'' \\ \langle\chi' \rangle_{\bm{\pi}} =  \langle\chi'' \rangle_{\bm{\pi}} = 0}}  \inf_{\eta} &\frac{\displaystyle\left[\sum_{k \neq m}W_{km} \pi_m d_{km} (\chi'_k+\chi'_m)\right]^2}
{\displaystyle \sum_{k \neq m}W_{km} \pi_m (\chi'_k-\chi'_m)^2
+\frac{4\omega^2\left[\sum_{k} \eta''_k \chi'_k \pi_k\right]^2}{\displaystyle \frac{\left[\sum_{k \neq m}W_{km} \pi_m (\eta''_k-\eta''_m)(\chi''_k+\chi''_m)\right]^2}{\displaystyle \sum_{k \neq m}W_{km} \pi_m (\chi''_k-\chi''_m)^2}+\sum_{k \neq m}W_{km} \pi_m (\eta''_k-\eta''_m)^2}}.
\end{aligned}
\end{equation}
where $D_0 = \frac{1}{2} \sum_{k \neq m} W_{km}\pi_m  d_{km}^2$.
The upper bound is monotonically increasing with the term
\begin{equation}
	Q = \displaystyle \frac{\left[\sum_{k \neq m}W_{km} \pi_m (\eta''_k-\eta''_m)(\chi''_k+\chi''_m)\right]^2}{\displaystyle \sum_{k \neq m}W_{km} \pi_m (\chi''_k-\chi''_m)^2},
\end{equation}
 which appears in the denominator of the bound. 
Using Eq.~\eqref{kappa_itur_before} and the definition of the symmetrized spectral gap $\lambda$, we have the upper bound on $Q$:
\begin{equation}
	Q \leq  \frac{ 2\kappa}{\lambda} \sum_{k \neq m} W_{km} \pi_m (\eta''_k-\eta''_m)^2 .
\end{equation}
Defining $\tilde{\lambda} = \sqrt{\lambda(\lambda +  2\kappa )}$.
Substituting this back into $S(\omega)$
\begin{equation}
	\begin{aligned}
S(\omega) &\leq 2D_0 + \sup_{\substack{\chi', \\ \langle\chi' \rangle_{\bm{\pi}} = 0}} \inf_{\eta} &\frac{\displaystyle\left[\sum_{k \neq m}W_{km} \pi_m d_{km} (\chi'_k+\chi'_m)\right]^2}
{\displaystyle \sum_{k \neq m}W_{km} \pi_m (\chi'_k-\chi'_m)^2
+\frac{4\omega^2 \lambda^2\left[\sum_{k} \eta''_k \chi'_k \pi_k\right]^2}{\tilde{\lambda}^2\sum_{k \neq m}W_{km} \pi_m (\eta''_k-\eta''_m)^2}}.
\end{aligned}
\end{equation}
Substituting $\eta''_k = \chi'_k$, we get:
\begin{equation}
	\begin{aligned}
S(\omega) &\leq 2D_0 + \sup_{\substack{\chi',  \\ \langle\chi' \rangle_{\bm{\pi}} = 0}}  \frac{\displaystyle\left[\sum_{k \neq m}W_{km}\pi_md_{km}(\chi'_k+\chi'_m)\right]^2}
{\displaystyle \sum_{k \neq m}W_{km}\pi_m(\chi'_k-\chi'_m)^2
+\frac{4\omega^2\lambda^2\left[\sum_{k}  (\chi'_k)^2\pi_k\right]^2}{\tilde{\lambda}^2\sum_{k \neq m}W_{km}\pi_m(\chi'_k-\chi'_m)^2}}.
\end{aligned}
\label{bound_with_eta_chi}
\end{equation}
Using Eq.~\eqref{before_itur_long}, the numerator is upper bounded by
\begin{equation}
	\qty(\sum_{k \neq m} d_{km} ( \chi'_k +  \chi'_m) W_{km} \pi_m)^2 \leq 4 \Bigl( \kappa f(d_{\text{max}}^2\sigma_\text{st},  2D_0) - j_{\text{st}}^2 \Bigr) \sum_k  (\chi')_k^2 \pi_k .
\label{given_inequality}
\end{equation}
We obtain:
\begin{equation}
S(\omega) \leq 2 D_0 +  \sup_{\substack{\chi',  \\ \langle\chi' \rangle_{\bm{\pi}} = 0}}  \frac{ 4\Bigl( \kappa f(d_{\text{max}}^2\sigma_\text{st},  2D_0) - j_{\text{st}}^2 \Bigr)\sum_k (\chi'_k)^2 \pi_k }
{ \sum_{k \neq m}W_{km}\pi_m(\chi'_k-\chi'_m)^2
+\frac{4\omega^2\lambda^2\left[\sum_{k}  (\chi'_k)^2\pi_k\right]^2}{\tilde{\lambda}^2\sum_{k \neq m}W_{km}\pi_m(\chi'_k-\chi'_m)^2}}.
\end{equation}
We define:
\begin{equation}
x := \frac{2\sum_k (\chi'_k)^2 \pi_k}{\sum_{k \neq m} W_{km} \pi_m (\chi'_k - \chi'_m)^2}.
\end{equation}
Since $x \leq \frac{1}{\lambda}$ by the variational definition of the spectral gap, we obtain:
\begin{equation}
 S(\omega) \leq 2D_0 + 2 \Bigl( \kappa f(d_{\text{max}}^2\sigma_\text{st},  2D_0) - j_{\text{st}}^2 \Bigr) \sup_{0 \leq x \leq \frac{1}{\lambda}} \frac{x}{1 + \frac{\omega^2\lambda^2}{\tilde{\lambda}^2}x^2}.
\end{equation}
We obtain the upper bound:
\begin{equation}
S(\omega) \leq 2D_0 + 2 \Bigl( \kappa f(d_{\text{max}}^2\sigma_\text{st},  2D_0) - j_{\text{st}}^2 \Bigr)
\begin{cases}
\displaystyle \frac{1}{\lambda}\frac{1}{1+(\omega/\tilde{\lambda})^2}, & \text{for } \omega < \tilde{\lambda}, \\[1mm]
\displaystyle \frac{\tilde{\lambda}}{2\lambda\omega}, & \text{for }  \omega \geq \tilde{\lambda}.
\end{cases}
\end{equation}

\subsection{Finite-time iTUR for Markov jump processes}
\label{Finite_time_iTUR_MK}
By the Wiener--Khinchin theorem, $D_\tau$ is expressed in terms of the spectral density:
\begin{equation}
D_\tau = \frac{1}{\pi} \int_0^\infty d\omega   S(\omega) \frac{1 - \cos(\omega \tau)}{\omega^2 \tau}.
\end{equation}
Substituting the upper bound for $S(\omega)$ into the integral expression for $D_\tau$, 
we arrive at the finite-time iTUR for Markov jump processes:
\begin{equation}
	D_\tau \leq D_0 + \frac{ g(\tilde{\lambda}\tau)}{\lambda} \Bigl( \kappa f(d_{\text{max}}^2\sigma_\text{st},  2D_0) - j_{\text{st}}^2 \Bigr).  
\end{equation}
Importantly, $ f(x,y)\leq \min\{x/2, y\}$ and $g(x) \leq 1$. As a consequence, our iTUR provides a stricter upper bound on $D_\tau$  compared to the previously established bound of $D_\tau \leq D_0 (1 + \frac{ 2\kappa}{\lambda})$ \cite{Garrahan_PRL_2023}.

\section{Variational Representation of the Spectral Gap of Symmetrized Operators}
\subsection{Spectral Gap of the Symmetrized Fokker-Planck Operator} 
\label{spectral_gap_overdamped}
The Fokker--Planck operator associated with Eq.~\eqref{langevin_intro} is defined as:
\begin{equation}\label{fp_operator}
\mathbb{L} \alpha(\bm{x}) := -\nabla \cdot \Bigl( \bm{A}(\bm{x}) \alpha(\bm{x}) \Bigr) + \nabla \cdot \Bigl( \bm{B} \nabla \alpha(\bm{x}) \Bigr),
\end{equation}
where  $\alpha(\bm{x})$ is any observable or probability density. Its adjoint $\mathbb{L}^\dagger$  with respect to the inner product:
\begin{equation}\label{adjoint_definition}
\langle\beta, \mathbb{L} \alpha \rangle_{p_\text{st}^{-1}} = \langle \mathbb{L}^\dagger\beta, \alpha \rangle_{p_\text{st}^{-1}},
\end{equation}
where the inner product is
\begin{equation}\label{inner_product}
\langle\beta, \alpha \rangle_{p_\text{st}^{-1}} := \int \frac{\beta(\bm{x}) \alpha(\bm{x})}{p_\text{st}(\bm{x}) }  d\bm{x}.
\end{equation}
The symmetrized operator, $\mathbb{L}^{\rm s} = (\mathbb{L} + \mathbb{L}^\dagger)/2$, simplifies spectral analysis due to its real eigenvalues and provides a lower bound on the real part of the spectral gap of $\mathbb{L}$.  In equilibrium, $\mathbb{L} = \mathbb{L}^{\rm s}$.  Explicitly, the symmetrized operator is
\begin{equation}\label{symmetrized_simplified}
\mathbb{L}^{\rm s} \alpha(\bm{x}) = \nabla \cdot \Bigl( \bm{B}  \alpha(\bm{x} ) \nabla \ln p_\text{st}+ \bm{B} \nabla \alpha(\bm{x}) \Bigr).
\end{equation}
This describes an equivalent equilibrium system with the same $p_\text{st}(\bm{x})$ and $\bm{B}$, and effective drift $\tilde{\bm{A}}(\bm{x}) = \bm{B} \nabla \ln p_\text{st}(\bm{x})$, resulting in zero local mean velocity, $\tilde{\bm{\nu}}_\text{st}(\bm{x}) = \tilde{\bm{A}}(\bm{x}) p_\text{st}(\bm{x}) - \bm{B} \nabla p_\text{st}(\bm{x}) = 0$.

The symmetrized spectral gap, $\lambda$, is the smallest non-zero eigenvalue of $-\mathbb{L}^{\rm s}$. The eigenvalue equation:
\begin{equation}
 \nabla\cdot(\tilde{\bm{A}}\psi-\bm{B}\nabla\psi)=\lambda\psi,
\end{equation}
 becomes:
\begin{equation}
-\nabla\cdot(p_\text{st}\bm{B}\nabla\chi)=\lambda\chi p_\text{st},
\end{equation}
 with $\psi=\chi p_\text{st}$.  Define:
\begin{equation}
\mathbb{H}\chi:=-\frac{\nabla\cdot(p_\text{st}\bm{B}\nabla\chi)}{p_\text{st}},
\end{equation}
so the eigenvalue equation is $\mathbb{H}\chi=\lambda\chi$.
To find the spectral gap, we can use the Rayleigh quotient, which provides a variational characterization of the eigenvalues. The Rayleigh quotient for $\mathbb{H}$ and any $\chi\neq 0$ is defined as
\begin{equation}
R(\chi):=\frac{\langle\chi,\mathbb{H}\chi\rangle_{p_\text{st}}}{\langle\chi,\chi\rangle_{p_\text{st}}},
\end{equation}
with the numerator given by
\begin{align*}
\langle\chi,\mathbb{H}\chi\rangle_{p_\text{st}}
&=\int \nabla\chi(\bm{x})\cdot\Bigl(p_\text{st}(\bm{x})\bm{B}\nabla\chi(\bm{x})\Bigr) d\bm{x}\\
&=\langle\nabla\chi\cdot\bm{B}\nabla\chi\rangle_{p_\text{st}}.
\end{align*}
The spectral gap can be expressed as
\begin{equation}
\lambda=\inf_{\chi\neq0} R(\chi)
=\inf_{\chi\neq0}\frac{\langle\nabla\chi\cdot\bm{B}\nabla\chi\rangle_{p_\text{st}}}{\langle\chi^2\rangle_{p_\text{st}}}
=\inf_{\chi}\left[\frac{\langle\nabla\chi\cdot\bm{B}\nabla\chi\rangle_{p_\text{st}}}{\text{Var}_{\text{st}}(\chi)}\right],
\label{O_spectral_gap}
\end{equation}

\subsection{Spectral Gap of the Symmetrized CTMJP Generator}
\label{Spectral_Gap_M}

The generator $\mathbb{L}$ has elements $\mathbb{L}_{mk} := W_{km} - \delta_{mk} \sum_j W_{jm}$.  The steady state $\bm{\pi}$ satisfies $\sum_m \pi_m  \mathbb{L}_{mk} = 0$.  The $\bm{\pi}$-weighted inner product is $\langle \bm{\alpha}, \bm{\beta} \rangle_{\bm{\pi}} := \sum_k \pi_k  \alpha_k  \beta_k^*$.  The adjoint $\mathbb{L}^\dagger$ satisfies $\langle \bm{\alpha}, \mathbb{L}\bm{\beta}\rangle_{\bm{\pi}} = \langle \mathbb{L}^\dagger \bm{\alpha}, \bm{\beta}\rangle_{\bm{\pi}}$, with elements $\mathbb{L}^\dagger_{ji} = \pi_i  \mathbb{L}_{ij} / \pi_j$.
To facilitate the analysis, particularly for characterizing the spectral gap, we introduce the symmetrized generator:
\begin{equation}
\mathbb{L}^{\rm s} := \frac{\mathbb{L} + \mathbb{L}^\dagger}{2}.
\end{equation}
This is reversible with respect to $\bm{\pi}$: $\pi_k  \mathbb{L}^{\rm s}_{km} = \pi_m  \mathbb{L}^{\rm s}_{mk}$.

We use a similarity transformation to a symmetric matrix $A$: $a_{km} := \pi_k^{1/2}  \mathbb{L}^{\rm s}_{km}  \pi_m^{-1/2}$.  Since $\mathbb{L}^{\rm s}$ is reversible, $A$ is symmetric.  $A$ has orthonormal eigenvectors $\bm{\psi}_i$: $A  \bm{\psi}_i = \mu_i  \bm{\psi}_i$, $\langle \bm{\psi}_i, \bm{\psi}_j \rangle = \delta_{ij}$.  Eigenfunctions of $\mathbb{L}^{\rm s}$ are $\bm{\varphi}_i := D^{-1/2}  \bm{\psi}_i$ (where $D_{kk}=\pi_k$), and are orthonormal with respect to the $\bm{\pi}$-weighted inner product: $\langle \bm{\varphi}_i, \bm{\varphi}_j \rangle_{\bm{\pi}} = \delta_{ij}$.
For ergodic processes, the largest eigenvalue of $\mathbb{L}^{\rm s}$ is $\mu_0 = 0$ (corresponding to steady state).  All others are negative. The spectral gap, $\lambda$, is the magnitude of the largest non-zero eigenvalue: $\lambda := -\mu_1 > 0$.

For any $\bm{\alpha}$ orthogonal to the steady state ($\langle \bm{\alpha}, \bm{\pi} \rangle = 0$), we can expand it in the eigenbasis: $\bm{\alpha} = \sum_{i\geq 1} \phi_i  \bm{\varphi}_i$. Then, $- \langle \mathbb{L}^{\rm s}  \bm{\alpha}, \bm{\alpha} \rangle_{\bm{\pi}} = \sum_{i\geq 1} (-\mu_i) |\phi_i|^2 \geq \lambda  \langle \bm{\alpha}, \bm{\alpha} \rangle_{\bm{\pi}}$. This leads to the variational representation:

\begin{equation}
\lambda = \inf_{\substack{\bm{\alpha} \\ \langle \bm{\alpha}, \bm{\pi} \rangle = 0}} \frac{-\langle \mathbb{L}^{\rm s}  \bm{\alpha}, \bm{\alpha} \rangle_{\bm{\pi}}}{\langle \bm{\alpha}, \bm{\alpha} \rangle_{\bm{\pi}}}= \inf_{\substack{\bm{\alpha} \\ \sum_k \alpha_k \pi_k = 0}} \frac{\frac{1}{2} \sum_{k \ne m} W_{km} \pi_m ( \alpha_k - \alpha_m)^2}{\sum_k \pi_k \alpha_k^2}.
\label{MK_spectral_gap}
\end{equation}

\subsection{Experimental Measurement of the Symmetrized Spectral Gap}
The Courant-Fischer-Weyl min-max principle states that $\lambda$ can be found by minimizing the Rayleigh quotient \cite{Horn_Johnson_1985}:
\[ \lambda = \inf_{\chi} \frac{\langle \chi, -\mathbb{L}_{\text{sym}} \chi \rangle_{\text{st}}}{\text{Var}_{\text{st}}(\chi)}, \]
where the infimum is taken over all suitable test functions $\chi$. The term in the numerator, $\mathcal{E}(\chi, \chi) := \langle \chi, -\mathbb{L}_{\text{sym}} \chi \rangle_{\text{st}}$, is known as the Dirichlet form. For overdamped Langevin dynamics, it has the simple representation:
\[ \mathcal{E}(\chi, \chi) = \langle \nabla \chi \cdot \mathbf{B} \nabla \chi \rangle_{\text{st}}. \]
This gives the practical expression for the Rayleigh quotient:
\[ \mathcal{R}[\chi] = \frac{\langle \nabla \chi \cdot \mathbf{B} \nabla \chi \rangle_{\text{st}}}{\text{Var}_{\text{st}}(\chi)}. \]
In practice, this infimum is approximated by minimizing the quotient over a finite-dimensional subspace spanned by a set of $M$ basis functions $\{\chi_i(\bm{x})\}_{i=1}^M$, which are chosen to capture the system's slow dynamics and are constructed to have zero mean. A trial function is formed as a linear combination $\chi(\bm{x}) = \sum_{i=1}^M c_i \chi_i(\bm{x}) = \bm{c}^{\top} \bm{\chi}(\bm{x})$. This transforms the Rayleigh quotient into a ratio of quadratic forms:
\[
\mathcal{R}(\bm{c}) = \frac{\bm{c}^{\top} K \bm{c}}{\bm{c}^{\top} S \bm{c}},
\]
where $S$ and $K$ are the $M \times M$ {mass matrix} and {stiffness matrix}, respectively:
\begin{align}
S_{ij} &:= \langle\chi_i \chi_j\rangle_{\text{st}}, \\
K_{ij} &:= \langle\nabla\chi_i \cdot \mathbf{B}\nabla\chi_j\rangle_{\text{st}}.
\end{align}
Minimizing $\mathcal{R}(\bm{c})$ is equivalent to solving the generalized eigenvalue problem $K\bm{v} = \lambda_{\mathrm{est}} S\bm{v}$. The smallest positive eigenvalue, $\lambda_{\mathrm{est}}^{(M)}$, provides a monotonically converging upper bound: $\lambda \le \dots \le \lambda_{\mathrm{est}}^{(M+1)} \le \lambda_{\mathrm{est}}^{(M)}$. If the basis is complete, $\lim_{M\to\infty} \lambda_{\mathrm{est}}^{(M)} = \lambda$.

In an experimental setting, the matrices $S$ and $K$ are estimated from a long trajectory of $N$ data points $\{\bm{x}_n\}$ by replacing the ensemble averages with time averages (denoted by a hat) and using an estimate of the diffusion matrix $\hat{\mathbf{B}}$:
\begin{align}
\hat{S}_{ij} &:= \frac{1}{N} \sum_{n=0}^{N-1} \chi_i(\bm{x}_n) \chi_j(\bm{x}_n), \\
\hat{K}_{ij} &:= \frac{1}{N} \sum_{n=0}^{N-1} (\nabla\chi_i(\bm{x}_n))^{\top} \hat{\mathbf{B}}(\bm{x}_n) (\nabla\chi_j(\bm{x}_n)).
\end{align}
The resulting empirical generalized eigenvalue problem $\hat{K}\bm{v} = \hat{\lambda}_{\mathrm{est}} \hat{S}\bm{v}$ yields the estimate $\hat{\lambda}_{\mathrm{est}}^{(M)}$ as its smallest positive eigenvalue.